\documentclass[11pt]{article}
\usepackage{lmodern}
\usepackage{lmodern}
\usepackage[T1]{fontenc}
\usepackage{geometry}
\usepackage{color}
\usepackage{array}
\usepackage{float}
\usepackage{mathtools}
\usepackage{multirow}
\usepackage{amsmath}
\usepackage{amsthm}
\usepackage{amssymb}
\usepackage{graphicx}
\usepackage[authoryear,round,nonamebreak]{natbib}
\usepackage[unicode=true,
 bookmarks=false,
 breaklinks=false,pdfborder={0 0 1},backref=section,colorlinks=false]
 {hyperref}

\makeatletter

\providecommand{\tabularnewline}{\\}
\floatstyle{ruled}
\newfloat{algorithm}{tbp}{loa}
\providecommand{\algorithmname}{Algorithm}
\floatname{algorithm}{\protect\algorithmname}

\theoremstyle{plain}
\newtheorem{thm}{\protect\theoremname}[section]
\theoremstyle{plain}
\newtheorem{lem}[thm]{\protect\lemmaname}

\@ifundefined{date}{}{\date{}}

\allowdisplaybreaks

\makeatother

\providecommand{\lemmaname}{Lemma}
\providecommand{\theoremname}{Theorem}

\begin{document}
\global\long\def\E{\mathbb{E}}%
\global\long\def\F{\mathcal{F}}%
\global\long\def\S{\mathbf{S}}%
\global\long\def\T{\mathbf{T}}%
\global\long\def\O{\mathbf{O}}%
\global\long\def\A{\mathbf{A}}%
\global\long\def\B{\mathbf{B}}%
\global\long\def\X{\mathbf{X}}%
\global\long\def\Y{\mathbf{Y}}%
\global\long\def\supp{\mathbf{supp}}%

\title{Online and Streaming Algorithms for Constrained $k$-Submodular Maximization}
\author{Fabian Spaeh\thanks{Department of Computer Science, Boston University. ${\tt fspaeh@bu.edu}$}
\and Alina Ene\thanks{Department of Computer Science, Boston University. ${\tt aene@bu.edu}$}
\and Huy L. Nguyen\thanks{Khoury College of Computer Sciences, Northeastern University. ${\tt hu.nguyen@northeastern.edu}$}}
\maketitle
\begin{abstract}
Constrained $k$-submodular maximization is a general framework that
captures many discrete optimization problems such as ad allocation,
influence maximization, personalized recommendation, and many others.
In many of these applications, datasets are large or decisions need
to be made in an online manner, which motivates the development of
efficient streaming and online algorithms. In this work, we develop
single-pass streaming and online algorithms for constrained $k$-submodular
maximization with both monotone and general (possibly non-monotone)
objectives subject to cardinality and knapsack constraints. Our algorithms
achieve provable constant-factor approximation guarantees which improve
upon the state of the art in almost all settings. Moreover, they are
combinatorial and very efficient, and have optimal space and running
time. We experimentally evaluate our algorithms on instances for ad
allocation and other applications, where we observe that our algorithms
are efficient and scalable, and construct solutions that are comparable
in value to offline greedy algorithms.

\end{abstract}

\section{Introduction}

We develop algorithms for maximizing a $k$-submodular function $f$
subject to cardinality or knapsack constraints. $k$-Submodular functions
capture the property of diminishing returns under an allocation of
elements from a ground set $V$ to $k$ parts. Specifically, we are
trying to find $k$ disjoint subsets $(S_{1},\dots,S_{k})$ of $V$
such that $f(S_{1},\dots,S_{k})$ is maximized. Each part $a\in\{1,\dots,k\}$
has a specified budget $n_{a}$ and we are only allowed to allocate
at most $\left|S_{a}\right|\le n_{a}$ items to it. 

This problem is a generalization of submodular maximization under
a cardinality constraint, and for $k=1$ both problems are identical.
However, $k$-submodular functions are able to capture several important
applications, such as ad allocation. In this problem, ad impressions
arrive online which we have to allocate immediately to one of $k$
advertisers \citep{feldman09}. Advertisers are willing to pay for
at most $n_{a}$ ad impressions (specified in advance via a contract),
but are happy to receive more impressions. The advertising platform
tries to make an allocation that maximizes advertiser satisfaction,
which could be measured through user exposure, which is naturally
submodular.

Another important application is in personalized recommendation, which
motivates the study of general objectives. Consider, for example,
a movie recommender system where users specify a set of genres they
are interested in. The recommender system then tries to find a set
of representative movies from all genres (note that a movie might
belong to multiple genres). A $k$-submodular function can measures
the coverage and diversity of a set of recommendations, e.g. through
movie dissimilarity that is derived from past ratings \citep{mirzasoleiman16}.
Specifically, given a complete graph of movie dissimilarities, we
want to find a set which cuts the graph such that dissimilarity across
the cut (coverage) is minimized and the dissimilarity inside the set
(diversity) is maximized. Related tasks such as document summarization
\citep{lin11} or image summarization \citep{gomes10} can be modeled
through similar objectives. For additional motivation on influence
maximization, sensor placement, and video summarization, we refer
the reader to the works of \citet{ohsaka15} and \citet{feldman18}. 

The datasets used in all of these applications are typically large
and even offline greedy algorithms are not practical. Furthermore,
applications such ad allocation require us to make decisions in an
online fashion as the impressions arrive. We thus develop algorithms
for the streaming and online settings where we inspect each item only
once and allocate it immediately. Our algorithms achieve provable
constant-factor approximation guarantees, and optimal space and running
time. Moreover, they are combinatorial and very efficient. Our algorithms
also apply to the related but more structured problem of submodular
maximization with a partition matroid constraint. Many problems, such
as ad allocation with linear valuations, can also be modeled through
a partition matroid. 

\subsection{Our Contributions and Techniques}

\begin{table*}
\caption{\label{tb:results-ksubmod}Comparison of algorithms for $k$-submodular
maximization with cardinality constraints. We let $n=\min_{a\in[k]}n_{a}$
denote the minimum budget, $r=\sum_{a\in[k]}n_{a}$ the total budget,
and $m=\left|V\right|$.}

\begin{centering}
{\scriptsize{}\medskip{}
}{\scriptsize\par}
\par\end{centering}
\centering{}{\scriptsize{}}%
\begin{tabular}{|c|c|c|c|c|c|}
\hline 
\textbf{\scriptsize{}Objective} & \textbf{\scriptsize{}Reference} & \textbf{\scriptsize{}Setting} & \textbf{\scriptsize{}Approx.} & \textbf{\scriptsize{}Time} & \textbf{\scriptsize{}Space}\tabularnewline
\hline 
\multirow{4}{*}{{\scriptsize{}monotone}} & \multirow{2}{*}{{\scriptsize{}\citep{ene22}}} & \multirow{2}{*}{{\scriptsize{}online,streaming}} & {\scriptsize{}$\geq\frac{1}{4}$} & \multirow{2}{*}{{\scriptsize{}$O(mk)$}} & \multirow{2}{*}{{\scriptsize{}$O(r)$}}\tabularnewline
 &  &  & {\scriptsize{}$\approx0.2953$ as $n\to\infty$} &  & \tabularnewline
\cline{2-6} \cline{3-6} \cline{4-6} \cline{5-6} \cline{6-6} 
 & \textbf{\scriptsize{}Theorem}{\scriptsize{} \ref{thm:ksubmod}} & \multirow{2}{*}{{\scriptsize{}online,streaming}} & {\scriptsize{}$\ge\frac{1}{4}$} & \multirow{2}{*}{{\scriptsize{}$O(mk)$}} & \multirow{2}{*}{{\scriptsize{}$O(r)$}}\tabularnewline
 & \textbf{\scriptsize{}(This paper)} &  & {\scriptsize{}$\approx0.3178$ as $n\to\infty$} &  & \tabularnewline
\hline 
\multirow{3}{*}{{\scriptsize{}general}} & {\scriptsize{}\citep{xiao23}} & {\scriptsize{}offline} & {\scriptsize{}$\frac{1}{4+\max_{a}n_{a}}$} & {\scriptsize{}$O(rmk)$} & {\scriptsize{}$O(r)$}\tabularnewline
\cline{2-6} \cline{3-6} \cline{4-6} \cline{5-6} \cline{6-6} 
 & \textbf{\scriptsize{}Theorem}{\scriptsize{} \ref{thm:ksubmod-nonmon}} & \multirow{2}{*}{{\scriptsize{}online,streaming}} & {\scriptsize{}$\ge\frac{1}{8}$} & \multirow{2}{*}{{\scriptsize{}$O(mk)$}} & \multirow{2}{*}{{\scriptsize{}$O(r)$}}\tabularnewline
 & \textbf{\scriptsize{}(This paper)} &  & {\scriptsize{}$\approx0.1589$ as $n\to\infty$} &  & \tabularnewline
\hline 
\end{tabular}{\scriptsize\par}
\end{table*}
\begin{table*}
\caption{\label{tb:results-submod-partition}Comparison of algorithms for submodular
maximization with a partition matroid. We set $n$, $r$, and $m$
as in Table \ref{tb:results-ksubmod}.}

\centering{}{\scriptsize{}\medskip{}
}%
\begin{tabular}{|c|c|c|c|c|c|}
\hline 
\textbf{\scriptsize{}Objective} & \textbf{\scriptsize{}Reference} & \textbf{\scriptsize{}Setting} & \textbf{\scriptsize{}Approx.} & \textbf{\scriptsize{}Time} & \textbf{\scriptsize{}Space}\tabularnewline
\hline 
\multirow{6}{*}{{\scriptsize{}monotone}} & \multirow{2}{*}{{\scriptsize{}\citep{ene22}}} & {\scriptsize{}online,streaming} & {\scriptsize{}$\geq\frac{1}{4}$} & \multirow{2}{*}{{\scriptsize{}$O(m)$}} & \multirow{2}{*}{{\scriptsize{}$O(r)$}}\tabularnewline
 &  & {\scriptsize{}discrete} & {\scriptsize{}$\approx0.3178$ as $n\to\infty$} &  & \tabularnewline
\cline{2-6} \cline{3-6} \cline{4-6} \cline{5-6} \cline{6-6} 
 & \multirow{2}{*}{{\scriptsize{}\citep{feldman22}}} & {\scriptsize{}streaming} & \multirow{2}{*}{{\scriptsize{}$\approx0.3178-\epsilon$}} & \multirow{2}{*}{{\scriptsize{}$O\left(\frac{mr\log^{2}r}{\epsilon^{2}}\right)$}} & \multirow{2}{*}{{\scriptsize{}$O\left(r\log^{O(1)}m\right)$}}\tabularnewline
 &  & {\scriptsize{}continuous} &  &  & \tabularnewline
\cline{2-6} \cline{3-6} \cline{4-6} \cline{5-6} \cline{6-6} 
 & \textbf{\scriptsize{}Theorem}{\scriptsize{} \ref{thm:partition}} & {\scriptsize{}online,streaming} & {\scriptsize{}$\ge\frac{1}{4}$} & \multirow{2}{*}{{\scriptsize{}$O(m)$}} & \multirow{2}{*}{{\scriptsize{}$O(r)$}}\tabularnewline
 & \textbf{\scriptsize{}(This paper)} & {\scriptsize{}discrete} & {\scriptsize{}$\approx0.3178$ as $n\to\infty$} &  & \tabularnewline
\hline 
\multirow{6}{*}{{\scriptsize{}general}} & \multirow{2}{*}{{\scriptsize{}\citep{feldman18}}} & {\scriptsize{}online, streaming} & \multirow{2}{*}{{\scriptsize{}$\approx0.1716$}} & \multirow{2}{*}{{\scriptsize{}$O(rm)$}} & \multirow{2}{*}{{\scriptsize{}$O(r)$}}\tabularnewline
 &  & {\scriptsize{}discrete} &  &  & \tabularnewline
\cline{2-6} \cline{3-6} \cline{4-6} \cline{5-6} \cline{6-6} 
 & \multirow{2}{*}{{\scriptsize{}\citep{feldman22}}} & {\scriptsize{}streaming} & \multirow{2}{*}{{\scriptsize{}$\approx0.1921$}} & \multirow{2}{*}{{\scriptsize{}$O\left(\frac{mr\log^{2}r}{\epsilon^{2}}\right)$}} & \multirow{2}{*}{{\scriptsize{}$O\left(r\log^{O(1)}m\right)$}}\tabularnewline
 &  & {\scriptsize{}continuous} &  &  & \tabularnewline
\cline{2-6} \cline{3-6} \cline{4-6} \cline{5-6} \cline{6-6} 
 & \textbf{\scriptsize{}Theorem}{\scriptsize{} \ref{thm:partition-nonmonotone}} & {\scriptsize{}online,streaming} & {\scriptsize{}$\ge0.175$} & \multirow{2}{*}{{\scriptsize{}$O(m)$}} & \multirow{2}{*}{{\scriptsize{}$O(r)$}}\tabularnewline
 & \textbf{\scriptsize{}(This paper)} & {\scriptsize{}discrete} & {\scriptsize{}$\approx0.1921$ as $n\to\infty$} &  & \tabularnewline
\hline 
\end{tabular}
\end{table*}

For monotone $k$-submodular objectives, we design a new algorithm
with an improved approximation guarantee (Table \ref{tb:results-ksubmod}).
Our algorithm is inspired by the works of \citet{feldman09} for linear
objectives and \citet{ene22} for $k$-submodular functions. As in
both of those works, we use a threshold for each part that decides
the allocation of a new item and evolves over time. The thresholds
used by \citet{ene22} depend on all previous items (even items that
were already disposed). We use stronger thresholds, formed as a linear
combination of the marginal gains of currently allocated items and
exponentially increasing coefficients. This is inspired by the exponential
averaging approach of \citet{feldman09}, but requires new techniques
for submodular objectives. Our analysis is a significant departure
from both prior works.  We also use a novel analytical approach to
choose the coefficients that go into the thresholds, tailored to the
specific budget in each part. This allows us obtain better approximation
guarantees in challenging settings such as when budgets are imbalanced.
This was not done in previous works but is important for applications
such as ad allocation. We provide a more detailed comparison in Section
\ref{subsec:ksubmod-mon-algo}. 

For general $k$-submodular objectives, we design novel algorithms
with provable constant factor approximation guarantees (Table \ref{tb:results-ksubmod}).
Prior to our work, constant factor approximation guarantees were not
known even in the offline setting. Standard techniques developed for
submodular functions such as sub-sampling do not apply to $k$-submodular
functions, and new techniques are needed. We are able to leverage
properties of $k$-submodular functions to obtain constant-factor
approximation guarantees. For the related but more structured problem
of submodular maximization with a partition matroid constraint, we
close the gap between the approximation ratios for discrete and continuous
algorithms (Table \ref{tb:results-submod-partition}). 

Rethinking our algorithm for cardinality constraints, we are able
to derive a generalization to packing (knapsack) constraints, another
important constraint setting. We give the first algorithms with constant
factor approximation guarantees when the item sizes are small compared
to the budgets, which is a relevant setting for applications such
as ad allocation. Our work readily extends to the setting where we
have a common budget for all parts. Here, we obtain improved running
time and space over previous streaming algorithms which store multiple
solutions in memory and are thus not suitable for the online setting.
Moreover, we obtain improved approximation guarantees in the online
setting. 

Our algorithms achieve provable constant factor approximation guarantees
that improve upon the state of the art in all settings we consider,
with the exception of monotone submodular maximization with a partition
matroid constraint where we match the best known guarantees. Moreover,
the approximation guarantees improve as the budgets increase. Additionally,
all of our algorithms are combinatorial and very efficient, and have
optimal space and running time.

\subsection{Additional Related Work}

\paragraph{Monotone $k$-submodular}

 \citet{nguyen20} generalize the threshold greedy approach of \citet{badanidiyuru14}
to $k$-submodular maximization under a common cardinality constraint
of size $r$, that works by guessing the value of the optimum solution.
Their method achieves a near-optimal $\frac{1}{2}-\epsilon$ approximation,
but keeps multiple solutions in memory, which requires space $O(\frac{r\log r}{\epsilon})$
and is not suited for the online setting.

\paragraph{Non-monotone $k$-submodular}

The only prior work that considers general $k$-submodular maximization
under individual cardinality constraints is due to \citet{xiao23}.
Their offline greedy approach obtains a $\frac{1}{4+\max_{a}n_{a}}$
approximation, which decreases with the maximum budget. Furthermore,
\citet{nguyen20} show that for non-monotone objectives subject to
a common cardinality constraint, their threshold greedy algorithm
achieves a $\frac{1}{3}-\epsilon$ approximation. However, their approach
requires a total enumeration over all partial solutions, and thus
requires $O(\frac{r\log r}{\epsilon})$ time to output a solution.

\paragraph{Partition matroid}

For general matroid constraints, \citet{feldman22} give a streaming
algorithm based on the continuous extension of a submodular function.
Their algorithm maintains multiple solutions at the same time and
is therefore not suited for the online setting. It turns out that
for partition matroids, our discrete algorithms achieve the same guarantees
when the minimum budget tends to infinity. \citet{feldman22} further
show how to use multiple passes to essentially recover the $1-\frac{1}{e}$
approximation guarantee of the offline setting. A discrete algorithm
for general objectives under more general $p$-matchoid constraints
was given by \citet{feldman18}. Their algorithm sub-samples items,
which is also a technique we employ. For the more specialized but
important constraint of a partition matroid, we obtain a slightly
improved approximation ratio. 

\paragraph{Knapsack}

We consider the setting where item sizes are small compared to the
budgets, which is necessary to achieve a constant-factor approximation
ratio \citep{feldman09} and well-motivated from applications such
as ad allocation. We are the first to obtain a guarantee for individual
knapsack constraints for $k$-submodular maximization. For a common
knapsack constraint, \citet{pham22} develop single and multi-pass
streaming algorithms for monotone $k$-submodular maximization. Their
single pass algorithm achieves an approximation ratio of $\frac{1}{10}$
while their multi-pass algorithm achieves $\frac{1}{4}-\epsilon$
in $O(\frac{1}{\epsilon})$ rounds. \citet{tang22} use an offline
greedy algorithm to obtain an approximation ratio of $\frac{1}{2}\left(1-\frac{1}{e}\right)$.
We are able to improve upon both guarantees when the size each item
is sufficiently small. For a submodular objective under a $k$-sparse
packing constraint, \citet{chan17} give a polynomial time online
algorithm that maintains a fractional solution.

\section{Preliminaries}

\paragraph{$k$-Submodular functions}

Let $(k+1)^{V}\coloneqq\left\{ (X_{1},\dots,X_{k}):X_{a}\subseteq V,X_{a}\cap X_{b}=\emptyset\text{ for all }a,b\in[k]\right\} $
be the set of all $k$-tuples of disjoint subsets, where $[k]\coloneqq\{1,2,\dots,k\}$.
For two $k$-tuples $\X,\Y\in(k+1)^{V}$, we define $\supp(\X)\coloneqq X_{1}\cup\cdots\cup X_{k}$
and write $\X\preceq\Y$ if $X_{a}\subseteq Y_{a}$ for all $a\in[k]$.
We also define the intersection $\X\sqcap\Y$ of two $k$-tuples through
$(\X\sqcap\Y)_{a}\coloneqq X_{a}\cap Y_{a}$ for all $a\in[k]$, and
the union as $(\X\sqcup\Y)_{a}\coloneqq(X_{a}\cup Y_{a})\setminus\bigcup_{b\not=a}(X_{b}\cup Y_{b})$.
Given these operations, we say $f$ is $k$-submodular if
\[
f(\X)+f(\Y)\ge f(\X\sqcap\Y)+f(\X\sqcup\Y)
\]
for all $\X,\Y\in(k+1)^{V}$. The function $f$ is monotone if $f(\X)\le f(\Y)$
if $\X\preceq\Y$. We define the marginal gain of adding element $t$
to part $a$ of $\X$ as
\[
\Delta_{t,a}f(\X)\coloneqq f\left((X_{1},\dots,X_{a}\cup\{t\},\dots,X_{k})\right)-f(\X).
\]
To obtain a notion of diminishing returns, we say that $f$ is orthant
submodular if
\[
\Delta_{t,a}f(\X)\ge\Delta_{t,a}f(\Y)
\]
for all $\X\preceq\Y$ with $t\notin\supp(\Y)$. Furthermore, $f$
is pairwise monotone if
\[
\Delta_{t,a}f(\X)+\Delta_{t,b}f(\X)\ge0
\]
for all $t\notin\supp(\X)$ and $a\not=b$. We know that $f$ is $k$-submodular
if and only if $f$ is orthant submodular and pairwise monotone \citep{ward16}. 

\paragraph{Problem definition}

In $k$-submodular maximization, we are given a $k$-submodular function
$f\colon(k+1)^{V}\to\mathbb{R}_{+}$ and budgets $n_{1},\dots,n_{k}$
for every part. The goal is to find a solution that maximizes $f$
while allocating at most $n_{a}$ items to every part $a$. We define
the optimum solution as $\S^{*}\coloneqq\arg\max\left\{ f(\S):\S\in(k+1)^{V}\text{ with }\left|S_{a}\right|\le n_{a}\text{ for all }a\in[k]\right\} $.
A related problem is submodular maximization with a partition matroid.
Here, we are given a submodular function $f\colon2^{V}\to\mathbb{R}_{+}$,
and a partition matroid ${\cal P}=(P_{1},\dots,P_{k})$ with budgets
$n_{1},\dots,n_{k}$. A set $S$ is an independent set of ${\cal P}$
if $\left|S\cap P_{a}\right|\le n_{a}$ for all $a\in[k]$. The goal
is to find an independent set $S$ maximizing $f$. We define $S^{*}\coloneqq\arg\max\left\{ f(S):S\subseteq V\text{ is an independent set of }{\cal P}\right\} $.
We consider both monotone and general (possibly non-monotone) objectives
in both settings. 

We consider both problems in the (single-pass) streaming model. Here,
all items of $V$ arrive in an arbitrary (possibly adversarial) order
and the task is to generate a solution to the problem at the end of
the stream, while using as little space as possible. Our algorithms
simultaneously apply to the online setting with free disposal \citep{feldman09}.
Here, items also arrive one at a time, but now we are required to
maintain a single solution to the problem after each arrival. Additionally,
we are only allowed to add the arriving item to the solution, or dispose
(i.e. remove) an item that is in the current solution.

We also consider the extension to packing constraints where we have
sizes $u_{t,a}$ for each item $t$ and each part $a$, and we defer
the definition to the appendix. 

\paragraph{Examples of $k$-submodular functions}

We now give examples of $k$-submodular functions that arise in the
applications to ad allocation and recommender systems discussed in
the introduction and our experimental evaluation. The well-studied
submodular welfare problem is a special case of $k$-submodular maximization.
Here we have a set $V$ of items and $k$ agents with valuation functions
$g_{a}:2^{V}\to\mathbb{R}_{+}$, and the goal is to allocate each
item to at most one agent to maximize the social welfare $f(\X)\coloneqq\sum_{a}g_{a}(X_{a})$,
where $X_{a}$ is the set of items allocated to $a$. If the functions
$g_{a}$ are submodular then $f$ is orthant submodular. If the $g_{a}$'s
are monotone, then $f$ is monotone. Such instances appear for ad
allocation where advertiser satisfaction can be modeled through a
function $g_{a}$ that expresses, for example, the coverage of an
ad campaign. If $g_{a}=g$ where $g$ is a submodular function that
is \emph{symmetric} (i.e., $g(X)=g(V\setminus X)$ for all $X\subseteq V$),
then $f$ is a general $k$-submodular function (i.e., it is pairwise
monotone and orthant submodular). Such instances arise from graph
cut functions in applications such as recommender systems. Other examples
of $k$-submodular functions include generalizations of influence
maximization and sensor placement that were introduced in the work
\citet{ohsaka15}.

\paragraph{Outline}

In the main body, we present our algorithms for $k$-submodular maximization
and an analysis overview. We defer the full analysis to the appendix
(Section \ref{subsec:ksubmod-appendix} for monotone and Sections
\ref{subsec:k-submod-nonmon-appendix} and \ref{subsec:k-submod-any-max-budget}
for general objectives). Algorithms and analysis for submodular maximization
with a partition matroid can also be found in the appendix (Section
\ref{subsec:partition-appendix} for monotone and Section \ref{subsec:partition-nonmon-appendix}
for general objectives). We also defer our discussion of knapsack
and a common constraint to the appendix.

\section{$k$-Submodular Maximization}

\subsection{\label{subsec:ksubmod-mon-algo} Monotone}

\begin{algorithm}
\textbf{Parameters}:  $\left\{ g_{a}(i)\right\} _{a\in[k],i\in[n_{a}]}$

\textbf{Input}:\textbf{ }monotone $k$-submodular function $f$, budgets
$\left\{ n_{a}\right\} _{a\in[k]}$

$\S=\left(S_{1},\dots,S_{k}\right)\gets\left(\emptyset,\dots,\emptyset\right)$

$\beta_{a}\gets0$ for all $a\in\left[k\right]$

\textbf{for} $t=1,2,\dots,\left|V\right|$:

$\quad$let $w_{t,a}=\Delta_{t,a}f\left({\bf S}\right)$ for all $a\in[k]$

$\quad$let $a=\arg\max_{a\in[k]}\left\{ w_{t,a}-\beta_{a}\right\} $

$\quad$\textbf{if} $w_{t,a}-\beta_{a}\ge0$:

$\quad\quad$\textbf{if} $\left|S_{a}\right|<n_{a}$:

$\quad\quad\quad$$S_{a}\gets S_{a}\cup\left\{ t\right\} $

$\quad\quad$\textbf{else}:

$\quad\quad\quad$let $t'=\arg\min_{i\in S_{a}}w_{i,a}$

$\quad\quad\quad$$S_{a}\gets\left(S_{a}\setminus\left\{ t'\right\} \right)\cup\left\{ t\right\} $

$\quad\quad$let $w_{a}(i)$ be the $i$-th largest weight in $\left\{ w_{t,a}\colon t\in S_{a}\right\} $
and $w_{a}(i)=0$ for $i>\left|S_{a}\right|$

$\quad\quad$$\beta_{a}\gets\sum_{i=1}^{n_{a}}w_{a}(i)g_{a}(i)$

\textbf{return} ${\bf S}$

\caption{\label{alg:ksubmod-mon} Monotone $k$-submodular maximization.}
\end{algorithm}
Our algorithm for maximizing a monotone $k$-submodular function is
shown in Algorithm \ref{alg:ksubmod-mon}. On arrival of each item
$t$, we evaluate its marginal gains for each part with respect to
the current solution ${\bf S}$. We denote these marginal gains as
weights $w_{t,a}$ and note that all subsequent decisions made by
our algorithm depend only on weights. We compare the discounted weights
$w_{t,a}-\beta_{a}$ among all parts $a\in[k]$ and allocate $t$
to $S_{a}$ if the discounted weight of $a$ is the largest among
all parts and non-negative. Thus, $\beta_{a}$ can be thought of as
a threshold that the weight of item $t$ has to pass in order to be
added to the solution. After adding $t$ to $S_{a}$, we may dispose
of an element that was previously allocated to $S_{a}$ in order to
make space for the new item and ensure feasibility. It is therefore
important that the value of $\beta_{a}$ represents the weights of
items in $S_{a}$. We achieve this by setting $\beta_{a}$ to a linear
combination over weights $\left\{ w_{t,a}:t\in S_{a}\right\} $with
coefficients $\left\{ g_{a}(i)\colon a\in[k],i\in[n_{a}]\right\} $,
where
\[
g_{a}(i)\coloneqq\frac{c_{a}}{n_{a}}\left(1+\frac{d_{a}}{n_{a}}\right)^{i-1}\qquad\mathrm{for}\qquad c_{a}\coloneqq\frac{1+d_{a}}{\left(1+\frac{d_{a}}{n_{a}}\right)^{n_{a}}-1},
\]
for all $i\in[n_{a}]$ with constants $d_{a}$ which we will specify
in Theorem \ref{thm:ksubmod} according to the budget $n_{a}$. 

\paragraph{Intuition}

Note that as in \citet{feldman09}, we choose to weigh items with
larger weight less to strike a balance between a greedy scheme, which
allocates to maximize the difference in weight between the added and
disposed item, and uniform weighting, which may ignore potential gain
in favor of saving space. However, our definition of $\beta_{a}$
is novel in that it is no longer a convex combination. This is necessary
to account for submodularity, as we may dispose of valuable items
that had little marginal gain when we added them. We therefore require
new items to clear a higher threshold, to make up for potential loss.
We control this behavior via the parameters $c_{a}$, for each part
$a\in[k]$, and we show later how to derive $c_{a}$ from the analysis.

\begin{table}
\centering{}\caption{Parameter choices and approximation guarantee for monotone $k$-submodular
maximization. \label{tb:partition-monotone-params}}
\medskip{}
{\scriptsize{}}%
\begin{tabular}{ccccc}
\hline 
{\scriptsize{}$n_{a}$} & {\scriptsize{}$1$} & {\scriptsize{}$2$} & {\scriptsize{}$3$} & {\scriptsize{}$\geq4$}\tabularnewline
\hline 
{\scriptsize{}$d_{a}$} & {\scriptsize{}$1$} & {\scriptsize{}$1.0642$} & {\scriptsize{}$1.0893$} & {\scriptsize{}$1.1461$}\tabularnewline
{\scriptsize{}$\frac{1}{Q_{a}}$} & {\scriptsize{}$0.25$} & {\scriptsize{}$\geq0.2781$} & {\scriptsize{}$\geq0.2896$} & {\scriptsize{}$\geq0.3178\left(1-\frac{0.7681}{n_{a}}\right)$}\tabularnewline
\hline 
\end{tabular}\textbf{\scriptsize{}~~~~~~}{\scriptsize{}}%
\begin{tabular}{ccc}
\hline 
\multicolumn{3}{c}{\textbf{\scriptsize{}Approximation guarantee $\min_{a}\frac{1}{Q_{a}}$}}\tabularnewline
\hline 
{\scriptsize{}$\min_{a}n_{a}$} & {\scriptsize{}$\leq3$} & {\scriptsize{}$\geq4$}\tabularnewline
{\scriptsize{}approx} & {\scriptsize{}$\geq0.25$} & {\scriptsize{}$\geq0.3178\left(1-\frac{0.7681}{\min_{a}n_{a}}\right)$}\tabularnewline
\hline 
\end{tabular}{\scriptsize{}}
\end{table}
We obtain the following approximation guarantee for Algorithm \ref{alg:ksubmod-mon}.

\begin{thm}
\label{thm:ksubmod} We make the following choices for the parameters
$\left\{ d_{a}\right\} _{a\in[k]}$. Let $d=1.1461$, which is an
approximate solution to the equation $e^{d}-d-2=0$. We set $d_{a}=d$
if $n_{a}>n_{0}:=3$, and we set $d_{a}$ as shown in Table \ref{tb:partition-monotone-params}
if $n_{a}\leq n_{0}$. We obtain the approximation guarantees shown
in Table \ref{tb:partition-monotone-params}. Note that the approximation
is at least $0.25$ for any minimum budget and it tends to $\geq0.3178$
as the minimum budget tends to infinity.
\end{thm}

\paragraph{Analysis}

We now provide a high-level overview of the analysis for the approximation
ratio of Algorithm \ref{alg:ksubmod-mon}. A complete analysis can
be found in Section \ref{subsec:ksubmod-appendix} of the appendix.
Analyses for all other algorithms in this work follow the same proof
framework, but require further non-trivial~modifications.

We denote with superscript $(t)$ all quantities of the algorithm
at the end of iteration $t$. We denote all quantities at the end
of the stream without superscript. Let $T_{a}^{(t)}=\bigcup_{i=1}^{t}S_{a}^{(i)}$
be the set of all items that were allocated to $a$ in the first $t$
iterations. 

Our goal is to relate $f(\S)$ to the optimum $f(\S^{*})$. However,
comparing both is difficult as there is no direct relationship between
the allocation $\S$ created by our algorithm and the optimum solution
$\S^{*}$. What we can do is to relate both to marginal gains (weights)
and thresholds used in the algorithm, and then leverage the algorithm's
structure to compare both. In particular, we can construct the following
lower bound on the value of the solution $\S$:
\begin{equation}
f(\S)\ge\sum_{a}\sum_{t\in S_{a}}w_{t,a}.\label{eq:1}
\end{equation}
We can see relatively easily how this follows from orthant submodularity
(Lemma \ref{lem:ksubmod-bound-sol}). An upper bound on the optimum
value is harder to obtain, since our marginal gains are with respect
to the current solution $\S^{(t)}$, and it is unclear how to relate
this to the optimum. For submodular functions ($k=1$), a common approach
is to upper bound $f(S^{*})$ by $f(S\cup S^{*})$ and analyze the
latter via the marginal gains. However, this strategy no longer works
for $k$-submodular functions since they are only defined on allocations
where each item appears in at most one part. The solution is to create
a set of intermediate solutions $\O^{(t)}$ that agree with $\T^{(t)}$
on items $\left\{ 1,\dots,t\right\} $ and with $\S^{*}$ on $\left\{ t+1,\dots,\left|V\right|\right\} $,
and analyze $f(\O^{(t)})$. To this end, we upper bound the decrease
in function value $f(\O^{(t-1)})-f(\O^{(t)})$ in each iteration.
With some additional care where we critically use the allocation choice
of Algorithm \ref{alg:ksubmod-mon}, we obtain the following guarantee
(Lemma \ref{lem:ksubmod-bound-opt}):
\begin{equation}
f(\S^{*})\le\sum_{a}\left(\sum_{t\in T_{a}}\left(2w_{t,a}-\beta_{a}^{(t-1)}\right)+n_{a}\beta_{a}\right).\label{eq:2}
\end{equation}
Due to Equations \eqref{eq:1} and \eqref{eq:2}, it is now sufficient
to bound, for all parts $a\in[k]$,
\begin{equation}
\sum_{t\in T_{a}}\left(2w_{t,a}-\beta_{a}^{(t-1)}\right)+n_{a}\beta_{a}\le Q_{a}\sum_{t\in S_{a}}w_{t,a}.\label{eq:3}
\end{equation}
This gives us that $f(\S^{*})\le Qf(\S)$ where we try to make $Q\coloneqq\max_{a\in[k]}Q_{a}$
as small as possible. Note that the RHS of \eqref{eq:3} has the weights
$\left\{ w_{t,a}:t\in T_{a}\right\} $ of all of the items ever allocated
to $a$, including the ones that were discarded, as well as the thresholds.
In contrast, the RHS of \eqref{eq:3} has only the weights $\left\{ w_{t,a}:t\in S_{a}\right\} $
in the final solution. Thus we will need to relate the weights of
the discarded items and the thresholds to the items in the final solution.
 To this end, we use a primal potential that tracks the lower bound
\eqref{eq:1} and a dual potential that tracks the upper bound \eqref{eq:2}:
\[
P_{t}\coloneqq\sum_{i\in S_{a}^{(t)}}w_{i},\qquad D_{t}\coloneqq\sum_{i\in T_{a}^{(t)}}\left(2w_{ai}-\beta_{a}^{(i-1)}\right)+n_{a}\beta_{a}^{(t)}.
\]
We interpret the dual $D_{t}$ as follows: $2w_{at}-\beta_{a}^{(t-1)}$
is the cost of reallocating an item to the part chosen by the optimum
solution, and we use $n_{a}\beta_{a}^{(t)}$ to account for items
in $S_{a}^{*}$ that have not arrived yet by paying the current threshold
$\beta_{a}^{(t)}$ for each of them. Our analysis relates the change
in the dual to the change in the primal, in each iteration. If $t\notin T_{a}$,
we experience no change in either primal nor dual. If $t\in T_{a}$,
the change is
\[
P_{t}-P_{t-1}=w_{t,a}-\min_{i\in S_{a}^{(t-1)}}w_{i,a},\qquad D_{t}-D_{t-1}=2w_{t,a}-\beta_{a}^{(t-1)}+n_{a}\big(\beta_{a}^{(t)}-\beta_{a}^{(t-1)}\big).
\]
To relate the two, we make use of several properties maintained by
the algorithm: we only allocate the item if the discounted gain is
non-negative (i.e., $w_{t,a}\geq\beta_{a}^{(t-1)}$) and our threshold
is a combination of the largest weights with exponential coefficients.
Using these properties, we can upper bound the change in thresholds
$\beta_{a}^{(t)}-\beta_{a}^{(t-1)}$ (Lemma \ref{lem:ksubmod-beta-change})
using only the weights of the new item $w_{t,a}$ and the disposed
item $\min_{i\in S_{a}^{(t-1)}}w_{i,a}$, with appropriate coefficients.
By setting $c_{a}$ appropriately, we make the two coefficients equal,
which gives us the desired comparison. This agrees with the intuition
that $c_{a}$ describes exactly how much additional gain we require
from new items in order to account for the potential loss through
the disposal, which is expressed in the dual potential. This gives
us 
\[
{\textstyle Q_{a}=\left(1+d_{a}\right)\left(1+\frac{1}{\left(1+d_{a}/n_{a}\right)^{n_{a}}-1}\right)}.
\]
Thus it only remains to choose the parameters $d_{a}$ to optimize
the approximation guarantee. In the large budget case, we can approximate
$(1+d_{a}/n_{a})^{n_{a}}\approx\exp(d_{a})$ which does not depend
on the budget. Thus we can use the same parameter $d$ for all parts
and set it to the value that maximizes the approximation guarantee.
In order to account for all budgets, including very small ones, we
analyze the error incurred from approximating $(1+d_{a}/n_{a})^{n_{a}}$
by $\exp(d_{a})$ (Lemma \ref{lem:exp-approx}) and derive appropriate
choices $d_{a}$ that are tailored to the budgets $n_{a}$. As a result,
we can handle the challenging setting where budgets can be very different,
and obtain approximations that improve with the budget.

\paragraph{Comparison to previous work}

Our algorithm is inspired by the works of \citet{feldman09} for linear
objectives and \citet{ene22} for $k$-submodular functions. Both
algorithms use a threshold for each part which determines the allocation
of new items and evolves over time. \citet{ene22} set thresholds
depending on the marginal gains of all previously allocated items,
even those that were already disposed. In contrast, we use a different
scheme for setting the thresholds using linear combinations of the
gains of only the items in the current solution with coefficients
that are exponentially growing. Our approach is similar to \citet{feldman09}
with the notable difference that we no longer use a convex combination
of the gains, which is crucial for submodular objectives as discussed
above.  Our analysis is a significant departure from both prior works.
The analysis of \citet{feldman09} strongly leverages the special
structure of linear functions, and does not apply to submodular objectives.
\citet{ene22} use a global analysis that is tailored to their specific
threshold update scheme. In contrast, we use a different approach
for updating the thresholds and analyze it via a novel local analysis
as outlined above. Our approach is general and flexible, and it allows
us to handle both monotone and non-monotone objectives as well as
more general packing constraints.

\subsection{Non-Monotone}

In this section, we consider the case $k\geq2$. The $k=1$ case is
the problem of maximizing a non-negative submodular function subject
to a cardinality constraint, and we obtain a result as a special case
of our result for a partition matroid constraint. We first consider
the regime when the maximum budget is not too large (i.e. $\max_{a}n_{a}\le\frac{1}{2}\sum_{a}n_{a}$)
where we leverage pairwise monotonicity in a delicate adaptation of
Algorithm \ref{alg:ksubmod-mon}. Based on this, we derive an algorithm
for all budgets.

\paragraph{Algorithm for $\max_{a}n_{a}\protect\leq\frac{1}{2}\sum_{a}n_{a}$}

When using Algorithm \ref{alg:ksubmod-mon} for non-monotone objective,
there is a serious complication: We can no longer bound the difference
in function value after re-allocating item $t$ according to the optimum
solution using a linear combination of weights and thresholds of a
single part. We also need to take thresholds of the other parts into
account (for more details, we refer the reader to the proof of Lemma
\ref{lem:ksubmod-nonmon-bound-opt} in the appendix), so we make the
following modification: In each iteration $t$, we choose the part
that maximizes the following modified discounted gain:
\[
a\gets\arg\max_{a\in[k]}\Big\{\Delta_{t,a}f(\S^{(t-1)})-\beta_{a}^{(t-1)}-\min_{a'\neq a}\beta_{a'}^{(t-1)}\Big\}.
\]
The full pseudocode and analysis can be found in Section \ref{subsec:k-submod-nonmon-appendix}
in the appendix. We obtain:
\begin{thm}
\label{thm:ksubmod-nonmon} When setting the parameters $\{d_{a}\}_{a\in[k]}$
to the choices of Theorem \ref{thm:ksubmod}, the adapted algorithm
achieves an approximation guarantee that is $\frac{1}{2}$ of the
approximation in Theorem \ref{thm:ksubmod}.
\end{thm}

\paragraph{Algorithm for All Budgets}

If $\max_{a}n_{a}>\frac{1}{2}\sum_{a}n_{a}$, we can still obtain
a constant-factor approximation (in expectation). Note that we either
extract a lot of value from the part with maximum budget, or we can
decrease the maximum budget and still obtain a good fraction of the
original value. We mimic this idea by creating two solutions. For
the first solution, we only allocate to the part with maximum budget
while not exceeding the respective budget constraint.  For the second
solution, we solve the original problem, but reduce the budget of
the maximum advertiser such that we can again apply Theorem \ref{thm:ksubmod-nonmon}.
We select the better of the two solutions. This is only a streaming
algorithm as we create multiple solutions, but we can also obtain
an online algorithm by choosing a solution randomly. We defer a full
description and analysis of this algorithm to Section \ref{subsec:k-submod-any-max-budget}
in the appendix.

\section{Experiments}

\begin{figure}[t]
\begin{centering}
{\scriptsize{}\vspace{-5pt}
}\includegraphics[width=0.77\textwidth]{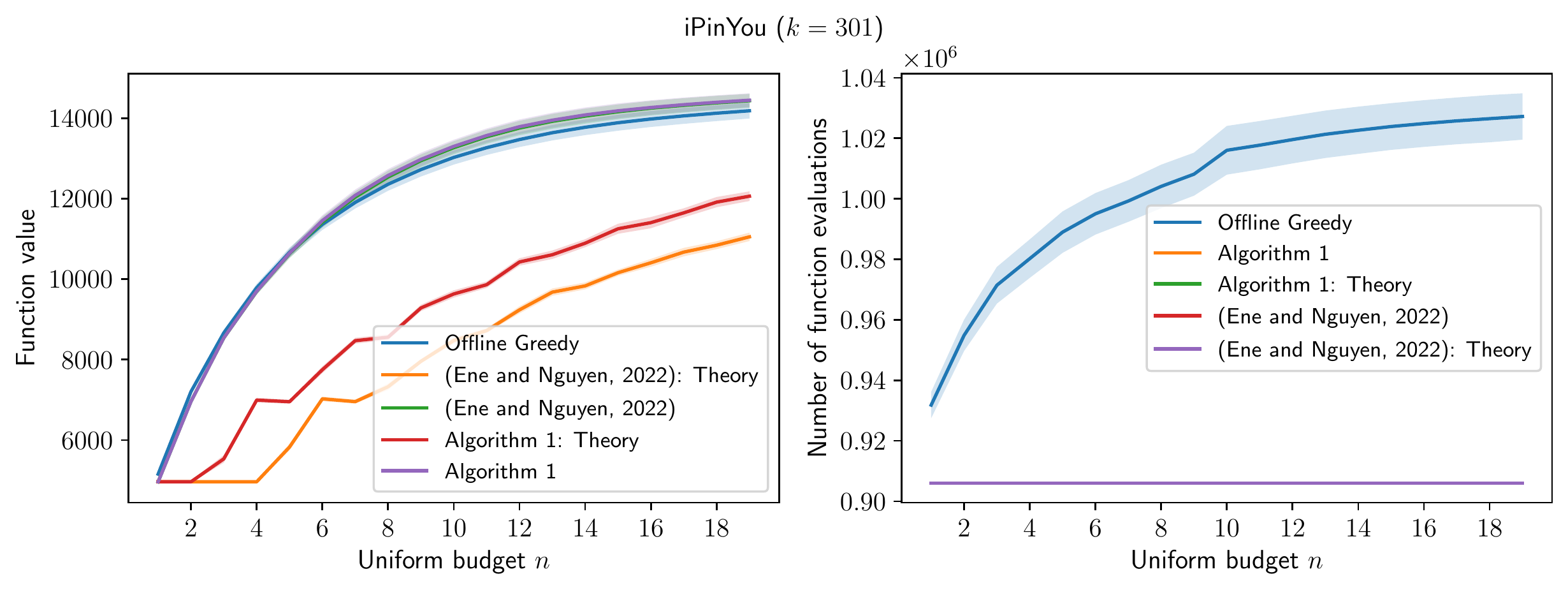}~~~~{\scriptsize{}\vspace{-5pt}
}{\scriptsize\par}
\par\end{centering}
\begin{centering}
\includegraphics[width=0.75\textwidth]{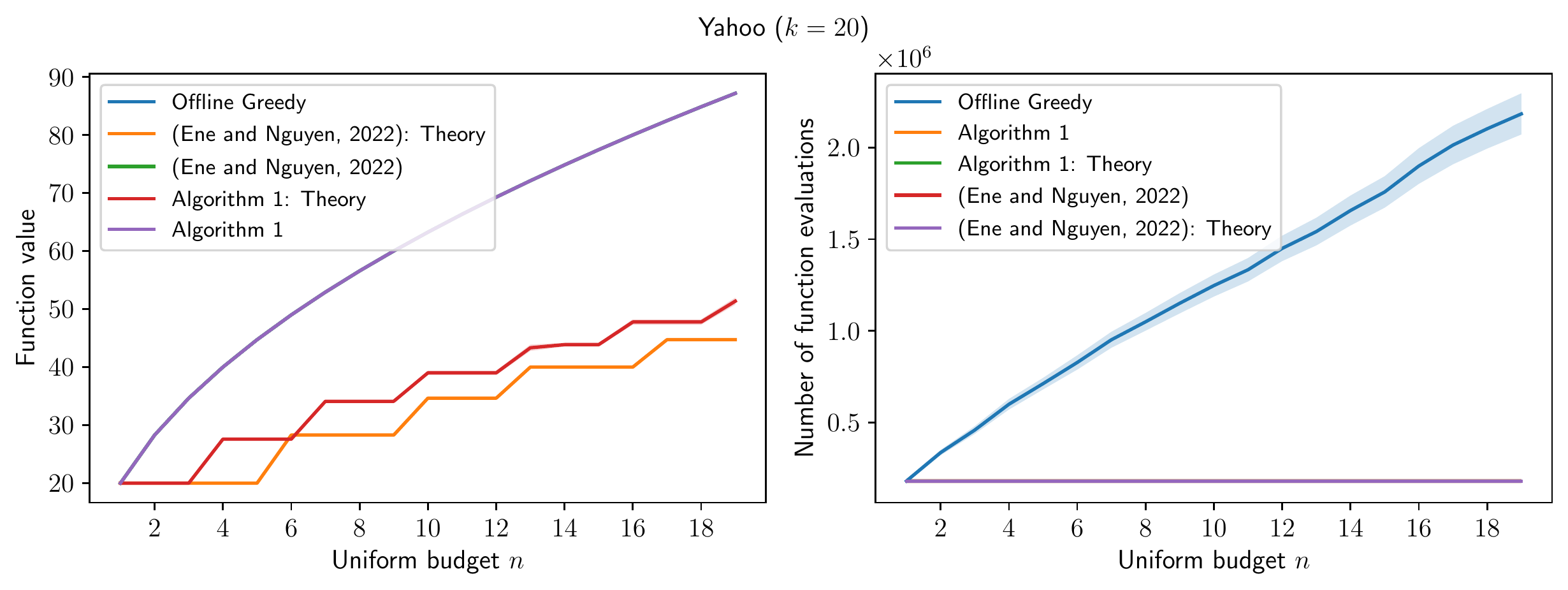}{\scriptsize{}\vspace{-5pt}
}{\scriptsize\par}
\par\end{centering}
\caption{\label{fig:yahoo} Ad allocation on the iPinYou (top) and Yahoo instance
(bottom). We report mean and standard deviation over all days in the
datasets, while varying a uniform budget $n_{a}=n$ for all $a\in[k]$.
Note that the online algorithms using modified parameter choices coincide
with offline greedy on the Yahoo instance. We indicate runs with the
theoretical parameters (e.g. \textquotedblleft Algorithm 1: Theory\textquotedblright{}
is Algorithm \ref{alg:ksubmod-mon} using the theoretically optimal
parameter choices). }
\end{figure}
\begin{table}[t]
\centering{}\caption{\label{tab:imbalanced} Ad allocation on the iPinYou instance with
imbalanced budgets. We report mean and standard deviation over 7 days.
We use theoretical and modified parameter choices.}
\medskip{}
{\scriptsize{}}%
\begin{tabular}{cccc}
\hline 
{\scriptsize{}Algorithm} & {\scriptsize{}Algorithm \ref{alg:ksubmod-mon}} & {\scriptsize{}\citep{ene22}} & {\scriptsize{}Offline Greedy}\tabularnewline
\hline 
{\scriptsize{}Theory} & {\scriptsize{}7499.13 $\pm$68.22} & {\scriptsize{}5698.33 $\pm$88.57} & \multirow{2}{*}{{\scriptsize{}10427.58 $\pm$214.04}}\tabularnewline
{\scriptsize{}Modified} & {\scriptsize{}10236.05 $\pm$220.22} & {\scriptsize{}9681.85 $\pm$152.87} & \tabularnewline
\hline 
\end{tabular}
\end{table}

In this section, we evaluate the practical applicability of our
algorithms for $k$-submodular maximization. We run experiments on
instances for ad allocation and max-cut, exemplifying the applications
mentioned in the introduction. We include further results in Appendix
\ref{sec:experiments-appendix}.

\paragraph{Instances}

Here, we briefly discuss our experiments with a more detailed description
in Appendix~\ref{sec:experiments-appendix}.
\begin{itemize}
\item \emph{Ad Allocation: }We consider the problem of allocating ad impressions
to $k$ advertisers \citep{mehta13}. Here, ad impressions $t\in V$
arrive online and have to be allocated immediately to budget-constrained
advertisers $a\in[k]$. Each advertiser $a$ derives a certain immediate
value $v_{t,a}\ge0$ from impression $t$, but its satisfaction is
only $g_{a}(S_{a})\coloneqq\sqrt{\sum_{t\in S_{a}}v_{t,a}}$.  Our
goal is to maximize total advertiser satisfaction $f(\S)\coloneqq\sum_{a}g_{a}(S_{a})$
while charging each advertiser for at most $\left|S_{a}\right|\le n_{a}$
impressions. We use data from the iPinYou ad exchange \citep{zhang14}
and a Yahoo dataset \citep{yahoo} where we replicate the setup of
\citet{spaeh23} and \citet{lavastida21} to obtain advertiser valuations.
The iPinYou dataset contains bids from $k=301$ advertisers, which
we use as advertiser valuations. We use the first $3000$ impressions,
for each of 7 days. For the Yahoo dataset, we consider only the first
7 days with $\approx8500$ instances per day for $k=20$ advertisers.
The results can be found in Figure \ref{fig:yahoo}. We further create
an imbalanced instance on the iPinYou dataset by sampling advertiser
budgets $n_{a}$ uniformly from $\{1,2,\dots,10\}$. We show results
in Table~\ref{tab:imbalanced}.
\end{itemize}

\begin{figure}
\begin{centering}
{\scriptsize{}\vspace{-5pt}
}\includegraphics[width=0.77\textwidth]{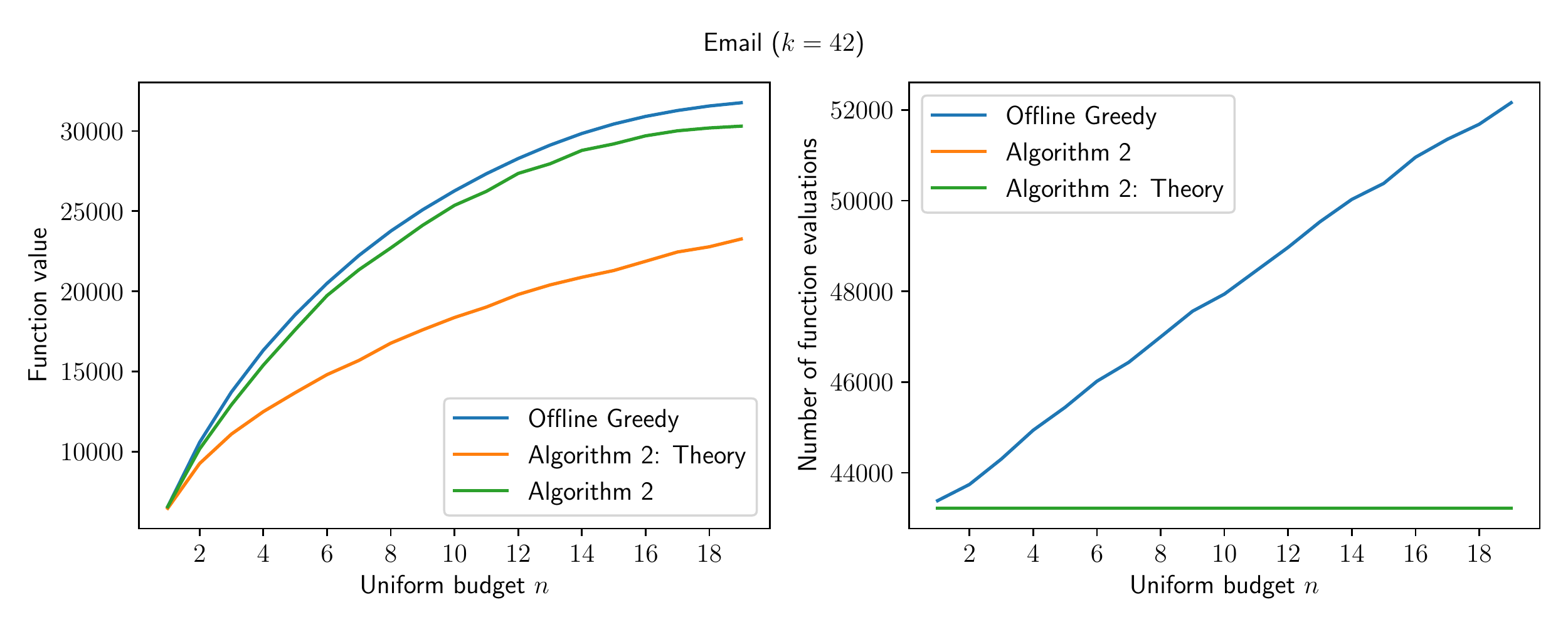}{\scriptsize{}\vspace{-5pt}
}{\scriptsize\par}
\par\end{centering}
\caption{\label{fig:maxcut} Max-$k$-cut on the Email instance: We vary a
uniform budget $n_{a}=n$ for all $a\in[k]$.}
\end{figure}

\begin{itemize}
\item \emph{Influence Maximization with $k$ Topics and Sensor Placement
with $k$ Measurements.} We use the same experimental setup as \citet{ene22}
to create instances for monotone $k$-submodular maximization. The
results for influence maximization and sensor placement are in Figure
\ref{fig:icml-experiments} and Figure \ref{fig:icml-experiments-1}
of Appendix \ref{sec:experiments-appendix}, respectively.
\item \emph{Max-$k$-Cut: }The max-$k$-cut problem asks, given a graph
$G=(V,E)$ and cardinality constraints $n_{1},\dots,n_{k}$ to find
$\S\in(k+1)^{V}$ maximizing the total cut size defined as $f(\S)\coloneqq\sum_{a\in[k]}\delta_{G}(S_{a})$
where $\delta_{G}(S)\coloneqq\left|\left\{ \{u,v\}\in E:u\in S,v\notin S\right\} \right|$.
We use the Email network from SNAP \citep{leskovec14} with $k=42$
parts. The network contains 1005 nodes and 16706 edges. We show the
results in Figure \ref{fig:maxcut}. 
\end{itemize}

\paragraph{Algorithms}

We use the algorithms developed in this work for monotone and general
$k$-submodular maximization. We use Algorithm \ref{alg:ksubmod-mon}
for the monotone instance ad allocation and Algorithm \ref{alg:ksubmod-nonmon}
for the general instance max-$k$-cut. We use two parameter choices
for the online algorithms: First, we set $\{d_{a}\}_{a\in[k]},\{c_{a}\}_{a\in[k]}$
to the optimal theoretical choice as the minimizer of $Q_{a}$ in
Lemma \ref{lem:ksubmod-qa}. Second, we modify these parameters by
reducing each $c_{a}$ to $\frac{1}{4}$ of the the previous choice
to make the algorithms less conservative. We compare our algorithms
with the greedy algorithms of \citet{ohsaka15} for monotone and \citet{xiao23}
for general objectives. We implement both using lazy evaluations.
We also run the algorithm of \citet{ene22} on monotone instances.
The theoretical and modified parameter choices coincide with the ones
used in their experiments.
\begin{figure}
\begin{centering}
{\scriptsize{}\vspace{-5pt}
}\includegraphics[width=0.75\textwidth]{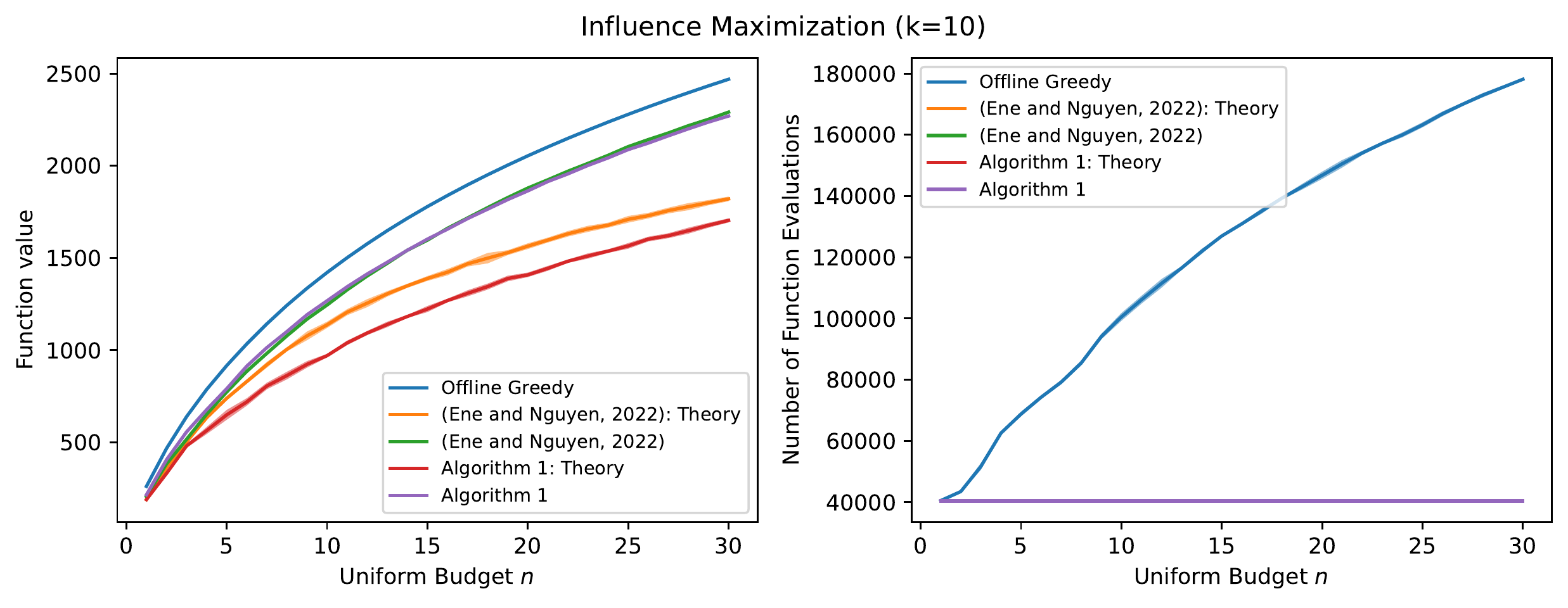}{\scriptsize{}\vspace{-5pt}
}{\scriptsize\par}
\par\end{centering}
\caption{\label{fig:icml-experiments} Influence maximization with $k$ topics.
We vary a uniform budget $n_{a}=n$ for all $a\in[k]$ and report
mean and standard deviation over 5 runs.}
\end{figure}

\section*{Conclusion}

We introduce novel online and streaming algorithms for constrained
$k$-submodular maximization and submodular maximization with a partition
matroid, both with monotone and general objectives. Our algorithms
are combinatorial and very efficient, and use optimal space and running
time. Our approximation guarantees improve with the minimum budget
and, in almost all settings, improve the state of the art. \textbf{ Limitations:}
There is still a gap between the approximation guarantee of our algorithms
and the offline setting, and we leave such improvements for future
work.

\bibliographystyle{plainnat}
\bibliography{ksubmodular-arxiv}

\begin{thebibliography}{21}
\providecommand{\natexlab}[1]{#1}
\providecommand{\url}[1]{\texttt{#1}}
\expandafter\ifx\csname urlstyle\endcsname\relax
  \providecommand{\doi}[1]{doi: #1}\else
  \providecommand{\doi}{doi: \begingroup \urlstyle{rm}\Url}\fi

\bibitem[Badanidiyuru et~al.(2014)Badanidiyuru, Mirzasoleiman, Karbasi, and
  Krause]{badanidiyuru14}
Ashwinkumar Badanidiyuru, Baharan Mirzasoleiman, Amin Karbasi, and Andreas
  Krause.
\newblock Streaming submodular maximization: massive data summarization on the
  fly.
\newblock In \emph{{KDD}}, pages 671--680. {ACM}, 2014.

\bibitem[Chan et~al.(2017)Chan, Jiang, Tang, and Wu]{chan17}
T.{-}H.~Hubert Chan, Shaofeng~H.{-}C. Jiang, Zhihao~Gavin Tang, and Xiaowei Wu.
\newblock Online submodular maximization problem with vector packing
  constraint.
\newblock In \emph{{ESA}}, volume~87 of \emph{LIPIcs}, pages 24:1--24:14.
  Schloss Dagstuhl - Leibniz-Zentrum f{\"{u}}r Informatik, 2017.

\bibitem[Ene and Nguyen(2022)]{ene22}
Alina Ene and Huy~L. Nguyen.
\newblock Streaming algorithm for monotone k-submodular maximization with
  cardinality constraints.
\newblock In \emph{{ICML}}, volume 162 of \emph{Proceedings of Machine Learning
  Research}, pages 5944--5967. {PMLR}, 2022.

\bibitem[Feldman et~al.(2009)Feldman, Korula, Mirrokni, Muthukrishnan, and
  P{\'{a}}l]{feldman09}
Jon Feldman, Nitish Korula, Vahab~S. Mirrokni, S.~Muthukrishnan, and Martin
  P{\'{a}}l.
\newblock Online ad assignment with free disposal.
\newblock In \emph{{WINE}}, volume 5929 of \emph{Lecture Notes in Computer
  Science}, pages 374--385. Springer, 2009.

\bibitem[Feldman et~al.(2018)Feldman, Karbasi, and Kazemi]{feldman18}
Moran Feldman, Amin Karbasi, and Ehsan Kazemi.
\newblock Do less, get more: Streaming submodular maximization with
  subsampling.
\newblock In \emph{NeurIPS}, pages 730--740, 2018.

\bibitem[Feldman et~al.(2022)Feldman, Liu, Norouzi{-}Fard, Svensson, and
  Zenklusen]{feldman22}
Moran Feldman, Paul Liu, Ashkan Norouzi{-}Fard, Ola Svensson, and Rico
  Zenklusen.
\newblock Streaming submodular maximization under matroid constraints.
\newblock In \emph{{ICALP}}, volume 229 of \emph{LIPIcs}, pages 59:1--59:20.
  Schloss Dagstuhl - Leibniz-Zentrum f{\"{u}}r Informatik, 2022.

\bibitem[Gomes and Krause(2010)]{gomes10}
Ryan Gomes and Andreas Krause.
\newblock Budgeted nonparametric learning from data streams.
\newblock In \emph{{ICML}}, pages 391--398. Omnipress, 2010.

\bibitem[Lavastida et~al.(2021)Lavastida, Moseley, Ravi, and Xu]{lavastida21}
Thomas Lavastida, Benjamin Moseley, R.~Ravi, and Chenyang Xu.
\newblock Using predicted weights for ad delivery.
\newblock In \emph{{ACDA}}, pages 21--31. {SIAM}, 2021.

\bibitem[Leskovec and Krevl(2014)]{leskovec14}
Jure Leskovec and Andrej Krevl.
\newblock {SNAP Datasets}: {Stanford} large network dataset collection.
\newblock \url{http://snap.stanford.edu/data}, June 2014.

\bibitem[Lin and Bilmes(2011)]{lin11}
Hui Lin and Jeff~A. Bilmes.
\newblock A class of submodular functions for document summarization.
\newblock In \emph{{ACL}}, pages 510--520. The Association for Computer
  Linguistics, 2011.

\bibitem[Mehta(2013)]{mehta13}
Aranyak Mehta.
\newblock Online matching and ad allocation.
\newblock \emph{Found. Trends Theor. Comput. Sci.}, 8\penalty0 (4):\penalty0
  265--368, 2013.

\bibitem[Mirzasoleiman et~al.(2016)Mirzasoleiman, Badanidiyuru, and
  Karbasi]{mirzasoleiman16}
Baharan Mirzasoleiman, Ashwinkumar Badanidiyuru, and Amin Karbasi.
\newblock Fast constrained submodular maximization: Personalized data
  summarization.
\newblock In \emph{{ICML}}, volume~48 of \emph{{JMLR} Workshop and Conference
  Proceedings}, pages 1358--1367. JMLR.org, 2016.

\bibitem[Nguyen and Thai(2020)]{nguyen20}
Lan Nguyen and My~T. Thai.
\newblock Streaming k-submodular maximization under noise subject to size
  constraint.
\newblock In \emph{{ICML}}, volume 119 of \emph{Proceedings of Machine Learning
  Research}, pages 7338--7347. {PMLR}, 2020.

\bibitem[Ohsaka and Yoshida(2015)]{ohsaka15}
Naoto Ohsaka and Yuichi Yoshida.
\newblock Monotone k-submodular function maximization with size constraints.
\newblock In \emph{{NIPS}}, pages 694--702, 2015.

\bibitem[Pham et~al.(2022)Pham, Ha, Hoang, and Tran]{pham22}
Canh~V. Pham, Dung K.~T. Ha, Huan~X. Hoang, and Tan~D. Tran.
\newblock Fast streaming algorithms for k-submodular maximization under a
  knapsack constraint.
\newblock In \emph{{DSAA}}, pages 1--10. {IEEE}, 2022.

\bibitem[Spaeh and Ene(2023)]{spaeh23}
Fabian Spaeh and Alina Ene.
\newblock Online ad allocation with predictions, 2023.

\bibitem[Tang et~al.(2022)Tang, Wang, and Chan]{tang22}
Zhongzheng Tang, Chenhao Wang, and Hau Chan.
\newblock On maximizing a monotone \emph{k}-submodular function under a
  knapsack constraint.
\newblock \emph{Oper. Res. Lett.}, 50\penalty0 (1):\penalty0 28--31, 2022.

\bibitem[Ward and Zivn{\'{y}}(2016)]{ward16}
Justin Ward and Stanislav Zivn{\'{y}}.
\newblock Maximizing \emph{k}-submodular functions and beyond.
\newblock \emph{{ACM} Trans. Algorithms}, 12\penalty0 (4):\penalty0
  47:1--47:26, 2016.

\bibitem[Xiao et~al.(2022)Xiao, Liu, Zhou, and Li]{xiao23}
Hao Xiao, Qian Liu, Yang Zhou, and Min Li.
\newblock Non-monotone k-submodular function maximization with individual size
  constraints.
\newblock In \emph{CSoNet}, volume 13831 of \emph{Lecture Notes in Computer
  Science}, pages 268--279. Springer, 2022.

\bibitem[Yahoo(2011)]{yahoo}
Yahoo.
\newblock Yahoo! webscope, 2011.
\newblock URL \url{https://webscope.sandbox.yahoo.com/}.
\newblock Accessed September 7, 2022.

\bibitem[Zhang et~al.(2014)Zhang, Yuan, and Wang]{zhang14}
Weinan Zhang, Shuai Yuan, and Jun Wang.
\newblock Real-time bidding benchmarking with ipinyou dataset.
\newblock \emph{CoRR}, abs/1407.7073, 2014.

\end{thebibliography}
\newpage

\appendix

\section{Omitted Algorithms and Analyses}

\subsection{Notation}

We use the following notation for the analysis of all of the algorithms.
For a $k$-tuple $\X\in(k+1)^{V}$, we denote with $\supp(\X)\coloneqq X_{1}\cup\cdots\cup X_{k}$
the support of $\X$. We say $\X,\Y\in(k+1)^{V}$ agree on item $t\in V$
if either $t\notin\supp(\X)\cup\supp(\Y)$ (the item is not allocated
in either allocation) or $t\in X_{a}\cap Y_{a}$ for some $a\in[k]$
(the item is allocated to the same part in both allocations). We denote
with superscript $(t)$ all quantities of the algorithm at the end
of iteration $t$. We denote all quantities at the end of the stream
without superscript. Let $T_{a}^{(t)}=\bigcup_{i=1}^{t}S_{a}^{(i)}$
be the set of all items that were allocated to $a$ in the first $t$
iterations, including items that were disposed. For $t\in\supp(\T)$,
let $a(t)$ be the part that $t$ is allocated to in $\T$ by our
algorithm, i.e. $t\in T_{a(t)}$. Let $a^{*}(t)$ be defined analogously
with respect to the optimal solution $\S^{*}$.

\subsection{\label{subsec:ksubmod-appendix} Monotone $k$-Submodular Maximization}

\subsubsection{Analysis}

The analysis of Algorithm \ref{alg:ksubmod-mon} and other algorithms
in this work follow the same proof outline. That is, to relate the
value of the solution created by Algorithm \ref{alg:ksubmod-mon}
$f({\bf S})$ to the optimum solution $f({\bf S}^{*})$, we first
obtain an appropriate lower bound on $f(\S)$ and an upper bound on
$f(\S^{*})$. We interpret the former as primal potential and the
latter as dual potential. Potentials are linear combinations of weights
$\{w_{t,a}\}_{t,a}$ and thresholds $\{\beta_{a}^{(t)}\}_{t,a}$.
With some additional work, we can to compare both bounds. In particular,
we bound the change in primal by the change in dual, in each iteration.
This is sufficient to establish our approximation guarantee.

Due to orthant submodularity, we can naturally lower bound $f({\bf S})$
as the sum over weights of items in ${\bf S}$:
\begin{lem}
\label{lem:ksubmod-bound-sol} The value of solution ${\bf S}$ is
at least
\[
f({\bf S})\ge\sum_{a}\sum_{t\in S_{a}}w_{t,a}.
\]
\end{lem}

\begin{proof}
We have
\begin{align*}
f(\S)-f(\S^{(0)}) & =\sum_{t\in\supp(\S)}\left(f\left({\bf S}\cap{\bf S}^{(t)}\right)-f\left({\bf S}\cap{\bf S}^{(t-1)}\right)\right)\\
 & =\sum_{t\in\supp(\S)}\Delta_{t,a(t)}f(\S\cap{\bf S}^{(t-1)})\\
 & \geq\sum_{t\in\supp(\S)}\Delta_{t,a(t)}f({\bf S}^{(t-1)})\\
 & =\sum_{t\in\supp(\S)}w_{t,a(t)}
\end{align*}
where the inequality is due to orthant submodularity.
\end{proof}

Next, we upper bound $f({\bf S}^{*})$ via a telescoping argument.
In particular, we are able to relate $\S^{*}$ to $\T$ by constructing
a series of intermediate solutions $\O^{(t)}$ that agree with $\T^{(t)}$
on items $\left\{ 1,\dots,t\right\} $ and with $\S^{*}$ on items
$\left\{ t+1,\dots,\left|V\right|\right\} $. For each $t$, we then
bound $f(\O^{(t-1)})-f(\O^{(t)})$, i.e. the difference in function
value after allocating item $t$ according to the optimum solution.
We show that if $t\in\supp(\T)\cap\supp(\S^{*})$, this difference
can be bounded by the marginal gain $\Delta_{t,a^{*}(t)}f(\S^{(t-1)})=w_{t,a^{*}(t)}$.
This holds due to submodularity and monotonicity, as changing the
allocation from one part to another cannot increase the function value
more than the marginal gain. If $t\in\supp(\S^{*})\setminus\supp(\T)$,
we did not allocate $t$ to any part as all weights were at most the
threshold in the respective part, and we can thus charge the difference
to the threshold. This allows us to obtain:
\begin{lem}
\label{lem:ksubmod-bound-opt} The value of the optimum solution ${\bf S}^{*}$
is at most
\[
f\left({\bf S}^{*}\right)\leq\sum_{a}\left(\sum_{t\in T_{a}}\left(2w_{t,a}-\beta_{a}^{(t-1)}\right)+n_{a}\beta_{a}\right).
\]
\end{lem}

\begin{proof}
Let $\O^{(t)}$ be the allocation that agrees with $\T^{(t)}$ on
items $\left\{ 1,\dots,t\right\} $, and it agrees with $\S^{*}$
on items $\left\{ t+1,\dots,\left|V\right|\right\} $. Let $\widetilde{\O}^{(t-1)}$
be the allocation obtained from $\O^{(t)}$ by dropping $t$ (i.e.,
$t$ is not assigned to any part under $\widetilde{\O}^{(t-1)}$).
For $t\in\supp(\T)$, let $a(t)$ be the part such that $t\in T_{a}$.
For $t\in\supp(\S^{*})$, let $a^{*}(t)$ be the part such that $t\in S_{a}^{*}$.

We have
\begin{align*}
 & f(\S^{*})-f(\T)\\
 & =f(\O^{(0)})-f(\O^{\left|V\right|})=\sum_{t=1}^{\left|V\right|}\left(f(\O^{(t-1)})-f(\O^{(t)})\right)\\
 & =\sum_{t\in\supp(\T)\cap\supp(\S^{*})}\left(f(\O^{(t-1)})-f(\O^{(t)})\right)+\sum_{t\notin\supp(\T)\cup\supp(\S^{*})}\left(f(\O^{(t-1)})-f(\O^{(t)})\right)\\
 & +\sum_{t\in\supp(\T)\setminus\supp(\S^{*})}\left(f(\O^{(t-1)})-f(\O^{(t)})\right)+\sum_{t\in\supp(\S^{*})\setminus\supp(\T)}\left(f(\O^{(t-1)})-f(\O^{(t)})\right).
\end{align*}
We analyze all four sums separately:
\begin{itemize}
\item Consider $t\in\supp(\T)\cap\supp(\S^{*})$. If $a(t)=a^{*}(t)$, we
have $\O^{(t-1)}=\O^{(t)}$, and thus
\[
f(\O^{(t-1)})-f(\O^{(t)})=0
\]
If $a(t)\neq a^{*}(t)$, we have
\begin{align*}
f(\O^{(t-1)})-f(\O^{(t)}) & =f(\O^{(t-1)})-f(\widetilde{\O}^{(t-1)})+f(\widetilde{\O}^{(t-1)})-f(\O^{(t)})\\
 & =\Delta_{t,a^{*}(t)}f(\widetilde{\O}^{(t-1)})-\Delta_{t,a(t)}f(\widetilde{\O}^{(t-1)})\\
 & \leq\Delta_{t,a^{*}(t)}f(\S^{(t-1)})-\underbrace{\Delta_{t,a(t)}f(\widetilde{\O}^{(t-1)})}_{\geq0}\\
 & \leq\Delta_{t,a^{*}(t)}f(\S^{(t-1)})
\end{align*}
In the first inequality, we used orthant submodularity since $\S^{(t-1)}\preceq\widetilde{\O}^{(t-1)}$.
In the second inequality, we used monotonicity.
\item Consider $t\notin\supp(\T)\cup\supp(\S^{*})$. We have $\O^{(t-1)}=\O^{(t)}$,
and thus
\[
f(\O^{(t-1)})-f(\O^{(t)})=0
\]
\item Consider $t\in\supp(\T)\setminus\supp(\S^{*})$. We have $\O^{(t-1)}\preceq\O^{(t)}$.
Since $f$ is monotone, we have
\begin{align*}
f(\O^{(t-1)})-f(\O^{(t)}) & \leq0
\end{align*}
\item Consider $t\in\supp(\S^{*})\setminus\supp(\T).$ We have 
\begin{align*}
f(\O^{(t-1)})-f(\O^{(t)}) & =\Delta_{t,a^{*}(t)}f(\O^{(t)})\leq\Delta_{t,a^{*}(t)}f(\S^{(t-1)})\leq\beta_{a^{*}(t)}^{(t-1)}
\end{align*}
where in the first inequality we used orthant submodularity since
$\S^{(t-1)}\preceq\O^{(t)}$, and in the second inequality we used
that all of the discounted gains are $\leq0$.
\end{itemize}
Putting everything together, we have
\[
f(\S^{*})\leq f(\T)+\sum_{t\in\supp(\T)\cap\supp(\S^{*})}w_{t,a^{*}(t)}+\sum_{t\in\supp(\S^{*})\setminus\supp(\T)}\beta_{a^{*}(t)}^{(t-1)}
\]
Using the fact that $\S^{(t)}\subseteq\T^{(t)}$ and orthant submodularity,
we can further upper bound
\begin{align*}
f(\T) & =\sum_{t\in\supp(\T)}\left(f(\T^{(t)})-f(\T^{(t-1)})\right)\\
 & =\sum_{t\in\supp(\T)}\Delta_{t,a(t)}f(\T^{(t-1)})\\
 & \leq\sum_{t\in\supp(\T)}\Delta_{t,a(t)}f(\S^{(t-1)})\\
 & =\sum_{t\in\supp(\T)}w_{t,a(t)}
\end{align*}
Thus,
\begin{align*}
f(\S^{*}) & \leq\sum_{t\in\supp(\T)}w_{t,a(t)}+\sum_{t\in\supp(\T)\cap\supp(\S^{*})}w_{t,a^{*}(t)}+\sum_{t\in\supp(\S^{*})\setminus\supp(\T)}\beta_{a^{*}(t)}^{(t-1)}\\
 & =\sum_{t\in\supp(\T)}w_{t,a(t)}+\sum_{t\in\supp(\T)\cap\supp(\S^{*})}\left(w_{t,a^{*}(t)}-\beta_{a^{*}(t)}^{(t-1)}\right)+\sum_{t\in\supp(\S^{*})}\beta_{a^{*}(t)}^{(t-1)}\\
 & \overset{(1)}{\leq}\sum_{t\in\supp(\T)}w_{t,a(t)}+\sum_{t\in\supp(\T)}\left(w_{t,a(t)}-\beta_{a(t)}^{(t-1)}\right)+\sum_{t\in\supp(\S^{*})}\beta_{a^{*}(t)}^{(t-1)}\\
 & =\sum_{t\in\supp(\T)}\left(2w_{t,a(t)}-\beta_{a(t)}^{(t-1)}\right)+\sum_{t\in\supp(\S^{*})}\beta_{a^{*}(t)}^{(t-1)}
\end{align*}
where in $(1)$ we used that $w_{t,a^{*}(t)}-\beta_{a^{*}(t)}^{(t-1)}\leq w_{t,a(t)}-\beta_{a(t)}^{(t-1)}$
for every $t\in\supp(\T)\cap\supp(\S^{*})$ due to the choice of $a(t)$,
and $w_{t,a(t)}-\beta_{a(t)}^{(t-1)}\geq0$ for every $t\in\supp(\T)$.

Finally, since the thresholds are non-decreasing and $\S^{*}$ is
a feasible allocation, we have
\[
\sum_{t\in\supp(\S^{*})}\beta_{a^{*}(t)}^{(t-1)}=\sum_{a=1}^{k}\sum_{t\in S_{a}^{*}}\beta_{a}^{(t-1)}\leq\sum_{a=1}^{k}n_{a}\beta_{a}
\]
\end{proof}

Due to Lemma \ref{lem:ksubmod-bound-sol} and Lemma \ref{lem:ksubmod-bound-opt},
it is sufficient to show that
\[
\sum_{a}\left(\sum_{t\in T_{a}}\left(2w_{t,a}-\beta_{a}^{(t-1)}\right)+n_{a}\beta_{a}\right)\le Q\sum_{a}\sum_{t\in S_{a}}w_{t,a}
\]
for $Q$ as small as we can make it. We will compare on a per-part
basis and show:
\begin{lem}
\label{lem:ksubmod-qa} For every part $a\in[k]$, we have
\[
\sum_{t\in T_{a}}\left(2w_{t,a}-\beta_{a}^{(t-1)}\right)+n_{a}\beta_{a}\le Q_{a}\sum_{t\in S_{a}}w_{t,a}
\]
where $d_{a}\ge1$ and
\[
Q_{a}\coloneqq\left(1+d_{a}\right)\left(1+\frac{1}{\left(1+\frac{d_{a}}{n_{a}}\right)^{n_{a}}-1}\right).
\]
\end{lem}

We can then set $Q=\max_{a}Q_{a}$. Let us now fix a part $a\in[k]$
to show Lemma \ref{lem:ksubmod-qa}. In each iteration, we consider
an evolving primal and dual, defined as
\begin{align*}
P_{t} & \coloneqq\sum_{i\in S_{a}^{(t)}}w_{i,a}\\
D_{t} & \coloneqq\sum_{i\in T_{a}^{(t)}}\left(2w_{i,a}-\beta_{a}^{(i-1)}\right)+n_{a}\beta_{a}^{(t)}.
\end{align*}
Note that we have $P_{0}=D_{0}=0$, $P_{T}=\sum_{t\in S_{a}}w_{t,a}$,
and $D_{T}=\sum_{t\in T_{a}}\left(2w_{t,a}-\beta_{a}^{(t-1)}\right)+n_{a}\beta_{a}$.
Thus it suffices to show that $D_{t}-D_{t-1}\leq Q_{a}$$\left(P_{t}-P_{t-1}\right)$
for all $t$ to show Lemma \ref{lem:ksubmod-qa}. To bound the change
in thresholds, we first need the following helper lemma. Here, we
merely use the definition of $\beta_{a}$ and implicitly that the
difference $\beta_{a}^{(t)}-\beta_{a}^{(t-1)}$ is maximized if $t$
becomes the most valuable item allocated to part $a(t)$.
\begin{lem}
\label{lem:ksubmod-beta-change}We have
\[
n_{a}\left(\beta_{a}^{(t)}-\beta_{a}^{(t-1)}\right)\leq d_{a}\beta_{a}^{(t-1)}+c_{a}w_{t,a}-c_{a}\left(1+\frac{d_{a}}{n_{a}}\right)^{n_{a}}\left(\min_{i\in S_{a}^{(t-1)}}w_{i,a}\right)
\]
\end{lem}

\begin{proof}
Fix a $t\in T$. Let $w(1)\geq w(2)\geq\dots\geq w(n_{a}+1)$ be the
$n_{a}+1$ largest weights among $\left\{ w_{i,a}\colon i\in T_{a}^{(t)}\right\} $;
if $T_{a}^{(t)}$ has less than $n_{a}+1$ items, we let $w(i)=0$
for $i>\left|T_{a}^{(t)}\right|$. Note that $\min_{i\in S_{a}^{(t-1)}}w_{i,a}=w(n_{a}+1)$.
Let $j$ be such that $w_{t,a}=w(j)$. We have

\begin{align*}
\beta_{a}^{(t)} & =\sum_{i=1}^{n_{a}}w(i)g_{a}(i)\\
\beta_{a}^{(t-1)} & =\sum_{i=1}^{j-1}w(i)g_{a}(i)+\sum_{i=j}^{n_{a}}w(i+1)g_{a}(i)=\sum_{i=1}^{j-1}w(i)g_{a}(i)+\sum_{i=j+1}^{n_{a}+1}w(i)g_{a}(i-1)
\end{align*}
Thus,
\begin{align*}
\beta_{a}^{(t)}-\beta_{a}^{(t-1)} & =\sum_{i=j}^{n_{a}}w(i)g_{a}(i)-\sum_{i=j+1}^{n_{a}+1}w(i)g_{a}(i-1)\\
 & =\sum_{i=j+1}^{n_{a}}w(i)\left(g_{a}(i)-g_{a}(i-1)\right)+w(j)g_{a}(j)-w(n_{a}+1)g_{a}(n_{a})\\
 & =\frac{d_{a}}{n_{a}}\sum_{i=j+1}^{n_{a}}w(i)g_{a}(i-1)+w(j)g_{a}(j)-w(n_{a}+1)g_{a}(n_{a})\\
 & =\frac{d_{a}}{n_{a}}\beta_{a}^{(t-1)}-\frac{d_{a}}{n_{a}}\sum_{i=1}^{j-1}w(i)g_{a}(i)+w(j)g_{a}(j)-\left(1+\frac{d_{a}}{n_{a}}\right)w(n_{a}+1)g_{a}(n_{a})\\
 & =\frac{d_{a}}{n_{a}}\beta_{a}^{(t-1)}-\frac{d_{a}}{n_{a}}\sum_{i=1}^{j-1}w(i)g_{a}(i)+w(j)g_{a}(j)-w(n_{a}+1)g_{a}(n_{a}+1)\\
 & \leq\frac{d_{a}}{n_{a}}\beta_{a}^{(t-1)}-\frac{d_{a}}{n_{a}}\sum_{i=1}^{j-1}w(j)g_{a}(i)+w(j)g_{a}(j)-w(n_{a}+1)g_{a}(n_{a}+1)\\
 & =\frac{d_{a}}{n_{a}}\beta_{a}^{(t-1)}+w(j)\underbrace{\left(\left(1+\frac{d_{a}}{n_{a}}\right)^{j-1}-\frac{d_{a}}{n_{a}}\sum_{i=1}^{j-1}\left(1+\frac{d_{a}}{n_{a}}\right)^{i-1}\right)}_{=1}g_{a}(1)\\
 & \quad-w(n_{a}+1)g_{a}(n_{a}+1)\\
 & =\frac{d_{a}}{n_{a}}\beta_{a}^{(t-1)}+w(j)g_{a}(1)-w(n_{a}+1)g_{a}(n_{a}+1)\\
 & =\frac{d_{a}}{n_{a}}\beta_{a}^{(t-1)}+\frac{c_{a}}{n_{a}}w_{t,a}-\frac{c_{a}}{n_{a}}\left(1+\frac{d_{a}}{n_{a}}\right)^{n_{a}}\left(\min_{i\in S_{a}^{(t-1)}}w_{i,a}\right)
\end{align*}
Using that $w(j)=w_{t,a}$, $\min_{i\in S_{a}^{(t-1)}}w_{i,a}=w(n_{a}+1)$,
the definition of $g_{a}(i)=\frac{c_{a}}{n_{a}}\left(1+\frac{d_{a}}{n_{a}}\right)^{i-1}$,
we obtain
\[
n_{a}\left(\beta_{a}^{(t)}-\beta_{a}^{(t-1)}\right)\le d_{a}\beta_{a}^{(t-1)}+c_{a}w_{t,a}-c_{a}\left(1+\frac{d_{a}}{n_{a}}\right)^{n_{a}}\left(\min_{i\in S_{a}^{(t-1)}}w_{i,a}\right)
\]
\end{proof}

We can now compare the change in primal to the change in dual to show
Lemma \ref{lem:ksubmod-qa}.\textcolor{blue}{{} }

\begin{proof}[Proof (Lemma \ref{lem:ksubmod-qa})]
If $t\notin T_{a}$, we have $\beta_{a}^{(t)}=\beta_{a}^{(t-1)}$
and thus $P_{t}-P_{t-1}=D_{t}-D_{t-1}=0$. Thus we may assume that
$t\in T_{a}$, and thus $w_{t,a}\geq\beta_{a}^{(t-1)}$. We have
\begin{align*}
D_{t}-D_{t-1} & =2w_{t,a}-\beta_{a}^{(t-1)}+n_{a}\left(\beta_{a}^{(t)}-\beta_{a}^{(t-1)}\right)\\
P_{t}-P_{t-1} & =w_{t,a}-\min_{i\in S_{a}^{(t-1)}}w_{i,a}.
\end{align*}
Recall that we set $d_{a}\geq1$. Using Lemma \ref{lem:ksubmod-beta-change},
we obtain
\begin{align*}
 & 2w_{t,a}-\beta_{a}^{(t-1)}+n_{a}\left(\beta_{a}^{(t)}-\beta_{a}^{(t-1)}\right)\\
 & \leq\underbrace{\left(d_{a}-1\right)}_{\geq0}\beta_{a}^{(t-1)}+\left(2+c_{a}\right)w_{t,a}-c_{a}\left(1+\frac{d_{a}}{n_{a}}\right)^{n_{a}}\left(\min_{i\in S_{a}^{(t-1)}}w_{i,a}\right)\\
 & \leq\left(d_{a}-1\right)w_{t,a}+\left(2+c_{a}\right)w_{t,a}-c_{a}\left(1+\frac{d_{a}}{n_{a}}\right)^{n_{a}}\left(\min_{i\in S_{a}^{(t-1)}}w_{i,a}\right)\\
 & =\left(1+d_{a}+c_{a}\right)w_{t,a}-c_{a}\left(1+\frac{d_{a}}{n_{a}}\right)^{n_{a}}\left(\min_{i\in S_{a}^{(t-1)}}w_{i,a}\right).
\end{align*}
Recall that by the definition of $c_{a}$,
\[
c_{a}=\frac{1+d_{a}}{\left(1+\frac{d_{a}}{n_{a}}\right)^{n_{a}}-1}\iff1+d_{a}+c_{a}=c_{a}\left(1+\frac{d_{a}}{n_{a}}\right)^{n_{a}}.
\]
We thus obtain
\[
Q_{a}=1+d_{a}+c_{a}=1+d_{a}+\frac{1+d_{a}}{\left(1+\frac{d_{a}}{n_{a}}\right)^{n_{a}}-1}=\left(1+d_{a}\right)\left(1+\frac{1}{\left(1+\frac{d_{a}}{n_{a}}\right)^{n_{a}}-1}\right).
\]
\end{proof}

\subsubsection{Setting the Parameters}

To complete the analysis, we show how to set the constants $\{d_{a}\}_{a\in[k]}$,
and derive the final approximation guarantee. Note that we can set
each $d_{a}$ to the value that minimizes $Q_{a}=\left(1+d_{a}\right)\left(1+\frac{1}{\left(1+\frac{d_{a}}{n_{a}}\right)^{n_{a}}-1}\right)$.
In the following, we give explicit choices for the $d_{a}$'s that
avoid this computation, and establish the approximation guarantee
for these explicit choices. 

Before proceeding, let us observe that, if the minimum budget $\min_{a\in[k]}n_{a}$
is sufficiently large, we have $\left(1+\frac{d_{a}}{n_{a}}\right)^{n_{a}}\approx e^{d_{a}}$
for all $a$. Suppose we set $d_{a}=d$ for some value $d$. Then
$Q_{a}=\left(1+d\right)\left(1+\frac{1}{\left(1+\frac{d}{n_{a}}\right)^{n_{a}}-1}\right)\approx\left(1+d\right)\left(1+\frac{1}{e^{d}-1}\right)$
and we obtain an approximation $\min_{a}\frac{1}{Q_{a}}\approx\frac{1}{1+d}\frac{1}{1+\frac{1}{e^{d}-1}}$.
We can then choose $d$ to be the value that maximizes the approximation
guarantee. By taking the derivative with respect to $d$ and setting
it to $0$, we obtain that $d$ should be set to the solution to the
equation $e^{d}-d-2=0$, which is $d\approx1.1461$. We obtain an
approximation $\geq0.3178$, matching the approximation of the streaming
continuous greedy algorithm of \citet{feldman22}. For budgets $n_{a}$
that are larger than an absolute constant $n_{0}$, we set $d_{a}$
to be equal to this value $d$. For smaller budgets, we give explicit
choices for $d_{a}$ that are good for that specific $n_{a}$. The
choices are given in Table \ref{tb:partition-monotone-params}. 

We start with the following helper lemma:
\begin{lem}
\label{lem:exp-approx}Let $n_{0}$ and $d$ be absolute constants
satisfying $n_{0}\geq d\geq0$. For every $n\geq n_{0}$, we have
\[
\frac{1}{1+\frac{1}{\left(1+\frac{d}{n}\right)^{n}-1}}\geq\frac{1}{1+\frac{1}{e^{d}-1}}\cdot\left(1-\frac{1}{n}\cdot\frac{n_{0}\left(\exp\left(\frac{d^{2}}{n_{0}}\right)-1\right)}{\exp\left(d\right)-1}\right)=\frac{1}{1+\frac{1}{e^{d}-1}}\cdot\left(1-O\left(\frac{1}{n}\right)\right).
\]
\end{lem}

\begin{proof}
Consider any $n\geq n_{0}$. We use the inequality $1+x\geq\exp\left(x-\frac{x^{2}}{2}\right)$,
which holds for $0\leq x\leq1$. Since $0\leq d\leq n_{0}\leq n$,
we have $0\leq\frac{d}{n}\leq1$. The inequality gives
\[
\left(1+\frac{d}{n}\right)^{n}\geq\exp\left(d-\frac{d^{2}}{n}\right)
\]
Thus
\begin{multline*}
\frac{1}{1+\frac{1}{\left(1+\frac{d}{n}\right)^{n}-1}}\geq\frac{1}{1+\frac{1}{\exp\left(d-\frac{d^{2}}{n}\right)-1}}=\frac{1}{1+\frac{1}{\exp\left(d\right)-1}}\cdot\frac{1+\frac{1}{\exp\left(d\right)-1}}{1+\frac{1}{\exp\left(d-\frac{d^{2}}{n}\right)-1}}\\
=\frac{1}{1+\frac{1}{\exp\left(d\right)-1}}\cdot\left(1-\frac{\exp\left(\frac{d^{2}}{n}\right)-1}{\exp\left(d\right)-1}\right).
\end{multline*}
Since $e^{x}$ is convex, for $0\leq x\leq a$, we have $e^{x}\leq\frac{x}{a}e^{a}+\left(1-\frac{x}{a}\right)e^{0}=\frac{x}{a}e^{a}+1-\frac{x}{a}$.
We use this inequality with $x=\frac{d^{2}}{n}$ and $a=\frac{d^{2}}{n_{0}}$.
Since $n\geq n_{0}$, we have $0\leq\frac{d^{2}}{n}\leq\frac{d^{2}}{n_{0}}$,
and the inequality gives 
\[
\exp\left(\frac{d^{2}}{n}\right)-1\leq\frac{n_{0}}{n}\left(\exp\left(\frac{d^{2}}{n_{0}}\right)-1\right)
\]
and thus
\[
\frac{1}{1+\frac{1}{\left(1+\frac{d}{n}\right)^{n}-1}}\geq\frac{1}{1+\frac{1}{\exp\left(d\right)-1}}\cdot\left(1-\frac{1}{n}\frac{n_{0}\left(\exp\left(\frac{d^{2}}{n_{0}}\right)-1\right)}{\exp\left(d\right)-1}\right).
\]
\end{proof}

We can now prove Theorem \ref{thm:ksubmod} that gives our final approximation
guarantee.

\begin{proof}[Proof (Theorem \ref{thm:ksubmod})]
 Let $n_{0}=3$. For $n_{a}\leq n_{0}$, we can verify that $\frac{1}{Q_{a}}$
is lower bounded by the values shown in Table \ref{tb:partition-monotone-params}

Consider any $n_{a}>n_{0}$. Recall that we set $d_{a}=d\le n_{0}$
in this case. Thus, by Lemma \ref{lem:ksubmod-qa} and Lemma \ref{lem:exp-approx},
we have
\begin{align*}
\frac{1}{Q_{a}} & =\frac{1}{\left(1+d\right)\left(1+\frac{1}{\left(1+\frac{d}{n_{a}}\right)^{n_{a}}-1}\right)}\geq\frac{1}{\left(1+d\right)\left(1+\frac{1}{\exp\left(d\right)-1}\right)}\left(1-\frac{1}{n_{a}}\cdot\frac{n_{0}\left(\exp\left(\frac{d^{2}}{n_{0}}\right)-1\right)}{\exp\left(d\right)-1}\right)
\end{align*}
Plugging in $d=1.1461$ and $n_{0}=3$, we obtain
\[
\frac{1}{Q_{a}}\geq0.3178\left(1-\frac{0.7681}{n_{a}}\right)
\]
Note that the above is $\geq0.25$ for all $n_{a}\geq4$. Overall,
we obtain that the approximation is $\geq0.25$ and it tends to $\geq0.3178$
as $\min_{a}n_{a}$ tends to infinity.
\end{proof}

\subsection{\label{subsec:k-submod-nonmon-appendix} Non-Monotone $k$-Submodular
Maximization: $\max_{a}n_{a}\le\frac{1}{2}\sum_{a}n_{a}$}

\begin{algorithm}
\textbf{Parameters}: $\left\{ g_{a}(i)\right\} _{a\in[k],i\in[n_{a}]}$

\textbf{Input}: $k$-submodular function $f$, budgets $\left\{ n_{a}\right\} _{a\in[k]}$

$\S=\left(S_{1},\dots,S_{k}\right)\gets\left(\emptyset,\dots,\emptyset\right)$

$\beta_{a}\gets0$ for all $a\in\left[k\right]$

\textbf{for} $t=1,2,\dots,\left|V\right|$:

$\quad$let $w_{t,a}=\Delta_{t,a}f\left({\bf S}\right)$ for all $a\in[k]$

$\quad$let $a=\arg\max_{a\in[k]}\left\{ \Delta_{t,a}f(\S)-\beta_{a}-\min_{a'\neq a}\beta_{a'}\right\} $

$\quad$\textbf{if} $w_{t,a}-\beta_{a}\ge0$:

$\quad\quad$\textbf{if} $\left|S_{a}\right|<n_{a}$:

$\quad\quad\quad$$S_{a}\gets S_{a}\cup\left\{ t\right\} $

$\quad\quad$\textbf{else}:

$\quad\quad\quad$let $t'=\arg\min_{i\in S_{a}}w_{i,a}$

$\quad\quad\quad$$S_{a}\gets\left(S_{a}\setminus\left\{ t'\right\} \right)\cup\left\{ t\right\} $

$\quad\quad$let $w_{a}(i)$ be the $i$-th largest weight in $\left\{ w_{t,a}\colon t\in S_{a}\right\} $
and $w_{a}(i)=0$ for $i>\left|S_{a}\right|$

$\quad\quad$$\beta_{a}\gets\sum_{i=1}^{n_{a}}w_{a}(i)g_{a}(i)$

\textbf{return} ${\bf S}$

\caption{\label{alg:ksubmod-nonmon} Non-monotone $k$-submodular maximization
for the case $\max_{a}n_{a}\le\frac{1}{2}\sum_{a}n_{a}$.}
\end{algorithm}

In this section, we present and analyze an algorithm (Algorithm \ref{alg:ksubmod-nonmon})
that works when the maximum budget is at most half the total budget,
i.e. $\max_{a}n_{a}\le\frac{1}{2}\sum_{a}n_{a}$. We show how to generalize
this approach to any budget in Section \ref{subsec:k-submod-any-max-budget}.
The algorithm uses the same choice of coefficients $\left\{ g_{a}(i)\right\} _{a\in[k],i\in[n_{a}]}$
as the monotone algorithm (Section \ref{subsec:ksubmod-mon-algo}).

\subsubsection{Analysis }

We follow the proof structure of Theorem \ref{thm:ksubmod} in the
monotone case. We start with suitable lower and upper bounds for $f(\S)$
and $f(\S^{*})$. 
\begin{lem}
\label{lem:ksubmod-nonmon-bound-sol} The value of solution ${\bf S}$
is at least
\[
f({\bf S})\ge\sum_{a}\sum_{t\in S_{a}}w_{t,a}.
\]
\end{lem}

\begin{proof}
This is the same as in the monotone analysis, since that proof only
relies on orthant submodularity of $f$.
\end{proof}

\begin{lem}
\label{lem:ksubmod-nonmon-bound-opt} The value of the optimum solution
${\bf S}^{*}$ is at most
\[
f({\bf S}^{*})\leq\sum_{a}\left(\sum_{t\in T_{a}}\left(3w_{t,a}-\beta_{a}^{(t-1)}\right)+2n_{a}\beta_{a}\right).
\]
\end{lem}

\begin{proof}
Let $\O^{(t)}$ be the allocation that agrees with $\T^{(t)}$ on
items $\left\{ 1,\dots,t\right\} $, and it agrees with $\S^{*}$
on items $\left\{ t+1,\dots,\left|V\right|\right\} $. Let $\widetilde{\O}^{(t-1)}$
be the allocation obtained from $\O^{(t)}$ by dropping $t$ (i.e.,
$t$ is not assigned to any part under $\widetilde{\O}^{(t-1)}$).
For $t\in\supp(\T)$, let $a(t)$ be the part such that $t\in T_{a}$.
For $t\in\supp(\S^{*})$, let $a^{*}(t)$ be the part such that $t\in S_{a}^{*}$.

We have
\begin{align*}
 & f(\S^{*})-f(\T)\\
 & =f(\O^{(0)})-f(\O^{\left|V\right|})=\sum_{t=1}^{\left|V\right|}\left(f(\O^{(t-1)})-f(\O^{(t)})\right)\\
 & =\sum_{t\in\supp(\T)\cap\supp(\S^{*})}\left(f(\O^{(t-1)})-f(\O^{(t)})\right)+\sum_{t\notin\supp(\T)\cup\supp(\S^{*})}\left(f(\O^{(t-1)})-f(\O^{(t)})\right)\\
 & +\sum_{t\in\supp(\T)\setminus\supp(\S^{*})}\left(f(\O^{(t-1)})-f(\O^{(t)})\right)+\sum_{t\in\supp(\S^{*})\setminus\supp(\T)}\left(f(\O^{(t-1)})-f(\O^{(t)})\right)\\
 & =\sum_{t\in\supp(\T)\setminus\supp(\S^{*})}\left(f(\O^{(t-1)})-f(\O^{(t)})\right)+\sum_{t\in\supp(\S^{*})\setminus\supp(\T)}\left(f(\O^{(t-1)})-f(\O^{(t)})\right)
\end{align*}
\begin{itemize}
\item Consider $t\in\supp(\T)\cap\supp(\S^{*})$. If $a(t)=a^{*}(t)$, we
have $\O^{(t-1)}=\O^{(t)}$, and thus
\[
f(\O^{(t-1)})-f(\O^{(t)})=0
\]
If $a(t)\neq a^{*}(t)$, we have
\begin{align*}
f(\O^{(t-1)})-f(\O^{(t)}) & =f(\O^{(t-1)})-f(\widetilde{\O}^{(t-1)})+f(\widetilde{\O}^{(t-1)})-f(\O^{(t)})\\
 & =\Delta_{t,a^{*}(t)}f(\widetilde{\O}^{(t-1)})-\Delta_{t,a(t)}f(\widetilde{\O}^{(t-1)})\\
 & \leq\Delta_{t,a^{*}(t)}f(\S^{(t-1)})-\Delta_{t,a(t)}f(\widetilde{\O}^{(t-1)})
\end{align*}
where the inequality is due to orthant submodularity since $\S^{(t-1)}\preceq\widetilde{\O}^{(t-1)}$.
\\
Let $a\in\arg\min_{a'\neq a(t)}\beta_{a'}^{(t-1)}$. We have
\[
-\Delta_{t,a(t)}f(\widetilde{\O}^{(t-1)})\leq\Delta_{t,a}f(\widetilde{\O}^{(t-1)})\le\Delta_{t,a}f(\S^{(t-1)})\leq\Delta_{t,a(t)}f(\S^{(t-1)})
\]
where the first inequality is by pairwise monotonicity, the second
is by orthant submodularity since $\S^{(t-1)}\preceq\widetilde{\O}^{(t-1)}$,
and the third is due to $a(t)$ having the largest modified discounted
gain:
\begin{align*}
\Delta_{t,a}f(\S^{(t-1)})-\beta_{a}^{(t-1)}-\min_{a'\neq a}\beta_{a'}^{(t-1)} & \leq\Delta_{t,a(t)}f(\S^{(t-1)})-\beta_{a(t)}^{(t-1)}-\min_{a'\neq a(t)}\beta_{a'}^{(t-1)}\\
 & =\Delta_{t,a(t)}f(\S^{(t-1)})-\beta_{a(t)}^{(t-1)}-\beta_{a}^{(t-1)}\\
\Rightarrow\Delta_{t,a}f(\S^{(t-1)}) & \leq\Delta_{t,a(t)}f(\S^{(t-1)})-\beta_{a(t)}^{(t-1)}+\min_{a'\neq a}\beta_{a'}^{(t-1)}\\
 & \leq\Delta_{t,a(t)}f(\S^{(t-1)})-\beta_{a(t)}^{(t-1)}+\beta_{a(t)}^{(t-1)}\\
 & =\Delta_{t,a(t)}f(\S^{(t-1)})
\end{align*}
Thus
\[
f(\O^{(t-1)})-f(\O^{(t)})\leq\Delta_{t,a^{*}(t)}f(\S^{(t-1)})+\Delta_{t,a(t)}f(\S^{(t-1)})=w_{t,a^{*}(t)}+w_{t,a(t)}
\]
\item Consider $t\in\supp(\T)\setminus\supp(\S^{*})$. We have
\begin{align*}
f(\O^{(t-1)})-f(\O^{(t)}) & =-\Delta_{t,a(t)}f(\O^{(t-1)})=-\Delta_{t,a(t)}f(\widetilde{\O}^{(t-1)})
\end{align*}
Using the same argument as above, we obtain
\[
-\Delta_{t,a(t)}f(\widetilde{\O}^{(t-1)})\leq\Delta_{t,a(t)}f(\S^{(t-1)})
\]
Thus
\[
f(\O^{(t-1)})-f(\O^{(t)})\leq\Delta_{t,a(t)}f(\S^{(t-1)})=w_{t,a(t)}
\]
\item Consider $t\in\supp(\S^{*})\setminus\supp(\T).$ We have 
\begin{multline*}
f(\O^{(t-1)})-f(\O^{(t)})=\Delta_{t,a^{*}(t)}f(\O^{(t)})=\Delta_{t,a^{*}(t)}f(\widetilde{\O}^{(t-1)})\\
\leq\Delta_{t,a^{*}(t)}f(\S^{(t-1)})\leq\beta_{a^{*}(t)}^{(t-1)}+\min_{a\neq a^{*}(t)}\beta_{a}^{(t-1)}
\end{multline*}
In the first inequality we used orthant submodularity since $\S^{(t-1)}\preceq\widetilde{\O}^{(t-1)}$.
In the second inequality, we used that $t\notin\supp(\T)$, and thus
\[
\Delta_{t,a}f(\S^{(t-1)}-\beta_{a}^{(t-1)}-\min_{a'\neq a}\beta_{a'}^{(t-1)}\leq0\quad\forall a\in[k]
\]
\item Consider $t\notin\supp(\T)\cup\supp(\S^{*})$. We have $\O^{(t-1)}=\O^{(t)}$,
and thus
\[
f(\O^{(t-1)})-f(\O^{(t)})=0
\]
\end{itemize}
Putting everything together, and using that $f(\T)\leq\sum_{t\in\supp(\T)}w_{t,a(t)}$,
we obtain
\begin{align*}
f(\S^{*}) & \leq\sum_{t\in\supp(\T)}w_{t,a(t)}+\sum_{t\in\supp(\T)\cap\supp(\S^{*})}\left(w_{t,a^{*}(t)}+w_{t,a(t)}\right)\\
 & \quad+\sum_{t\in\supp(\T)\setminus\supp(\S^{*})}w_{t,a(t)}+\sum_{t\in\supp(\S^{*})\setminus\supp(\T)}\left(\beta_{a^{*}(t)}^{(t-1)}+\min_{a\neq a^{*}(t)}\beta_{a}^{(t-1)}\right)\\
 & =\sum_{t\in\supp(\T)}2w_{t,a(t)}+\sum_{t\in\supp(\T)\cap\supp(\S^{*})}w_{t,a^{*}(t)}\\
 & \quad+\sum_{t\in\supp(\S^{*})\setminus\supp(\T)}\left(\beta_{a^{*}(t)}^{(t-1)}+\min_{a\neq a^{*}(t)}\beta_{a}^{(t-1)}\right)\\
 & =\sum_{t\in\supp(\T)}2w_{t,a(t)}+\sum_{t\in\supp(\T)\cap\supp(\S^{*})}\left(w_{t,a^{*}(t)}-\beta_{a^{*}(t)}^{(t-1)}-\min_{a\neq a^{*}(t)}\beta_{a}^{(t-1)}\right)\\
 & \quad+\sum_{t\in\supp(\S^{*})}\left(\beta_{a^{*}(t)}^{(t-1)}+\min_{a\neq a^{*}(t)}\beta_{a}^{(t-1)}\right)\\
 & \overset{(1)}{\leq}\sum_{t\in\supp(\T)}2w_{t,a(t)}+\sum_{t\in\supp(\T)\cap\supp(\S^{*})}\left(w_{t,a(t)}-\beta_{a(t)}^{(t-1)}-\min_{a\neq a(t)}\beta_{a}^{(t-1)}\right)\\
 & \quad+\sum_{t\in\supp(\S^{*})}\left(\beta_{a^{*}(t)}^{(t-1)}+\min_{a\neq a^{*}(t)}\beta_{a}^{(t-1)}\right)\\
 & \overset{(2)}{\leq}\sum_{t\in\supp(\T)}2w_{t,a(t)}+\sum_{t\in\supp(\T)}\left(w_{t,a(t)}-\beta_{a(t)}^{(t-1)}-\min_{a\neq a(t)}\beta_{a}^{(t-1)}\right)\\
 & \quad+\sum_{t\in\supp(\S^{*})}\left(\beta_{a^{*}(t)}^{(t-1)}+\min_{a\neq a^{*}(t)}\beta_{a}^{(t-1)}\right)\\
 & =\sum_{t\in\supp(\T)}\left(3w_{t,a(t)}-\beta_{a(t)}^{(t-1)}-\min_{a\neq a(t)}\beta_{a}^{(t-1)}\right)+\sum_{t\in\supp(\S^{*})}\left(\beta_{a^{*}(t)}^{(t-1)}+\min_{a\neq a^{*}(t)}\beta_{a}^{(t-1)}\right)\\
 & \overset{(3)}{\leq}\sum_{t\in\supp(\T)}\left(3w_{t,a(t)}-\beta_{a(t)}^{(t-1)}\right)+\sum_{t\in\supp(\S^{*})}\left(\beta_{a^{*}(t)}^{(t-1)}+\min_{a\neq a^{*}(t)}\beta_{a}^{(t-1)}\right)
\end{align*}
where $(1)$ follows from the choice of $a(t)$, $(2)$ follows from
the fact that every $t\in\supp(\T)$ has non-negative modified discounted
gain, and $(3)$ follows from the thresholds being non-negative.

Next, we relate $(\star):=\sum_{t\in\supp(\S^{*})}\min_{a\neq a^{*}(t)}\beta_{a}^{(t-1)}$
to $\sum_{a}n_{a}\beta_{a}$. By relabeling the parts, we may assume
without loss of generality that the \emph{final} thresholds satisfy
$\beta_{1}\leq\beta_{2}\leq\dots\leq\beta_{k}$. Using that the thresholds
are non-decreasing and $\left|S_{a}^{*}\right|\leq n_{a}$ for all
$a\in[k]$, we can show that
\[
(\star):=\sum_{t\in\supp(\S^{*})}\min_{a\neq a^{*}(t)}\beta_{a}^{(t-1)}=\sum_{a=1}^{k}\sum_{t\in S_{a}^{*}}\min_{a'\neq a}\beta_{a'}^{(t-1)}\leq n_{1}\beta_{2}+\sum_{a=2}^{k}n_{a}\beta_{1}
\]
For every $t\in S_{1}^{*}$, we have $\min_{a'\neq1}\beta_{a'}^{(t-1)}\leq\beta_{2}^{(t-1)}\leq\beta_{2}$.
Thus $\sum_{t\in S_{1}^{*}}\min_{a'\neq1}\beta_{a'}^{(t-1)}\leq n_{1}\beta_{2}$.
Consider any $a\geq2$. For every $t\in S_{a}^{*}$, we have $\min_{a'\neq a}\beta_{a'}^{(t-1)}\leq\beta_{1}^{(t-1)}\leq\beta_{1}$.
Thus $\sum_{t\in S_{a}^{*}}\min_{a'\neq a}\beta_{a'}^{(t-1)}\leq n_{a}\beta_{1}$.

Let $\alpha$ be such that $\max_{a}n_{a}=\left(1-\alpha\right)\left(\sum_{a=1}^{k}n_{a}\right)$.
Thus we have $n_{1}\leq\frac{1-\alpha}{\alpha}\left(\sum_{a=2}^{k}n_{a}\right)$.
We have
\begin{align*}
(\star) & \leq n_{1}\beta_{2}+\sum_{a=2}^{k}n_{a}\beta_{1}\\
 & \leq\frac{n_{1}}{\sum_{a=2}^{k}n_{a}}\left(\sum_{a=2}^{k}n_{a}\beta_{a}\right)+\sum_{a=2}^{k}n_{a}\beta_{1}\\
 & =\frac{n_{1}}{\sum_{a=2}^{k}n_{a}}\left(\sum_{a=1}^{k}n_{a}\beta_{a}\right)+\underbrace{\left(\sum_{a=2}^{k}n_{a}-\frac{n_{1}^{2}}{\sum_{a=2}^{k}n_{a}}\right)}_{(\diamond)}\beta_{1}
\end{align*}
If $(\diamond)\leq0$, we have
\[
(\star)\leq\frac{n_{1}}{\sum_{a=2}^{k}n_{a}}\left(\sum_{a=1}^{k}n_{a}\beta_{a}\right)\leq\frac{1-\alpha}{\alpha}\left(\sum_{a=1}^{k}n_{a}\beta_{a}\right)
\]
If $(\diamond)\geq0$, we have
\[
(\star)\leq\frac{n_{1}}{\sum_{a=2}^{k}n_{a}}\left(\sum_{a=1}^{k}n_{a}\beta_{a}\right)+\left(\sum_{a=2}^{k}n_{a}-\frac{n_{1}^{2}}{\sum_{a=2}^{k}n_{a}}\right)\frac{1}{\sum_{a=1}^{k}n_{a}}\left(\sum_{a=1}^{k}n_{a}\beta_{a}\right)=\sum_{a=1}^{k}n_{a}\beta_{a}
\]
Thus
\[
(\star)\leq\max\left\{ \frac{1-\alpha}{\alpha},1\right\} \left(\sum_{a=1}^{k}n_{a}\beta_{a}\right)
\]
Plugging into the previous inequality, we obtain
\begin{align*}
f(\S^{*}) & \leq\sum_{t\in\supp(\T)}\left(3w_{t,a(t)}-\beta_{a(t)}^{(t-1)}\right)+\sum_{t\in\supp(\S^{*})}\beta_{a^{*}(t)}^{(t-1)}+\max\left\{ \frac{1-\alpha}{\alpha},1\right\} \left(\sum_{a=1}^{k}n_{a}\beta_{a}\right)\\
 & \le\sum_{t\in\supp(\T)}\left(3w_{t,a(t)}-\beta_{a(t)}^{(t-1)}\right)+\max\left\{ \frac{1}{\alpha},2\right\} \left(\sum_{a=1}^{k}n_{a}\beta_{a}\right)
\end{align*}
\end{proof}

In light of Lemma \ref{lem:ksubmod-nonmon-bound-sol} and Lemma \ref{lem:ksubmod-nonmon-bound-opt},
it is sufficient to compare on a per-part basis, as we have done it
for the monotone case. In particular, we show:
\begin{lem}
\label{lem:ksubmod-nonmon-qa} For every part $a\in[k]$, we have
\[
\sum_{t\in T_{a}}\left(3w_{t,a}-\beta_{a}^{(t-1)}\right)+2n_{a}\beta_{a}\le Q_{a}\sum_{t\in S_{a}}w_{t,a}
\]
where $d_{a}\ge\frac{1}{2}$ and
\[
Q_{a}=2\left(1+d_{a}\right)\left(1+\frac{1}{\left(1+\frac{d_{a}}{n_{a}}\right)^{n_{a}}-1}\right).
\]
\end{lem}

\begin{proof}
We define our primal and dual potential as
\begin{align*}
P_{t} & \coloneqq\sum_{i\in S_{a}^{(t)}}w_{i,a}\\
D_{t} & \coloneqq\sum_{i\in T_{a}^{(t)}}\left(3w_{i,a}-\beta_{a}^{(i-1)}\right)+2n_{a}\beta_{a}^{(t)}.
\end{align*}
Note that we have $P_{0}=D_{0}=0$, $P_{T}=\sum_{t\in S_{a}}w_{t}$,
and $D_{T}=\sum_{t\in T_{a}}\sum_{t\in T_{a}}\left(3w_{t,a}-\beta_{a}^{(t-1)}\right)+2n_{a}\beta_{a}$.
Thus it suffices to show that $D_{t}-D_{t-1}\leq Q_{a}$$\left(P_{t}-P_{t-1}\right)$
for all $t$.

If $t\notin T_{a}$, we have $\beta_{a}^{(t)}=\beta_{a}^{(t-1)}$
and thus $P_{t}-P_{t-1}=D_{t}-D_{t-1}=0$. Thus we may assume that
$t\in T_{a}$, and thus $w_{t,a}\geq\beta_{a}^{(t-1)}$. We have
\begin{align*}
D_{t}-D_{t-1} & =3w_{t,a}-\beta_{a}^{(t-1)}+2n_{a}\left(\beta_{a}^{(t)}-\beta_{a}^{(t-1)}\right)\\
P_{t}-P_{t-1} & =w_{t,a}-\min_{i\in S_{a}^{(t-1)}}w_{ai}
\end{align*}
Suppose that we choose $d_{a}$ so that $2d_{a}-1\geq0$. Using Lemma
\ref{lem:ksubmod-beta-change} and $\beta_{a}^{(t-1)}\leq w_{t,a}$,
and obtain:
\begin{align*}
 & 3w_{t,a}-\beta_{a}^{(t-1)}+2n_{a}\left(\beta_{a}^{(t)}-\beta_{a}^{(t-1)}\right)\\
 & \leq\underbrace{\left(2d_{a}-1\right)}_{\geq0}\beta_{a}^{(t-1)}+\left(3+2c_{a}\right)w_{t,a}-2c_{a}\left(1+\frac{d_{a}}{n_{a}}\right)^{n_{a}}\left(\min_{i\in S_{a}^{(t-1)}}w_{i,a}\right)\\
 & \leq\left(2+2d_{a}+2c_{a}\right)w_{t,a}-2c_{a}\left(1+\frac{d_{a}}{n_{a}}\right)^{n_{a}}\left(\min_{i\in S_{a}^{(t-1)}}w_{i,a}\right)
\end{align*}
We set $c_{a}$ so that
\[
2+2d_{a}+2c_{a}=2c_{a}\left(1+\frac{d_{a}}{n_{a}}\right)^{n_{a}}\iff c_{a}=\frac{2+2d_{a}}{2\left(\left(1+\frac{d_{a}}{n_{a}}\right)^{n_{a}}-1\right)}
\]
and obtain
\[
Q_{a}=2+2d_{a}+2c_{a}=\left(2+2d_{a}\right)\left(1+\frac{1}{\left(1+\frac{d_{a}}{n_{a}}\right)^{n_{a}}-1}\right)
\]
\end{proof}

We thus get $Qf(\S)\ge f(\S^{*})$ for $Q\coloneqq\max_{a}Q_{a}$.

\subsubsection{Setting the Parameters}

As shown in Lemma \ref{lem:ksubmod-nonmon-qa}, $Q_{a}$ is exactly
twice as large as in \ref{lem:ksubmod-qa}. We can thus use the same
parameters as in the monotone case (cf. Theorem \ref{thm:ksubmod}),
and obtain an approximation that is $\frac{1}{2}$ of the monotone
approximation.

Note that the condition $\max_{a}n_{a}\le\frac{1}{2}\sum_{a}n_{a}$
is only for simplicity of presentation. Indeed, we can obtain guarantees
for any $0<\alpha<\frac{1}{2}$ with $\max_{a\in[k]}n_{a}\le(1-\alpha)\sum_{a}n_{a}$.
In this case,
\[
Q_{a}=\left(2+\frac{1}{\alpha}d_{a}\right)\left(1+\frac{1}{\left(1+\frac{d_{a}}{n_{a}}\right)^{n_{a}}-1}\right)
\]
which we can optimize independently of Theorem \ref{thm:ksubmod}. 

\subsection{\label{subsec:k-submod-any-max-budget} Non-Monotone $k$-Submodular
Maximization: Any Budget}

In this section, we show how to derive an algorithm for any budget
from Algorithms \ref{alg:ksubmod-nonmon} and \ref{alg:partition-nonmon}.
Our algorithm for any budget case works as follows. Without loss of
generality, suppose that the first part has the maximum budget. We
construct two solutions. For the first solution, we solve the submodular
maximization problem with a cardinality constraint $\max_{\left|S\right|\leq n_{1}}g(S)$,
where $g(S):=f(S,\emptyset,\dots,\emptyset)$ (i.e., we only allocate
to part $1$, which is the one with maximum budget) using Algorithm
\ref{alg:partition-nonmon}. Let $\A=\left(S,\emptyset,\dots,\emptyset\right)$
be the solution obtained. Let $\widehat{n}_{1}=\sum_{a=2}^{k}n_{a}$.
For the second solution, we solve the problem of maximizing $f$ but
subject to the lower budget $\widehat{n}_{1}$ for part $1$ (i.e.,
we lower the budget of part $1$, and we keep the budgets of the other
parts the same) using Algorithm \ref{alg:ksubmod-nonmon}. Let $\B$
be the solution obtained. We output the better of the two solutions.

We can show the following guarantee:
\begin{thm}
The algorithm for non-monotone $k$-submodular maximization with cardinality
constraints for any budget achieves an approximation guarantee of
\[
\mathbb{E}\left[\max\left\{ f(\A),f(\B)\right\} \right]\ge\frac{1}{\frac{1}{\alpha_{\mathrm{nmp}}}+\frac{2}{\alpha_{\mathrm{mon}}}}f(\S^{*})
\]
where $\alpha_{\mathrm{nmp}}$ is the approximation guarantee we derive
for submodular maximization with a partition matroid constraint (Theorem
\ref{thm:partition-nonmonotone}) and $\frac{1}{2}\alpha_{\mathrm{mon}}$
is the approximation guarantee we derived for $k$-submodular maximization
when the maximum budget is at most $\frac{1}{2}$ of the total budget
(Theorem \ref{thm:ksubmod}).
\end{thm}

\begin{proof}
For the first solution, Theorem \ref{thm:partition-nonmonotone} gives
an approximation guarantee $\alpha_{\mathrm{nmp}}$. For the second
solution, Theorem \ref{thm:ksubmod} gives an approximation guarantee
$\frac{1}{2}\alpha_{\mathrm{mon}}$. Thus we have
\begin{align*}
\E\left[f(\A)\right] & \geq\alpha_{\mathrm{nmp}}\cdot f\left(S_{1}^{*},\emptyset,\dots,\emptyset\right)\\
f(\B) & \geq\frac{1}{2}\alpha_{\mathrm{mon}}\cdot f\left(\emptyset,S_{2}^{*},\dots,S_{k}^{*}\right).
\end{align*}
 Recall that the definition of $k$-submodularity is that
\[
f(\X)+f(\Y)\geq f(\X\sqcap\Y)+f(\X\sqcup\Y).
\]
Applying the above with $\X=\left(S_{1}^{*},\emptyset,\dots,\emptyset\right)$
and $\Y=\left(\emptyset,S_{2}^{*},\dots,S_{k}^{*}\right)$ and noting
that $f(\X\sqcap\Y)\geq0$ and $f(\X\sqcup\Y)=f(\S^{*})$, we obtain
\begin{align*}
\E\left[\frac{1}{\alpha_{\mathrm{nmp}}}f(\A)+\frac{2}{\alpha_{\mathrm{mon}}}f(\B)\right] & \geq f\left(S_{1}^{*},\emptyset,\dots,\emptyset\right)+f\left(\emptyset,S_{2}^{*},\dots,S_{k}^{*}\right)\geq f(\S^{*}).
\end{align*}
Thus
\[
\E\left[\max\left\{ f(\A),f(\B)\right\} \right]\geq\frac{1}{\frac{1}{\alpha_{\mathrm{nmp}}}+\frac{2}{\alpha_{\mathrm{mon}}}}\E\left[\frac{1}{\alpha_{\mathrm{nmp}}}f(\A)+\frac{2}{\alpha_{\mathrm{mon}}}f(\B)\right]\geq\frac{1}{\frac{1}{\alpha_{\mathrm{nmp}}}+\frac{2}{\alpha_{\mathrm{mon}}}}f(\S^{*}).
\]
\end{proof}

The above gives a streaming algorithm since we construct two solutions
instead of one. We can also get an online algorithm in the oblivious
adversary setting by randomly choosing between the two solutions,
where with probability $q=\frac{\frac{1}{\alpha_{\mathrm{nmp}}}}{\frac{1}{\alpha_{\mathrm{nmp}}}+\frac{2}{\alpha_{\mathrm{mon}}}}$
we construct $\A$. We get the same guarantee in expectation.

\subsection{\label{subsec:partition-appendix} Monotone Submodular Maximization
with a Partition Matroid Constraint}

\begin{algorithm}
\textbf{Parameters}: $\left\{ g_{a}(i)\right\} _{a\in[k],i\in[n_{a}]}$

\textbf{Input}:\textbf{ }monotone submodular function $f$, partition
${\cal P}=\left(P_{1},\dots,P_{k}\right)$, budgets $n_{1},\dots,n_{k}$.

$S\gets\emptyset$

$\beta_{a}\gets0$ for all $a\in\left[k\right]$

\textbf{for} $t=1,2,\dots,\left|V\right|$:

$\quad$let $a$ be such that $t\in P_{a}$

$\quad$let $w_{t}=f\left(S\cup\left\{ t\right\} \right)-f\left(S\right)$

$\quad$\textbf{if} $w_{t}-\beta_{a}\ge0$:

$\quad\quad$\textbf{if} $\left|S\cap P_{a}\right|<n_{a}$:

$\quad\quad\quad$$S\gets S\cup\left\{ t\right\} $

$\quad\quad$\textbf{else}:

$\quad\quad\quad$let $t'=\arg\min_{i\in S\cap P_{a}}w_{i}$

$\quad\quad\quad$$S\gets\left(S\setminus\left\{ t'\right\} \right)\cup\left\{ t\right\} $

$\quad\quad$let $w_{a}(i)$ be the $i$-th largest weight in $\left\{ w_{t}\colon t\in S\cap P_{a}\right\} $
and $w_{a}(i)=0$ for $i>\left|S\cap P_{a}\right|$

$\quad\quad$$\beta_{a}\gets\sum_{i=1}^{n_{a}}w_{a}(i)g_{a}(i)$

\textbf{return} ${\bf S}$

\caption{\label{alg:partition-mon} Monotone submodular maximization with a
partition matroid constraint.}
\end{algorithm}
We immediately obtain a guarantee for monotone submodular maximization
under a partition matroid constraint through our algorithm for monotone
$k$-submodular maximization. In particular, given a monotone submodular
function $f$ and a partition matroid ${\cal P}=(P_{1},\dots,P_{k})$
with associated budgets $n_{1},\dots,n_{k}$, we can create an instance
of $k$-submodular maximization with the same budgets using
\[
g(\X)\coloneqq f\left({\textstyle \bigcup_{a}}(P_{a}\cap X_{a})\right).
\]
We can easily verify that $g$ is indeed $k$-submodular: For all
$k$-sets $\X,\Y\in(k+1)^{V}$,
\begin{align*}
g(\X)+g(\Y) & =f\left({\textstyle \bigcup_{a}}(P_{a}\cap X_{a})\right)+f\left({\textstyle \bigcup_{a}}(P_{a}\cap Y_{a})\right)\\
 & \ge f\left({\textstyle \bigcup_{a}}\left(P_{a}\cap(X_{a}\cap Y_{a})\right)\right)+f\left({\textstyle \bigcup_{a}}\left(P_{a}\cap(X_{a}\cup Y_{a})\right)\right)\\
 & \ge f\left({\textstyle \bigcup_{a}}\left(P_{a}\cap(X_{a}\cap Y_{a})\right)\right)+f\left({\textstyle \bigcup_{a}}\left(P_{a}\cap(X_{a}\cup Y_{a})\setminus{\textstyle \bigcup_{b\not=a}}(X_{b}\cup Y_{b})\right)\right)\\
 & =g(\X\sqcap\Y)+g(\X\sqcup\Y)
\end{align*}
where the first and second inequalities are due to submodularity and
monotonicity of $f$, respectively. 

For completeness, we state the algorithm for monotone submodular maximization
with a partition matroid in Algorithm \ref{alg:partition-mon}. We
use the same choice of coefficients $\left\{ g_{a}(i)\right\} _{a\in[k],i\in[n_{a}]}$
and obtain the same guarantee as for the monotone $k$-submodular
problem.
\begin{thm}
\label{thm:partition} When setting the parameters $\{d_{a}\}_{a\in[k]}$
according to the choices of Theorem \ref{thm:ksubmod}, Algorithm
\ref{alg:partition-mon} achieves the same approximation guarantee
as in Theorem \ref{thm:ksubmod}.
\end{thm}

\subsection{\label{subsec:partition-nonmon-appendix} Non-Monotone Submodular
Maximization with a Partition Matroid Constraint}

\begin{algorithm}
\textbf{Parameters}: $\left\{ g_{a}(i)\right\} _{a\in[k],i\in[n_{a}]}$

\textbf{Input}: submodular function $f$, partition ${\cal P}=\left(P_{1},\dots,P_{k}\right)$,
budgets $n_{1},\dots,n_{k}$.

$S\gets\emptyset$

$\beta_{a}\gets0$ for all $a\in\left[k\right]$

\textbf{for} $t=1,2,\dots,\left|V\right|$:

$\quad$let $a$ be such that $t\in P_{a}$

$\quad$let $w_{t}=f\left(S\cup\left\{ t\right\} \right)-f\left(S\right)$

$\quad$let $Z_{t}\sim\mathrm{Ber}(p)$

$\quad$\textbf{if} $w_{t}-\beta_{a}\ge0$ and $Z_{t}=1$:

$\quad\quad$\textbf{if} $\left|S\cap P_{a}\right|<n_{a}$:

$\quad\quad\quad$$S\gets S\cup\left\{ t\right\} $

$\quad\quad$\textbf{else}:

$\quad\quad\quad$let $t'=\arg\min_{i\in S\cap P_{a}}w_{i}$

$\quad\quad\quad$$S\gets\left(S\setminus\left\{ t'\right\} \right)\cup\left\{ t\right\} $

$\quad\quad$let $w_{a}(i)$ be the $i$-th largest weight in $\left\{ w_{t}\colon t\in S\cap P_{a}\right\} $
and $w_{a}(i)=0$ for $i>\left|S\cap P_{a}\right|$

$\quad\quad$$\beta_{a}\gets\sum_{i=1}^{n_{a}}w_{a}(i)g_{a}(i)$

\textbf{return} ${\bf S}$

\caption{\label{alg:partition-nonmon} Non-monotone submodular maximization
with a partition matroid constraint.}
\end{algorithm}
We use the standard approach of subsampling to extend our monotone
algorithm for submodular maximization with a partition matroid setting
to non-monotone objectives. Specifically, we sub-sample each element
with probability $p$ before adding it to the solution. 

Our algorithm is described in Algorithm \ref{alg:partition-nonmon}
and as before, we define, for all $a\in[k]$, 
\[
g_{a}(i)\coloneqq\frac{c_{a}}{n_{a}}\left(1+\frac{d_{a}}{n_{a}}\right)^{i-1}\qquad\mathrm{for}\qquad c_{a}\coloneqq\frac{1+d_{a}}{\left(1+\frac{d_{a}}{n_{a}}\right)^{n_{a}}-1}
\]
for $i\in[n_{a}]$, and positive constants positive constants $\{d_{a}\}_{a\in[k]}$
that we specify in Theorem \ref{thm:partition-nonmonotone}.

We note that, although subsampling is a well-known approach for deriving
an algorithm for non-monotone objectives, integrating the subsampling
into our analysis framework requires new insights. Additionally, we
obtain approximation guarantees that improve upon the previously best
guarantees for discrete algorithms due to \citet{feldman18}. Similarly
to \citet{feldman18}, we are able to show that the subsampling is
beneficial on two fronts: it reduces the number of evaluations while
achieving improved approximation guarantees. In particular, there
is an intricate interplay between the subsampling parameter $p$ and
the parameters $c_{a}$ and $d_{a}$ that we use to set the coefficients
for the thresholds. We refer the reader to the proof of Theorem \ref{thm:partition-nonmonotone}
below for more details.

\begin{table}
\begin{centering}
\caption{Parameter choices and approximation guarantee for non-monotone submodular
maximization with a partition matroid constraint.}
\setlength{\tabcolsep}{2.5pt}\medskip{}
{\scriptsize{}}%
\begin{tabular}{cccccccccccc}
\hline 
\multicolumn{12}{c}{\textbf{\scriptsize{}Small budget case $\min_{a}n_{a}\leq10$: $p=0.3$}}\tabularnewline
\hline 
{\scriptsize{}$n_{a}$} & {\scriptsize{}$1$} & {\scriptsize{}$2$} & {\scriptsize{}$3$} & {\scriptsize{}$4$} & {\scriptsize{}$5$} & {\scriptsize{}$6$} & {\scriptsize{}$7$} & {\scriptsize{}$8$} & {\scriptsize{}$9$} & {\scriptsize{}$10$} & {\scriptsize{}$\geq11$}\tabularnewline
{\scriptsize{}$d_{a}$} & {\scriptsize{}$1$} & {\scriptsize{}$1.7961$} & {\scriptsize{}$2.0654$} & {\scriptsize{}$2.1627$} & {\scriptsize{}$2.2107$} & {\scriptsize{}$2.2387$} & {\scriptsize{}$2.2567$} & {\scriptsize{}$2.2692$} & {\scriptsize{}$2.2783$} & {\scriptsize{}$2.2852$} & {\scriptsize{}$\frac{1-p}{p}=\frac{7}{3}$}\tabularnewline
{\scriptsize{}$\frac{1-p}{Q_{a}}$} & {\scriptsize{}$\geq\!0.175$} & {\scriptsize{}$\geq\!0.18$} & {\scriptsize{}$\geq\!0.18$} & {\scriptsize{}$\geq\!0.182$} & {\scriptsize{}$\geq\!0.183$} & {\scriptsize{}$\geq\!0.183$} & {\scriptsize{}$\geq\!0.184$} & {\scriptsize{}$\geq\!0.185$} & {\scriptsize{}$\geq\!0.185$} & {\scriptsize{}$\geq\!0.185$} & {\scriptsize{}$\geq\!0.1896\left(1-\frac{0.7771}{n_{a}}\right)$}\tabularnewline
\hline 
\end{tabular}{\scriptsize\par}
\par\end{centering}
\medskip{}

\begin{centering}
\setlength{\tabcolsep}{5pt}
\par\end{centering}
\begin{centering}
{\scriptsize{}}%
\begin{tabular}{cc}
\hline 
\multicolumn{2}{c}{\textbf{\scriptsize{}Large budget case $\min_{a}n_{a}\geq11$: $p\approx0.3386$}}\tabularnewline
\hline 
{\scriptsize{}$n_{a}$} & {\scriptsize{}$\geq11$}\tabularnewline
{\scriptsize{}$d_{a}$} & {\scriptsize{}$\frac{1-p}{p}=1.9532$}\tabularnewline
{\scriptsize{}$\frac{1-p}{Q_{a}}$} & {\scriptsize{}$\geq0.1921\left(1-\frac{0.7676}{n_{a}}\right)$}\tabularnewline
\hline 
\end{tabular}\textbf{\scriptsize{}\hspace{3.5em}}{\scriptsize{}}%
\begin{tabular}{ccc}
\hline 
\multicolumn{3}{c}{\textbf{\scriptsize{}Approximation guarantee $\min_{a}\frac{1-p}{Q_{a}}$}}\tabularnewline
\hline 
{\scriptsize{}$\min_{a}n_{a}$} & {\scriptsize{}$\leq10$} & {\scriptsize{}$\geq11$}\tabularnewline
{\scriptsize{}approx} & {\scriptsize{}$\geq0.175$} & {\scriptsize{}$\geq0.1921\left(1-\frac{0.7676}{\min_{a}n_{a}}\right)$}\tabularnewline
\hline 
\end{tabular}{\scriptsize\par}
\par\end{centering}
\label{tb:partition-nonmonotone-params}
\end{table}

\begin{thm}
\label{thm:partition-nonmonotone} We make the following choices for
the parameters $p$ and $\{d_{a}\}_{a\in[k]}$.
\begin{enumerate}
\item \textbf{Small budget case:} Suppose that $\min_{a\in[k]}n_{a}\leq n_{0}:=10$.
We set $p=0.3$. For every $a$ such that $n_{a}\geq11$, we set $d_{a}=\frac{1-p}{p}$.
For every $a$ such that $n_{a}\leq10$, we set $d_{a}$ as shown
in Table \ref{tb:partition-nonmonotone-params}.
\item \textbf{Large budget case:} Suppose that $\min_{a\in[k]}n_{a}>n_{0}:=10$.
Let $d=1.9532$, which is an approximate solution to the equation
$e^{d}\left(d-1\right)-d^{2}-2d+1=0$. We set $p=\frac{1}{d+1}$ and
$d_{a}=d$ for all $a\in[k]$.
\end{enumerate}
We obtain the approximation guarantees shown in Table \ref{tb:partition-nonmonotone-params}.
Note that the approximation is at least $0.175$ for any minimum budget,
and it tends to $\geq0.1921$ as the minimum budget tends to infinity.
\end{thm}

\subsubsection{Analysis }

We follow the proof structure of Theorem \ref{thm:ksubmod}. As before,
we start with appropriate lower and upper bounds on $f(S)$ and $f(S^{*})$,
respectively.
\begin{lem}
\label{lem:ksubmod-bound-sol-1} The value of solution $S$ is at
least
\[
f(S)\geq\sum_{a=1}^{k}\sum_{t\in S_{a}}w_{t,a}.
\]
\end{lem}

\begin{proof}
We calculate
\begin{align*}
\sum_{a=1}^{k}\sum_{t\in S_{a}}w_{t,a} & =\sum_{t\in S}w_{t,a(t)}\\
 & =\sum_{t\in S}\left(f\left(S^{(t-1)}\cup\left\{ t\right\} \right)-f\left(S^{(t-1)}\right)\right)\\
 & \le\sum_{t\in S}\left(f\left(S\cap S^{(t-1)}\cup\left\{ t\right\} \right)-f\left(S\cap S^{(t-1)}\right)\right)\\
 & =f(S)
\end{align*}
where the inequality is due to submodularity.
\end{proof}

We will use the following standard lemma that was shown in previous
work, and we include its proof for completeness.
\begin{lem}
The value of the optimum solution $S^{*}$ is at most
\[
\left(1-p\right)f(S^{*})\leq\E\left[f(S^{*}\cup T)\right].
\]
\end{lem}

\begin{proof}
We define the Lovasz extension $\hat{f}\colon\mathbb{R}^{V}\to\mathbb{R}$
as
\[
\hat{f}(x)=\E_{\lambda}\left[f\left(\left\{ i:x_{i}>\lambda\right\} \right)\right]
\]
where $\lambda$ is uniformly random from $[0,1]$. It is well known
that the Lovasz extension is convex if and only if $f$ is submodular.
We use this fact to bound
\begin{align*}
\E_{T}\left[f(S^{*}\cup T)\right] & =\E_{T}\left[\widehat{f}\left(1_{S^{*}\cup T}\right)\right]\\
 & \geq\widehat{f}\left(\E_{T}\left[1_{S^{*}\cup T}\right]\right)\\
 & =\E_{\lambda}\left[f\left(\left\{ i:{\textstyle \Pr_{T}}\left[i\in S^{*}\cup T\right]>\lambda\right\} \right)\right]\\
 & =\E_{\lambda}\left[f\left(S^{*}\cup\left\{ i\not\in S^{*}:{\textstyle \Pr_{T}}\left[i\in T\right]>\lambda\right\} \right)\right].
\end{align*}
where the inequality is due to Jensen's inequality. Since every element
$i\not\in S^{*}$ is in $T$ with probability at most $p$ and $f$
is non-negative,
\[
\E_{\lambda}\left[f\left(S^{*}\cup\left\{ i\not\in S^{*}:{\textstyle \Pr_{T}}\left[i\in T\right]>\lambda\right\} \right)\right]=\left(1-p\right)f\left(S^{*}\right).
\]
\end{proof}

\begin{lem}
We can further bound
\[
f(S^{*}\cup T)\leq\sum_{a=1}^{k}\left(\sum_{t\in T_{a}\setminus S_{a}^{*}}w_{t,a}+\sum_{t\in S_{a}^{*}\colon w_{t}\geq\beta_{a}^{(t-1)}}w_{t,a}+\sum_{t\in S_{a}^{*}\colon w_{t}<\beta_{a}^{(t-1)}}\beta_{a}^{(t-1)}\right).
\]
\end{lem}

\begin{proof}
Using submodularity, we can bound
\begin{align*}
f\left(S^{*}\cup T\right) & =\sum_{t\in T}\underbrace{\left(f\left(T^{(t)}\right)-f\left(T^{(t-1)}\right)\right)}_{\le w_{t,a(t)}}\\
 & +\sum_{t\in S^{*}\setminus T}\underbrace{\left(f\left(T\cup\left(S^{*}\cap\left\{ 1,\dots,t\right\} \right)\right)-f\left(T\cup\left(S^{*}\cap\left\{ 1,\dots,t-1\right\} \right)\right)\right)}_{\le f\left(T\cup\left\{ t\right\} \right)-f\left(T\right)\le f\left(S^{(t-1)}\cup\left\{ t\right\} \right)-f\left(S^{(t-1)}\right)=w_{t,a(t)}}\\
 & \leq\sum_{t\in T\cup S^{*}}w_{t,a(t)}\\
 & =\sum_{t\in T\setminus S^{*}}w_{t,a(t)}+\sum_{t\in S^{*}}w_{t,a(t)}\\
 & \leq\sum_{t\in T\setminus S^{*}}w_{t,a(t)}+\sum_{t\in S^{*}\colon w_{t,a(t)}\geq\beta_{a(t)}^{(t-1)}}w_{t(t)}+\sum_{t\in S^{*}\colon w_{t,a(t)}<\beta_{a(t)}^{(t-1)}}\beta_{a(t)}^{(t-1)}\\
 & =\sum_{a}\left(\sum_{t\in T_{a}\setminus S_{a}^{*}}w_{t,a}+\sum_{t\in S_{a}^{*}\colon w_{t,a}\geq\beta_{a}^{(t-1)}}w_{t,a}+\sum_{t\in S_{a}^{*}\colon w_{t,a}<\beta_{a}^{(t-1)}}\beta_{a}^{(t-1)}\right).
\end{align*}
\end{proof}

Thus we need to show that
\[
\E\left[\sum_{a=1}^{k}\left(\sum_{t\in T_{a}\setminus S_{a}^{*}}w_{t,a}+\sum_{t\in S_{a}^{*}\colon w_{t,a}\geq\beta_{a}^{(t-1)}}w_{t,a}+\sum_{t\in S_{a}^{*}\colon w_{t}<\beta_{a}^{(t-1)}}\beta_{a}^{(t-1)}\right)\right]\leq Q\cdot\E\left[\sum_{a=1}^{k}\sum_{t\in S_{a}}w_{t,a}\right]
\]
and obtain an approximation of $\frac{1-p}{Q}$. We will compare on
a per-part basis and show:
\begin{lem}
\label{lem:partition-nonmonotone-coefficient}For every part $a\in[k]$,
we have
\[
\E\left[\sum_{t\in T_{a}\setminus S_{a}^{*}}w_{t,a}+\sum_{t\in S_{a}^{*}\colon w_{t,a}\geq\beta_{a}^{(t-1)}}w_{t,a}+\sum_{t\in S_{a}^{*}\colon w_{t,a}<\beta_{a}^{(t-1)}}\beta_{a}^{(t-1)}\right]\leq Q_{a}\cdot\E\left[\sum_{t\in S_{a}}w_{t,a}\right]
\]
where
\[
Q_{a}=\max\left\{ 1+c_{a}+d_{a},\left(1-\frac{1}{n_{a}}\right)c_{a}+\frac{1}{p}\right\} .
\]
Thus we obtain, for $Q=\max_{a\in[k]}Q_{a}$,
\[
\E\left[f(S)\right]\geq\frac{1-p}{Q}\cdot f(S^{*}).
\]
\end{lem}

Fix a part $a$. We will analyze the change in the LHS and the RHS
of the inequality in the lemma statement with each iteration. To this
end, we define the following:

\begin{align*}
P_{t} & =\sum_{i\in S_{a}^{(t)}}w_{i,a}\\
D_{t} & =\sum_{i\in T_{a}^{(t)}\setminus S_{a}^{*}}w_{i,a}+\sum_{i\in S_{a}^{*}\cap\left\{ 1,\dots,t\right\} \colon w_{i,a}\geq\beta_{a}^{(i-1)}}w_{i,a}+\sum_{i\in S_{a}^{*}\cap\left\{ 1,\dots,t\right\} \colon w_{i,a}<\beta_{a}^{(i-1)}}\beta_{a}^{(i-1)}\\
 & \quad+\left|S_{a}^{*}\cap\left\{ t+1,\dots,T\right\} \right|\beta_{a}^{(t)}
\end{align*}
Note that $D_{t}$ is accounting for the items in $S_{a}^{*}\cap\left\{ t+1,\dots,T\right\} $
that have not arrived yet by paying the current threshold $\beta_{a}^{(t)}$
for each of them. Note that we have $P_{0}=D_{0}=0$ and $P_{T}$
and $D_{T}$ are equal to the RHS and LHS of the inequality, respectively.
Thus it suffices to relate the changes $\E\left[P_{t}-P_{t-1}\right]$
and $\E\left[D_{t}-D_{t-1}\right]$ with each iteration. We will show
that $\E_{Z_{t}}\left[D_{t}-D_{t-1}\vert Z_{1},\dots,Z_{t-1}\right]\leq Q_{a}\cdot\E_{Z_{t}}\left[P_{t}-P_{t-1}\vert Z_{1},\dots,Z_{t-1}\right]$
for all iterations $t$.
\begin{lem}
\label{lem:partition-nonmonotone-pd-change} Let $Q_{a}=\max\left\{ 1+c_{a}+d_{a},\left(1-\frac{1}{n_{a}}\right)c_{a}+\frac{1}{p}\right\} $
be as in Lemma \ref{lem:partition-nonmonotone-coefficient}. For each
iteration $t$, we have 
\[
\E_{Z_{t}}\left[D_{t}-D_{t-1}\vert Z_{1},\dots,Z_{t-1}\right]\leq Q_{a}\cdot\E_{Z_{t}}\left[P_{t}-P_{t-1}\vert Z_{1},\dots,Z_{t-1}\right]
\]
and thus
\[
\E\left[D_{t}-D_{t-1}\right]\leq Q_{a}\cdot\E\left[P_{t}-P_{t-1}\right]
\]
Summing up over all iterations and using that $P_{0}=D_{0}=0$, we
obtain
\[
\E\left[D_{T}\right]\leq Q_{a}\cdot\E\left[P_{T}\right]
\]
and thus
\[
\E\left[\sum_{t\in T_{a}\setminus S_{a}^{*}}w_{t,a}+\sum_{t\in S_{a}^{*}\colon w_{t,a}\geq\beta_{a}^{(t-1)}}w_{t,a}+\sum_{t\in S_{a}^{*}\colon w_{t,a}<\beta_{a}^{(t-1)}}\beta_{a}^{(t-1)}\right]\leq Q_{a}\cdot\E\left[\sum_{t\in S_{a}}w_{t,a}\right]
\]
\end{lem}

We fix an iteration $t$ and bound the expected changes in $P_{t}$
and $D_{t}$. In the following, we condition on $Z_{1},\dots,Z_{t-1}$.
Let $\widehat{\beta}_{a}^{(t)}=\sum_{i=1}^{n_{a}}w(i)g_{a}(i)$ where
$\left\{ w(i)\right\} _{1\leq i\leq n_{a}}$ are the $n_{a}$ largest
weights in $\left\{ w_{i,a}\colon i\in S_{a}^{(t-1)}\cup\left\{ t\right\} \right\} $;
if $S_{a}^{(t-1)}\cup\left\{ t\right\} $has less than $n_{a}$ items,
we let $w(i)=0$ for $i>\left|S_{a}^{(t-1)}\cup\left\{ t\right\} \right|$.
Note that $\widehat{\beta}_{a}^{(t)}$ is deterministic conditioned
on $Z_{1},\dots,Z_{t-1}$. Moreover, conditioned on $Z_{t}=1$, we
have $\beta_{a}^{(t)}=\widehat{\beta}_{a}^{(t)}$.

We start with the following helper lemma.
\begin{lem}
\label{lem:partition-nonmonotone-beta-change}We have
\[
\widehat{\beta}_{a}^{(t)}-\beta_{a}^{(t-1)}\leq\frac{d_{a}}{n_{a}}\beta_{a}^{(t-1)}+\frac{c_{a}}{n_{a}}w_{t}-\frac{c_{a}}{n_{a}}\left(1+\frac{d_{a}}{n_{a}}\right)^{n_{a}}\left(\min_{i\in S_{a}^{(t-1)}}w_{i}\right)
\]
\end{lem}

\begin{proof}
Let $w(1)\geq w(2)\geq\dots\geq w(n_{a}+1)$ be the $n_{a}+1$ largest
weights among $\left\{ w_{i,a}\colon i\in S_{a}^{(t-1)}\cup\left\{ t\right\} \right\} $;
if $S_{a}^{(t-1)}\cup\left\{ t\right\} $ has less than $n_{a}+1$
items, we let $w(i)=0$ for $i>\left|S_{a}^{(t-1)}\cup\left\{ t\right\} \right|$.
Note that $\min_{i\in S_{a}^{(t-1)}}w_{i,a}=w(n_{a}+1)$. Let $j$
be such that $w_{t,a}=w(j)$. We have
\begin{align*}
\widehat{\beta}_{a}^{(t)} & =\sum_{i=1}^{n_{a}}w(i)g_{a}(i)\\
\beta_{a}^{(t-1)} & =\sum_{i=1}^{j-1}w(i)g_{a}(i)+\sum_{i=j}^{n_{a}}w(i+1)g_{a}(i)=\sum_{i=1}^{j-1}w(i)g_{a}(i)+\sum_{i=j+1}^{n_{a}+1}w(i)g_{a}(i-1)
\end{align*}
Thus
\begin{align*}
\widehat{\beta}_{a}^{(t)}-\beta_{a}^{(t-1)} & =\sum_{i=j}^{n_{a}}w(i)g_{a}(i)-\sum_{i=j+1}^{n_{a}+1}w(i)g_{a}(i-1)\\
 & =\sum_{i=j+1}^{n_{a}}w(i)\left(g_{a}(i)-g_{a}(i-1)\right)+w(j)g_{a}(j)-w(n_{a}+1)g_{a}(n_{a})\\
 & =\frac{d_{a}}{n_{a}}\sum_{i=j+1}^{n_{a}}w(i)g_{a}(i-1)+w(j)g_{a}(j)-w(n_{a}+1)g_{a}(n_{a})\\
 & =\frac{d_{a}}{n_{a}}\beta_{a}^{(t-1)}-\frac{d_{a}}{n_{a}}\sum_{i=1}^{j-1}w(i)g_{a}(i)+w(j)g_{a}(j)-\left(1+\frac{d_{a}}{n_{a}}\right)w(n_{a}+1)g_{a}(n_{a})\\
 & =\frac{d_{a}}{n_{a}}\beta_{a}^{(t-1)}-\frac{d_{a}}{n_{a}}\sum_{i=1}^{j-1}w(i)g_{a}(i)+w(j)g_{a}(j)-w(n_{a}+1)g_{a}(n_{a}+1)\\
 & \leq\frac{d_{a}}{n_{a}}\beta_{a}^{(t-1)}-\frac{d_{a}}{n_{a}}\sum_{i=1}^{j-1}w(j)g_{a}(i)+w(j)g_{a}(j)-w(n_{a}+1)g_{a}(n_{a}+1)\\
 & =\frac{d_{a}}{n_{a}}\beta_{a}^{(t-1)}+w(j)\underbrace{\left(\left(1+\frac{d_{a}}{n_{a}}\right)^{j-1}-\frac{d_{a}}{n_{a}}\sum_{i=1}^{j-1}\left(1+\frac{d_{a}}{n_{a}}\right)^{i-1}\right)}_{=1}g_{a}(1)\\
 & \quad-w(n_{a}+1)g_{a}(n_{a}+1)\\
 & =\frac{d_{a}}{n_{a}}\beta_{a}^{(t-1)}+w(j)g_{a}(1)-w(n_{a}+1)g_{a}(n_{a}+1)\\
 & =\frac{d_{a}}{n_{a}}\beta_{a}^{(t-1)}+\frac{c_{a}}{n_{a}}w_{t}-\frac{c_{a}}{n_{a}}\left(1+\frac{d_{a}}{n_{a}}\right)^{n_{a}}\left(\min_{i\in S_{a}^{(t-1)}}w_{i}\right)
\end{align*}
where we used that $w(j)=w_{t}$, $\min_{i\in S_{a}^{(t-1)}}w_{i}=w(n_{a}+1)$,
and the definition of $g_{a}(i)=\frac{c_{a}}{n_{a}}\left(1+\frac{d_{a}}{n_{a}}\right)^{i-1}$
for all $i\geq1$.
\end{proof}

With the above lemma in hand, we proceed with the main analysis and
show Lemma \ref{lem:partition-nonmonotone-pd-change}. 

\begin{proof}[Proof (Lemma \ref{lem:partition-nonmonotone-pd-change})]
We have the following cases:
\begin{enumerate}
\item $w_{t,a}\geq\beta_{a}^{(t-1)}$ and $t\in S_{a}^{*}$ : If $Z_{t}=1$,
we have
\begin{align*}
P_{t}-P_{t-1} & =w_{t,a}-\min_{i\in S_{a}^{(t-1)}}w_{i,a}\\
D_{t}-D_{t-1} & =w_{t,a}+\left|S_{a}^{*}\cap\left\{ t+1,\dots,T\right\} \right|\beta_{a}^{(t)}-\left|S_{a}^{*}\cap\left\{ t,\dots,T\right\} \right|\beta_{a}^{(t-1)}\\
 & =w_{t,a}+\left|S_{a}^{*}\cap\left\{ t+1,\dots,T\right\} \right|\left(\beta_{a}^{(t)}-\beta_{a}^{(t-1)}\right)-\beta_{a}^{(t-1)}\\
 & \leq w_{t,a}+\left(n_{a}-1\right)\left(\beta_{a}^{(t)}-\beta_{a}^{(t-1)}\right)-\beta_{a}^{(t-1)}\\
 & =w_{t,a}+\left(n_{a}-1\right)\left(\widehat{\beta}_{a}^{(t)}-\beta_{a}^{(t-1)}\right)-\beta_{a}^{(t-1)}
\end{align*}
If $Z_{t}=0$, we have $\beta_{a}^{(t)}=\beta_{a}^{(t-1)}$, and thus
\begin{align*}
P_{t}-P_{t-1} & =0\\
D_{t}-D_{t-1} & =w_{t,a}+\left|S_{a}^{*}\cap\left\{ t+1,\dots,T\right\} \right|\beta_{a}^{(t)}-\left|S_{a}^{*}\cap\left\{ t,\dots,T\right\} \right|\beta_{a}^{(t-1)}\\
 & =w_{t,a}+\left|S_{a}^{*}\cap\left\{ t+1,\dots,T\right\} \right|\left(\beta_{a}^{(t)}-\beta_{a}^{(t-1)}\right)-\beta_{a}^{(t-1)}\\
 & \leq w_{t,a}+\left(n_{a}-1\right)\left(\beta_{a}^{(t)}-\beta_{a}^{(t-1)}\right)-\beta_{a}^{(t-1)}\\
 & =w_{t,a}-\beta_{a}^{(t-1)}
\end{align*}
 Thus
\begin{align*}
\E_{Z_{t}}\left[P_{t}-P_{t-1}\right] & =p\left(w_{t,a}-\min_{i\in S_{a}^{(t-1)}}w_{i,a}\right)\\
\E_{Z_{t}}\left[D_{t}-D_{t-1}\right] & \leq w_{t,a}-\beta_{a}^{(t-1)}+\left(n_{a}-1\right)p\left(\widehat{\beta}_{a}^{(t)}-\beta_{a}^{(t-1)}\right)
\end{align*}
Thus it suffices to show that
\[
\left(n_{a}-1\right)\left(\widehat{\beta}_{a}^{(t)}-\beta_{a}^{(t-1)}\right)+\frac{1}{p}\left(w_{t,a}-\beta_{a}^{(t-1)}\right)\leq Q_{a}\cdot\left(w_{t,a}-\min_{i\in S_{a}^{(t-1)}}w_{i,a}\right)
\]
Using Lemma \ref{lem:partition-nonmonotone-beta-change}, we obtain
\begin{align*}
 & \left(n_{a}-1\right)\left(\widehat{\beta}_{a}^{(t)}-\beta_{a}^{(t-1)}\right)+\frac{1}{p}\left(w_{t,a}-\beta_{a}^{(t-1)}\right)\\
 & \leq\left(n_{a}-1\right)\left(\frac{d_{a}}{n_{a}}\beta_{a}^{(t-1)}+\frac{c_{a}}{n_{a}}w_{t,a}-\frac{c_{a}}{n_{a}}\left(1+\frac{d_{a}}{n_{a}}\right)^{n_{a}}\left(\min_{i\in S_{a}^{(t-1)}}w_{i,a}\right)\right)+\frac{1}{p}\left(w_{t,a}-\beta_{a}^{(t-1)}\right)\\
 & =\left(1-\frac{1}{n_{a}}\right)\left(d_{a}\beta_{a}^{(t-1)}+c_{a}w_{t,a}-c_{a}\left(1+\frac{d_{a}}{n_{a}}\right)^{n_{a}}\left(\min_{i\in S_{a}^{(t-1)}}w_{i,a}\right)\right)+\frac{1}{p}\left(w_{t,a}-\beta_{a}^{(t-1)}\right)\\
 & =\left(\left(1-\frac{1}{n_{a}}\right)d_{a}-\frac{1}{p}\right)\beta_{a}^{(t-1)}+\left(\left(1-\frac{1}{n_{a}}\right)c_{a}+\frac{1}{p}\right)w_{t,a}\\
 & \quad-\left(1-\frac{1}{n_{a}}\right)c_{a}\left(1+\frac{d_{a}}{n_{a}}\right)^{n_{a}}\left(\min_{i\in S_{a}^{(t-1)}}w_{i,a}\right)
\end{align*}
We now consider two cases depending on whether the coefficient of
$\beta_{a}^{(t-1)}$ above is non-negative or negative.
\begin{enumerate}
\item If $\left(1-\frac{1}{n_{a}}\right)d_{a}-\frac{1}{p}\geq0$: We use
that $\beta_{a}^{(t-1)}\leq w_{t,a}$, and obtain
\begin{align*}
 & \left(n_{a}-1\right)\left(\beta_{a}^{(t)}-\beta_{a}^{(t-1)}\right)+\frac{1}{p}\left(w_{t,a}-\beta_{a}^{(t-1)}\right)\\
 & \leq\underbrace{\left(\left(1-\frac{1}{n_{a}}\right)d_{a}-\frac{1}{p}\right)}_{\geq0}\beta_{a}^{(t-1)}+\left(\left(1-\frac{1}{n_{a}}\right)c_{a}+\frac{1}{p}\right)w_{t,a}\\
 & \quad-\left(1-\frac{1}{n_{a}}\right)c_{a}\left(1+\frac{d_{a}}{n_{a}}\right)^{n_{a}}\left(\min_{i\in S_{a}^{(t-1)}}w_{i,a}\right)\\
 & \leq\left(1-\frac{1}{n_{a}}\right)\left(c_{a}+d_{a}\right)w_{t,a}-\left(1-\frac{1}{n_{a}}\right)c_{a}\left(1+\frac{d_{a}}{n_{a}}\right)^{n_{a}}\left(\min_{i\in S_{a}^{(t-1)}}w_{i,a}\right)\\
 & \leq\left(1-\frac{1}{n_{a}}\right)\left(c_{a}+d_{a}\right)\left(w_{t,a}-\min_{i\in S_{a}^{(t-1)}}w_{i,a}\right)\\
 & \leq Q_{a}\cdot\left(w_{t,a}-\min_{i\in S_{a}^{(t-1)}}w_{i,a}\right)
\end{align*}
where we used that the choice $c_{a}=\frac{1+d_{a}}{\left(1+\frac{d_{a}}{n_{a}}\right)^{n_{a}}-1}$
ensures that
\[
\left(1-\frac{1}{n_{a}}\right)\left(c_{a}+d_{a}\right)\leq\left(1-\frac{1}{n_{a}}\right)c_{a}\left(1+\frac{d_{a}}{n_{a}}\right)^{n_{a}}\Leftrightarrow c_{a}\geq\frac{d_{a}}{\left(1+\frac{d_{a}}{n_{a}}\right)^{n_{a}}-1}
\]
and
\[
\left(1-\frac{1}{n_{a}}\right)\left(c_{a}+d_{a}\right)\leq1+c_{a}+d_{a}\leq Q_{a}
\]
\item If $\left(1-\frac{1}{n_{a}}\right)d_{a}-\frac{1}{p}\leq0$: We use
that
\[
\beta_{a}^{(t-1)}\geq\left(\min_{i\in S_{a}^{(t-1)}}w_{i,a}\right)\sum_{i=1}^{n_{a}}g_{a}(i)=\left(\min_{i\in S_{a}^{(t-1)}}w_{i,a}\right)\frac{1+d_{a}}{d_{a}}
\]
and thus
\begin{align*}
 & \left(n_{a}-1\right)\left(\widehat{\beta}_{a}^{(t)}-\beta_{a}^{(t-1)}\right)+\frac{1}{p}\left(w_{t,a}-\beta_{a}^{(t-1)}\right)\\
 & \leq-\underbrace{\left(\frac{1}{p}-\left(1-\frac{1}{n_{a}}\right)d_{a}\right)}_{\geq0}\beta_{a}^{(t-1)}+\left(\left(1-\frac{1}{n_{a}}\right)c_{a}+\frac{1}{p}\right)w_{t,a}\\
 & \quad-\left(1-\frac{1}{n_{a}}\right)c_{a}\left(1+\frac{d_{a}}{n_{a}}\right)^{n_{a}}\left(\min_{i\in S_{a}^{(t-1)}}w_{i,a}\right)\\
 & \leq\left(\left(1-\frac{1}{n_{a}}\right)c_{a}+\frac{1}{p}\right)w_{t,a}\\
 & \quad-\left(\left(1-\frac{1}{n_{a}}\right)c_{a}\left(1+\frac{d_{a}}{n_{a}}\right)^{n_{a}}-\left(1-\frac{1}{n_{a}}\right)\left(1+d_{a}\right)+\frac{1}{p}\frac{1+d_{a}}{d_{a}}\right)\left(\min_{i\in S_{a}^{(t-1)}}w_{i,a}\right)\\
 & =\left(\left(1-\frac{1}{n_{a}}\right)c_{a}+\frac{1}{p}\right)w_{t,a}\\
 & \quad-\left(\left(1-\frac{1}{n_{a}}\right)c_{a}\left(1+\frac{d_{a}}{n_{a}}\right)^{n_{a}}-\left(1-\frac{1}{n_{a}}\right)c_{a}\left(\left(1+\frac{d_{a}}{n_{a}}\right)^{n_{a}}-1\right)+\frac{1}{p}\frac{1+d_{a}}{d_{a}}\right)\left(\min_{i\in S_{a}^{(t-1)}}w_{i,a}\right)\\
 & =\left(\left(1-\frac{1}{n_{a}}\right)c_{a}+\frac{1}{p}\right)w_{t,a}-\left(\left(1-\frac{1}{n_{a}}\right)c_{a}+\frac{1}{p}\frac{1+d_{a}}{d_{a}}\right)\left(\min_{i\in S_{a}^{(t-1)}}w_{i,a}\right)\\
 & \leq\left(\left(1-\frac{1}{n_{a}}\right)c_{a}+\frac{1}{p}\right)\left(w_{t,a}-\min_{i\in S_{a}^{(t-1)}}w_{i,a}\right)\\
 & \leq Q_{a}\cdot\left(w_{t,a}-\min_{i\in S_{a}^{(t-1)}}w_{i,a}\right)
\end{align*}
as needed.
\end{enumerate}
\item $w_{t,a}\geq\beta_{a}^{(t-1)}$ and $t\notin S_{a}^{*}$ : If $Z_{t}=1$,
we have $t\in T_{a}^{(t)}\setminus S_{a}^{*}$ and thus
\begin{align*}
P_{t}-P_{t-1} & =w_{t,a}-\min_{i\in S_{a}^{(t-1)}}w_{i,a}\\
D_{t}-D_{t-1} & =w_{t,a}+\left|S_{a}^{*}\cap\left\{ t+1,\dots,T\right\} \right|\beta_{a}^{(t)}-\left|S_{a}^{*}\cap\left\{ t,\dots,T\right\} \right|\beta_{a}^{(t-1)}\\
 & =w_{t,a}+\left|S_{a}^{*}\cap\left\{ t+1,\dots,T\right\} \right|\left(\beta_{a}^{(t)}-\beta_{a}^{(t-1)}\right)\\
 & \leq w_{t,a}+n_{a}\left(\beta_{a}^{(t)}-\beta_{a}^{(t-1)}\right)\\
 & =w_{t,a}+n_{a}\left(\widehat{\beta}_{a}^{(t)}-\beta_{a}^{(t-1)}\right)
\end{align*}
 If $Z_{t}=0$, we have $t\notin T_{a}^{(t)}\cup S_{a}^{*}$ and $\beta_{a}^{(t)}=\beta_{a}^{(t-1)}$,
and thus
\begin{align*}
P_{t}-P_{t-1} & =0\\
D_{t}-D_{t-1} & =\left|S_{a}^{*}\cap\left\{ t+1,\dots,T\right\} \right|\beta_{a}^{(t)}-\left|S_{a}^{*}\cap\left\{ t,\dots,T\right\} \right|\beta_{a}^{(t-1)}\\
 & =\left|S_{a}^{*}\cap\left\{ t+1,\dots,T\right\} \right|\left(\beta_{a}^{(t)}-\beta_{a}^{(t-1)}\right)\\
 & =0
\end{align*}
 Thus
\begin{align*}
\E_{Z_{t}}\left[P_{t}-P_{t-1}\right] & =p\left(w_{t,a}-\min_{i\in S_{a}^{(t-1)}}w_{i,a}\right)\\
\E_{Z_{t}}\left[D_{t}-D_{t-1}\right] & =p\left(w_{t,a}+n_{a}\left(\widehat{\beta}_{a}^{(t)}-\beta_{a}^{(t-1)}\right)\right)
\end{align*}
Thus it suffices to show that
\[
w_{t,a}+n_{a}\left(\widehat{\beta}_{a}^{(t)}-\beta_{a}^{(t-1)}\right)\leq Q_{a}\cdot\left(w_{t,a}-\min_{i\in S_{a}^{(t-1)}}w_{i,a}\right)
\]
Using Lemma \ref{lem:partition-nonmonotone-beta-change} and that
$\beta_{a}^{(t-1)}\leq w_{t,a}$, we obtain
\begin{align*}
w_{t,a}+n_{a}\left(\widehat{\beta}_{a}^{(t)}-\beta_{a}^{(t-1)}\right) & \leq w_{t,a}+d_{a}\beta_{a}^{(t-1)}+c_{a}w_{t,a}-c_{a}\left(1+\frac{d_{a}}{n_{a}}\right)^{n_{a}}\left(\min_{i\in S_{a}^{(t-1)}}w_{i,a}\right)\\
\leq & \left(1+d_{a}+c_{a}\right)w_{t,a}-c_{a}\left(1+\frac{d_{a}}{n_{a}}\right)^{n_{a}}\left(\min_{i\in S_{a}^{(t-1)}}w_{i,a}\right)\\
 & =\left(1+d_{a}+c_{a}\right)\left(w_{t,a}-\min_{i\in S_{a}^{(t-1)}}w_{i,a}\right)\\
 & \leq Q_{a}\cdot\left(w_{t,a}-\min_{i\in S_{a}^{(t-1)}}w_{i,a}\right)
\end{align*}
where we have used that the choice of $c_{a}$ ensures
\[
1+d_{a}+c_{a}=c_{a}\left(1+\frac{d_{a}}{n_{a}}\right)^{n_{a}}\Leftrightarrow c_{a}=\frac{1+d_{a}}{\left(1+\frac{d_{a}}{n_{a}}\right)^{n_{a}}-1}
\]
\item $w_{t,a}<\beta_{a}^{(t-1)}$: We have $t\notin T_{a}$ and $\beta_{a}^{(t)}=\beta_{a}^{(t-1)}$,
and thus
\begin{align*}
P_{t}-P_{t-1} & =0\\
D_{t}-D_{t-1} & =\sum_{i\in S_{a}^{*}\cap\left\{ 1,\dots,t\right\} \colon w_{i,a}<\beta_{a}^{(i-1)}}\beta_{a}^{(i-1)}-\sum_{i\in S_{a}^{*}\cap\left\{ 1,\dots,t-1\right\} \colon w_{i,a}<\beta_{a}^{(i-1)}}\beta_{a}^{(i-1)}\\
 & \quad+\left|S_{a}^{*}\cap\left\{ t+1,\dots,T\right\} \right|\beta_{a}^{(t)}-\left|S_{a}^{*}\cap\left\{ t,\dots,T\right\} \right|\beta_{a}^{(t-1)}\\
 & =\beta_{a}^{(t-1)}\cdot1_{\left[t\in S_{a}^{*}\right]}-\beta_{a}^{(t-1)}\cdot1_{\left[t\in S_{a}^{*}\right]}\\
 & =0
\end{align*}
\end{enumerate}
\end{proof}

\subsubsection{Setting the Parameters}

To complete the analysis, we show how to set $p$ and the constants
$\{d_{a}\}_{a\in[k]}$, and derive the final approximation guarantee.
Note that, once we have chosen $p$, we can set each $d_{a}$ to the
value that minimizes $Q_{a}$, which amounts to the value that balances
the two terms in the maximum in the definition of $Q_{a}$. Thus one
approach is to computationally choose $p$ and the $d_{a}$'s by iterating
over values for $p$ and, for a given $p$, iterate over values for
$d_{a}$ to find one that approximately minimizes $Q_{a}$. In the
following, we give explicit choices for $p$ and the $d_{a}$'s that
avoid this computation, and establish the approximation guarantee
for these explicit choices. We note that we have emphasized obtaining
simpler choices for $p$ and the $d_{a}$'s, and one can derive better
approximations by using our approach with a more involved case analysis.

Before proceeding, let us observe that, if the minimum budget $\min_{a\in[k]}n_{a}$
is sufficiently large, we have $\left(1+\frac{d_{a}}{n_{a}}\right)^{n_{a}}\approx e^{d_{a}}$
for all $a$. Suppose we set $d_{a}=d$ and $p=\frac{1}{1+d}$ for
some value $d$. Then $Q_{a}=\left(1+d\right)\left(1+\frac{1}{\left(1+\frac{d}{n_{a}}\right)^{n_{a}}-1}\right)\approx\left(1+d\right)\left(1+\frac{1}{e^{d}-1}\right)$
and we obtain an approximation $\frac{1-p}{\max_{a}Q_{a}}\approx\frac{d}{\left(1+d\right)^{2}}\frac{1}{1+\frac{1}{e^{d}-1}}$.
We can then choose $d$ to be the value that maximizes the approximation
guarantee. By taking the derivative with respect to $d$ and setting
it to $0$, we obtain that $d$ should be set to the solution to the
equation $e^{d}\left(d-1\right)-d^{2}-2d+1=0$, which is $d\approx1.9532$.
We obtain an approximation $\geq0.1921$, matching the approximation
of the streaming continuous greedy algorithm of \citet{feldman22}.
This is the choice we make if the minimum budget is larger than an
absolute constant $n_{0}$ (we use $n_{0}=10$ below). If the minimum
budget is small, this setting of $p$ and $\{d_{a}\}$ gives weaker
approximations than the state of the art for discrete algorithms \citet{feldman18}.
In this regime, we use a simple choice of $p=0.3$. For small values
of $n_{a}$, we give explicit choices for $d_{a}$ that are good for
that specific $n_{a}$. For values of $n_{a}$ that are larger than
an absolute constant $n_{0}$, we set all of the $d_{a}$s to the
same value $d=\frac{1-p}{p}$, similarly to the large budget case.
We have chosen an absolute constant $n_{0}=10$ so that the number
of explicit values $d_{a}$ that we list is small (we list $n_{0}$
different values, one for each $n_{a}\leq n_{0}$) while still obtaining
an approximation guarantee that improves upon the state of the art
for discrete algorithms \citet{feldman18}. One can obtain better
approximation guarantees by considering a different value of $p$
in the small budget case and a larger $n_{0}$. 

We now prove Theorem \ref{thm:partition-nonmonotone}.

\begin{proof}[Proof (Theorem \ref{thm:partition-nonmonotone})]
 We consider each case in turn.
\begin{enumerate}
\item For $n_{a}\leq n_{0}$, we can verify that $\frac{1-p}{Q_{a}}$ is
lower bounded by the values shown in Table \ref{tb:partition-nonmonotone-params}.\\
Consider any $n_{a}>n_{0}.$ Let $d=\frac{1-p}{p}=\frac{7}{3}$. Recall
that we set $d_{a}=d=\frac{1-p}{p}\leq n_{0}$ in this case. Since
$d_{a}=\frac{1-p}{p}$, we have $Q_{a}=\left(1+d\right)\left(1+\frac{1}{\left(1+\frac{d}{n_{a}}\right)^{n_{a}}-1}\right)$.
Thus, by Lemma \ref{lem:exp-approx}, we have
\begin{multline*}
\frac{1-p}{Q_{a}}=\frac{d}{\left(1+d\right)^{2}\left(1+\frac{1}{\left(1+\frac{d}{n_{a}}\right)^{n_{a}}-1}\right)}\\
\geq\frac{d}{\left(1+d\right)^{2}\left(1+\frac{1}{\exp\left(d\right)-1}\right)}\left(1-\frac{1}{n_{a}}\cdot\frac{n_{0}\left(\exp\left(\frac{d^{2}}{n_{0}}\right)-1\right)}{\exp\left(d\right)-1}\right).
\end{multline*}
Plugging in $d=\frac{7}{3}$ and $n_{0}=10$, we obtain
\[
\frac{1-p}{Q_{a}}\geq0.1896\left(1-\frac{0.7771}{n_{a}}\right).
\]
Note that the above is $\geq0.175$ for all $n_{a}\geq11$. Overall,
we obtain that the approximation is $\geq0.175$.
\item For all $a\in[k]$, we set $d_{a}=d=\frac{1-p}{p}\leq n_{0}$. Thus,
as above, Lemma \ref{lem:exp-approx} gives
\begin{multline*}
\frac{1-p}{Q_{a}}=\frac{d}{\left(1+d\right)^{2}\left(1+\frac{1}{\left(1+\frac{d}{n_{a}}\right)^{n_{a}}-1}\right)}\\
\geq\frac{d}{\left(1+d\right)^{2}\left(1+\frac{1}{\exp\left(d\right)-1}\right)}\left(1-\frac{1}{n_{a}}\cdot\frac{n_{0}\left(\exp\left(\frac{d^{2}}{n_{0}}\right)-1\right)}{\exp\left(d\right)-1}\right).
\end{multline*}
Plugging in $d=1.9532$ and $n_{0}=10$, we obtain
\[
\frac{1-p}{Q_{a}}\geq0.1921\left(1-\frac{0.7676}{n_{a}}\right).
\]
Note that the above is $\geq0.175$ for all $n_{a}$, since we have
$n_{a}\geq11$ for all $a$. Overall, we obtain that the approximation
is $\geq0.175$ and it tends to $\geq0.1921$ as $\min_{a}n_{a}$
tends to infinity.
\end{enumerate}
\end{proof}

\subsection{Monotone $k$-Submodular Maximization with Knapsack Constraints}

\begin{algorithm}
\textbf{Parameters}: $g(u)\coloneqq ce^{du}$ for parameters $c,d\ge0$ 

\textbf{Input}:\textbf{ }monotone $k$-submodular function $f$

$\S=\left(S_{1},\dots,S_{k}\right)\gets\left(\emptyset,\dots,\emptyset\right)$

$\tilde{\S}=\left(\tilde{S}_{1},\dots,\tilde{S}_{k}\right)\gets\left(\emptyset,\dots,\emptyset\right)$

$\beta_{a}\gets0$ for all $a\in\left[k\right]$

\textbf{for} $t=1,2,\dots,\left|V\right|$:

$\quad$let $\rho_{t,a}=\frac{\Delta_{t,a}f\left({\bf S}\right)}{u_{t,a}}$
for all $a\in[k]$

$\quad$let $a=\arg\max_{a\in[k]}\left\{ u_{t,a}\left(\rho_{t,a}-\beta_{a}\right)\right\} $

$\quad$\textbf{if} $\rho_{t,a}-\beta_{a}\ge0$:

$\quad\quad$$S_{a}\gets S_{a}\cup\left\{ t\right\} $

$\quad\quad$\textbf{while} $\sum_{i\in S_{a}}u_{i,a}>1$:

$\quad\quad\quad$remove $t'=\arg\min_{i\in S_{a}}\rho_{i,a}$ from
$S_{a}$

$\quad\quad$let $t'$ be the last removed item and set $\tilde{S}_{a}\gets S_{a}\cup\left\{ t'\right\} $;
if no item was removed, set $\tilde{S}_{a}\gets S_{a}$

$\quad\quad$let $\rho_{a}(u)=\max\left\{ \rho:\sum_{i\in\tilde{S}_{a}:\rho_{i,a}\ge\rho}u_{i,a}>u\right\} $
for $u<\sum_{i\in\tilde{S}_{a}}u_{i,a}$ and $\rho_{a}(u)=0$, otherwise

$\quad\quad$$\beta_{a}\gets\int_{0}^{1}\rho_{a}(u)g(u)du$

\textbf{return} ${\bf S}$

\caption{\label{alg:ksubmod-knapsack} Monotone $k$-submodular maximization
under individual knapsack constraints. We assume without loss of generality
that each part has a budget of $1$.}
\end{algorithm}

We now study the problem of maximizing a $k$-submodular function
under individual knapsack constraints. For simplicity, we only present
the monotone case. The extension for general $k$-submodular functions
and submodular maximization with a partition matroid constraint follow
analogously to the previous sections.

Formally, each item $t$ has a size $u_{t,a}\ge0$ associated with
each part $a$, and the goal is to find a solution $\S$ with maximum
$f(\S)$ such that $\sum_{t\in S_{a}}u_{t,a}\le1$ for all $a\in[k]$.
Note that we assume that the budget of each part is equal to $1$;
this is without loss of generality, as we can rescale the item sizes
by the budgets.

We denote with $\epsilon\coloneqq\max_{t,a}u_{t,a}$ the maximum size
of any item in the stream. Our algorithm achieves provable constant
factor approximations if $\epsilon$ is sufficiently small. This assumption
is motivated by applications such as ad-allocation where bids are
small compared to an advertiser's total budget. Furthermore, assuming
that sizes are small is necessary to achieve a constant-factor approximation
ratio \citep{feldman09}.

Our algorithm is described in Algorithm \ref{alg:ksubmod-knapsack},
where we allocate items according to their densities $\rho$, the
fraction of item weight and size. We now define $g(u)$ continuously
as
\[
g(u)\coloneqq ce^{du}\qquad\mathrm{where}\qquad c\coloneqq\frac{e^{d\epsilon}-1+\epsilon}{\epsilon e^{d}-\frac{1}{d}\left(e^{d\epsilon}-1\right)}
\]
for all sizes $u\in[0,1]$ and $d$ specified later in Theorem \ref{thm:ksubmod-knapsack}.
Note that in each iteration $t$, $\beta_{a}^{(t)}$ can be efficiently
evaluated: Let $\left\{ t_{1},t_{2},\dots,t_{\ell},t_{\ell+1}\right\} =\tilde{S}_{a}^{(t)}$
be such that $\rho_{t_{1},a}\ge\rho_{t_{2},a}\ge\cdots\ge\rho_{t_{\ell},a}$
and define the intervals
\[
U_{1}\coloneqq[0,u_{t_{1}}),U_{2}\coloneqq[u_{t_{1}},u_{t_{1}}+u_{t_{2}}),\dots,U_{i}\coloneqq\Big[{\textstyle \sum_{j<i}u_{t_{j},a},\sum_{j\le i}u_{t_{j},a}\Big)},\dots
\]
By definition, $\rho_{a}^{(t)}(u)$ is a step function with $\rho_{a}^{(t)}(u)=\rho_{t_{i},a}$
if $u\in U_{i}$. Furthermore, $t_{\ell+1}=t'$ is the disposed item
with minimum density among items in $\tilde{S}_{a}^{(t)}$ (if we
disposed in iteration $t$). Thus,
\[
\beta_{a}^{(t)}=\int_{0}^{1}\rho_{a}^{(t)}(u)g(u)du=\sum_{i=1}^{\ell}\rho_{t_{i},a}\int_{U_{i}}g(u)du+\rho_{t_{\ell+1},a}\int_{U_{\ell+1}\cap[0,1]}g(u)du
\]
and all integrals can be computed explicitly through integration of
$g$.
\begin{thm}
\label{thm:ksubmod-knapsack} As $\epsilon\to0$, Algorithm \ref{alg:ksubmod-knapsack}
achieves an approximation guarantee of
\[
\frac{f(\S)}{f(\S^{*})}\ge\frac{1-e^{-d}}{d+1}\geq0.3178
\]
when choosing $d$ as the solution of the equation $e^{d}-d-2=0$,
which is $d\approx1.1461$.
\end{thm}

Note that this recovers the guarantee of \ref{thm:ksubmod} when the
budgets tend to infinity. 

\subsubsection{Analysis}
\begin{lem}
\label{lem:knapsack-bound-sol} The value of solution $\S$ is at
least
\[
f\left({\bf S}\right)\ge\left(1-\epsilon\right)\sum_{a}\int_{0}^{1}\rho_{a}(u)du.
\]
\end{lem}

\begin{proof}
As in the cardinality-constrained case, we have
\begin{align*}
f(\S)-f(\S^{(0)}) & =\sum_{t\in\supp(\S)}\left(f\left({\bf S}\cap{\bf S}^{(t)}\right)-f\left({\bf S}\cap{\bf S}^{(t-1)}\right)\right)\\
 & =\sum_{t\in\supp(\S)}\Delta_{t,a(t)}f(\S\cap{\bf S}^{(t-1)})\\
 & \geq\sum_{t\in\supp(\S)}\Delta_{t,a(t)}f({\bf S}^{(t-1)})\\
 & =\sum_{t\in\supp(\S)}u_{t,a(t)}\rho_{t,a(t)}\\
 & =\sum_{a}\sum_{t\in S_{a}}u_{t,a}\rho_{t,a}
\end{align*}
where the inequality is due to orthant submodularity. Let $\left\{ t_{1},t_{2},\dots,t_{m}\right\} =S_{a}$
be ordered such that $\rho_{t_{1},a}\ge\rho_{t_{2},a}\ge\cdots\ge\rho_{t_{m},a}$.
Let $t_{m+1}$ be the single impression disposed of last. Recall that
$\rho_{a}(u)=\rho_{t_{i}}$ on $u\in[\sum_{j<i}u_{t_{j},a},\sum_{j\le i}u_{t_{j},a})$
and thus
\begin{align*}
\sum_{t\in S_{a}}u_{t,a}\rho_{t,a} & =\sum_{i=1}^{m}u_{t_{i},a}\rho_{t_{i},a}\\
 & =\int_{0}^{\sum_{t\in S_{a}}u_{t,a}}\rho_{a}(u)du\\
 & \ge\int_{0}^{1-u_{t_{m+1},a}}\rho_{a}(u)du\\
 & =\int_{0}^{1}\rho_{a}(u)du-\int_{1-u_{t_{m+1},a}}^{1}\rho_{a}(u)du\\
 & \ge\int_{0}^{1}\rho_{a}(u)du-u_{t_{m+1},a}\int_{0}^{1}\rho_{a}(u)du\\
 & \ge\left(1-\epsilon\right)\int_{0}^{1}\rho_{a}(u)du
\end{align*}
where the first inequality is due to $\sum_{t\in S_{a}}u_{t,a}>1-u_{t_{m+1},a}$
and the second inequality holds since $\rho_{a}$ is decreasing.
\end{proof}

\begin{lem}
\label{lem:knapsack-bound-opt} The value of the optimum solution
$\S^{*}$ is at most
\[
f\left({\bf S}^{*}\right)\leq\sum_{a}\left(\sum_{t\in T_{a}}u_{t,a}\left(2\rho_{t,a}-\beta_{a}^{(t-1)}\right)+\beta_{a}\right)
\]
\end{lem}

\begin{proof}
Let $\O^{(t)}$ be the allocation that agrees with $\T^{(t)}$ on
items $\left\{ 1,\dots,t\right\} $, and it agrees with $\S^{*}$
on items $\left\{ t+1,\dots,\left|V\right|\right\} $. Let $\widetilde{\O}^{(t-1)}$
be the allocation obtained from $\O^{(t)}$ by dropping $t$ (i.e.,
$t$ is not assigned to any part under $\widetilde{\O}^{(t-1)}$).
For $t\in\supp(\T)$, let $a(t)$ be the part such that $t\in T_{a}$.
For $t\in\supp(\S^{*})$, let $a^{*}(t)$ be the part such that $t\in S_{a}^{*}$.

We have
\begin{align*}
 & f(\S^{*})-f(\T)\\
 & =f(\O^{(0)})-f(\O^{\left|V\right|})=\sum_{t=1}^{\left|V\right|}\left(f(\O^{(t-1)})-f(\O^{(t)})\right)\\
 & =\sum_{t\in\supp(\T)\cap\supp(\S^{*})}\left(f(\O^{(t-1)})-f(\O^{(t)})\right)+\sum_{t\notin\supp(\T)\cup\supp(\S^{*})}\left(f(\O^{(t-1)})-f(\O^{(t)})\right)\\
 & +\sum_{t\in\supp(\T)\setminus\supp(\S^{*})}\left(f(\O^{(t-1)})-f(\O^{(t)})\right)+\sum_{t\in\supp(\S^{*})\setminus\supp(\T)}\left(f(\O^{(t-1)})-f(\O^{(t)})\right)
\end{align*}
\begin{itemize}
\item Consider $t\in\supp(\T)\cap\supp(\S^{*})$. If $a(t)=a^{*}(t)$, we
have $\O^{(t-1)}=\O^{(t)}$, and thus
\[
f(\O^{(t-1)})-f(\O^{(t)})=0
\]
If $a(t)\neq a^{*}(t)$, we have
\begin{align*}
f(\O^{(t-1)})-f(\O^{(t)}) & =f(\O^{(t-1)})-f(\widetilde{\O}^{(t-1)})+f(\widetilde{\O}^{(t-1)})-f(\O^{(t)})\\
 & =\Delta_{t,a^{*}(t)}f(\widetilde{\O}^{(t-1)})-\Delta_{t,a(t)}f(\widetilde{\O}^{(t-1)})\\
 & \leq\Delta_{t,a^{*}(t)}f(\S^{(t-1)})-\underbrace{\Delta_{t,a(t)}f(\widetilde{\O}^{(t-1)})}_{\geq0}\\
 & \leq\Delta_{t,a^{*}(t)}f(\S^{(t-1)})
\end{align*}
In the first inequality, we used orthant submodularity since $\S^{(t-1)}\preceq\widetilde{\O}^{(t-1)}$.
In the second inequality, we used monotonicity.
\item Consider $t\notin\supp(\T)\cup\supp(\S^{*})$. We have $\O^{(t-1)}=\O^{(t)}$,
and thus
\[
f(\O^{(t-1)})-f(\O^{(t)})=0
\]
\item Consider $t\in\supp(\T)\setminus\supp(\S^{*})$. We have $\O^{(t-1)}\preceq\O^{(t)}$.
Since $f$ is monotone, we have
\begin{align*}
f(\O^{(t-1)})-f(\O^{(t)}) & \leq0
\end{align*}
\item Consider $t\in\supp(\S^{*})\setminus\supp(\T).$ We have 
\begin{align*}
f(\O^{(t-1)})-f(\O^{(t)}) & =\Delta_{t,a^{*}(t)}f(\O^{(t)})\leq\Delta_{t,a^{*}(t)}f(\S^{(t-1)})\leq u_{t,a}\beta_{a^{*}(t)}^{(t-1)}
\end{align*}
where in the first inequality we used orthant submodularity since
$\S^{(t-1)}\preceq\O^{(t)}$, and in the second inequality we used
that all of the discounted gains are $\leq0$.
\end{itemize}
Putting everything together, we have
\[
f(\S^{*})\leq f(\T)+\sum_{t\in\supp(\T)\cap\supp(\S^{*})}u_{t,a^{*}(t)}\rho_{t,a^{*}(t)}+\sum_{t\in\supp(\S^{*})\setminus\supp(\T)}u_{t,a^{*}(t)}\beta_{a^{*}(t)}^{(t-1)}
\]
Using the fact that $\S^{(t)}\subseteq\T^{(t)}$ and orthant submodularity,
we can further upper bound
\begin{align*}
f(\T) & =\sum_{t\in\supp(\T)}\left(f(\T^{(t)})-f(\T^{(t-1)})\right)\\
 & =\sum_{t\in\supp(\T)}\Delta_{t,a(t)}f(\T^{(t-1)})\\
 & \leq\sum_{t\in\supp(\T)}\Delta_{t,a(t)}f(\S^{(t-1)})\\
 & =\sum_{t\in\supp(\T)}u_{t,a(t)}\rho_{t,a(t)}
\end{align*}
Thus
\begin{align*}
f(\S^{*}) & \leq\sum_{t\in\supp(\T)}u_{t,a(t)}\rho_{t,a(t)}+\sum_{t\in\supp(\T)\cap\supp(\S^{*})}u_{t,a^{*}(t)}\rho_{t,a^{*}(t)}\\
 & \quad+\sum_{t\in\supp(\S^{*})\setminus\supp(\T)}u_{a^{*}(t)}\beta_{a^{*}(t)}^{(t-1)}\\
 & =\sum_{t\in\supp(\T)}u_{t,a(t)}\rho_{t,a(t)}+\sum_{t\in\supp(\T)\cap\supp(\S^{*})}u_{t,a^{*}(t)}\left(\rho_{t,a^{*}(t)}-\beta_{a^{*}(t)}^{(t-1)}\right)\\
 & \quad+\sum_{t\in\supp(\S^{*})}u_{a^{*}(t)}\beta_{a^{*}(t)}^{(t-1)}\\
 & \overset{(1)}{\leq}\sum_{t\in\supp(\T)}u_{t,a(t)}\rho_{t,a(t)}+\sum_{t\in\supp(\T)}u_{t,a(t)}\left(\rho_{t,a(t)}-\beta_{a(t)}^{(t-1)}\right)+\sum_{t\in\supp(\S^{*})}u_{a^{*}(t)}\beta_{a^{*}(t)}^{(t-1)}\\
 & =\sum_{t\in\supp(\T)}u_{t,a(t)}\left(2\rho_{t,a(t)}-\beta_{a(t)}^{(t-1)}\right)+\sum_{t\in\supp(\S^{*})}u_{t,a^{*}(t)}\beta_{a^{*}(t)}^{(t-1)}
\end{align*}
where in $(1)$ we used that $u_{t,a^{*}(t)}\left(\rho_{t,a^{*}(t)}-\beta_{a^{*}(t)}^{(t-1)}\right)\leq u_{t,a(t)}\left(\rho_{t,a(t)}-\beta_{a(t)}^{(t-1)}\right)$
for every $t\in\supp(\T)\cap\supp(\S^{*})$ due to the choice of $a(t)$,
and $u_{t,a(t)}\left(\rho_{t,a(t)}-\beta_{a(t)}^{(t-1)}\right)\geq0$
for every $t\in\supp(\T)$.

Finally, since the thresholds are non-decreasing and $\S^{*}$ is
a feasible allocation, we have
\[
\sum_{t\in\supp(\S^{*})}u_{t,a^{*}(t)}\beta_{a^{*}(t)}^{(t-1)}=\sum_{a=1}^{k}\sum_{t\in S_{a}^{*}}u_{t,a}\beta_{a}^{(t-1)}\leq\sum_{a=1}^{k}\beta_{a}.
\]
\end{proof}

Due to Lemma \ref{lem:knapsack-bound-sol} and \ref{lem:knapsack-bound-opt},
it is sufficient to show that
\[
\sum_{a}\left(\sum_{t\in T_{a}}u_{t,a}\left(2\rho_{t,a}-\beta_{a}^{(t-1)}\right)+\beta_{a}\right)\le Q\sum_{a}\int_{0}^{1}\rho_{a}(u)du
\]
for $Q$ as small as we can make it. We will compare on a per-part
basis and show:
\begin{lem}
\label{lem:knapsack-q} For all parts $a\in[k]$,
\[
\sum_{t\in T_{a}}u_{t,a}\left(2\rho_{t,a}-\beta_{a}^{(t-1)}\right)+\beta_{a}\le Q\int_{0}^{1}\rho_{a}(u)du
\]
for
\[
Q\coloneqq\frac{e^{d\epsilon}-1+\epsilon}{\epsilon e^{d}-\frac{1}{d}\left(e^{d\epsilon}-1\right)}e^{d}.
\]
\end{lem}

This gives us an approximation ratio of $\frac{f(\S)}{f(\S^{*})}\ge\frac{1-\epsilon}{Q}$.
To prove this lemma, we fix a part $a$. Let
\begin{align*}
P_{t} & \coloneqq\int_{0}^{1}\rho_{a}^{(t)}(u)du\\
D_{t} & \coloneqq\sum_{i\in T_{a}^{(t)}}u_{t,a}\left(2\rho_{ai}-\beta_{a}^{(i-1)}\right)+\beta_{a}^{(t)}
\end{align*}
where $\rho_{a}^{(t)}(u)=\max\left\{ \rho:\sum_{i\in T_{a}^{(t)}:\rho_{i,a}\ge\rho}u_{i,a}>u\right\} $
for $u<\sum_{i\in T_{a}^{(t)}}u_{i,a}$ and $\rho_{a}^{(t)}(u)=0$,
otherwise. Note that we have $P_{0}=D_{0}=0$, $P_{T}=\int_{0}^{1}\rho_{a}(u)du$,
and $D_{T}=\sum_{t\in T_{a}}u_{t,a}\left(2\rho_{t,a}-\beta_{a}^{(t-1)}\right)+\beta_{a}$.
Thus it suffices to show that $D_{t}-D_{t-1}\leq Q$$\left(P_{t}-P_{t-1}\right)$
for all $t$. 

If $t\notin T_{a}$, we have $\beta_{a}^{(t)}=\beta_{a}^{(t-1)}$
and thus $P_{t}-P_{t-1}=D_{t}-D_{t-1}=0$. Thus we may assume that
$t\in T_{a}$, and thus $\rho_{t,a}\geq\beta_{a}^{(t-1)}$. Let $u'\coloneqq\sum_{i\in T_{a}^{(t-1)}:\rho_{i,a}\ge\rho_{t,a}}u_{i,a}\in[0,1]$
be the position at which we add item $t$. We have
\[
\rho_{a}^{(t)}(u)=\begin{cases}
\rho_{a}^{(t-1)}(u) & \text{for }u<u'\\
\rho_{t,a} & \text{for }u\in[u',u'+u_{t,a})\\
\rho_{a}^{(t-1)}(u-u_{t,a}) & \text{for }u\ge u'+u_{t,a}
\end{cases}
\]
We thus have
\begin{align*}
P_{t}-P_{t-1} & =\rho_{t,a}u_{t,a}-\int_{1-u_{t,a}}^{1}\rho_{a}^{(t-1)}(u)du\\
D_{t}-D_{t-1} & =u_{t,a}\left(2\rho_{t,a}-\beta_{a}^{(t-1)}\right)+\beta_{a}^{(t)}-\beta_{a}^{(t-1)}.
\end{align*}
The primal change is the change in $\rho_{a}$ after allocating $t$
to $a$: Recall the interpretation of $\rho_{a}(u)$ through consecutive
intervals $U_{i}$ of size $u_{t_{i}}$, where $t_{i}$ is the item
with $i$-th largest density currently allocated to $S_{a}$, such
that $\rho_{a}(u)=\rho_{t_{i},a}$ if $u\in U_{i}$. After allocating
$t$ to $a$, we introduce a new interval for item $t$ of size $u_{t,a}$,
which pushes all intervals corresponding to items with lower density
to the right. We thus gain $\rho_{t,a}u_{t,a}$ in the primal but
loose the densities belonging to intervals which are pushed out of
the range $[0,1]$ which is exactly $\int_{1-u_{t,a}}^{1}\rho_{a}^{(t-1)}(du)du$. 

\begin{lem}
\label{lem:ksubmod-beta-change-1}We have
\begin{align*}
\beta_{a}^{(t)}-\beta_{a}^{(t-1)} & \le\left(e^{du_{t,a}}-1\right)\beta_{a}^{(t-1)}+\rho_{t,a}\frac{c}{d}\left(e^{du_{t,a}}-1\right)-g(1)\int_{1-u_{t,a}}^{1}\rho_{a}^{(t-1)}(u)du
\end{align*}
\end{lem}

\begin{proof}
We have

\begin{align*}
\beta_{a}^{(t)} & =\int_{0}^{1}\rho_{a}^{(t)}(u)g(u)du\\
\beta_{a}^{(t-1)} & =\int_{0}^{u'}\rho_{a}^{(t)}(u)g(u)du+\int_{u'}^{1}\rho_{a}^{(t-1)}(u)g(u)du.
\end{align*}
Thus,
\begin{align*}
 & \beta_{a}^{(t)}-\beta_{a}^{(t-1)}\\
 & =\int_{u'}^{1}\rho_{a}^{(t)}(u)g(u)du-\int_{u'}^{1}\rho_{a}^{(t-1)}(u)g(u)du\\
 & =\int_{u'}^{u'+u_{t,a}}\rho_{a}^{(t)}(u)g(u)du+\int_{u'}^{1-u_{t,a}}\rho_{a}^{(t-1)}(u)g(u+u_{t,a})du\\
 & \quad-\int_{u'}^{1-u_{t,a}}\rho_{a}^{(t-1)}(u)g(u)du-\int_{1-u_{t,a}}^{1}\rho_{a}^{(t-1)}(u)g(u)du\\
 & =\int_{u'}^{1-u_{t,a}}\rho_{a}^{(t-1)}(u)\left(g(u+u_{t,a})-g(u)\right)du+\int_{u'}^{u'+u_{t,a}}\rho_{a}^{(t)}(u)g(u)du\\
 & \quad-\int_{1-u_{t,a}}^{1}\rho_{a}^{(t-1)}(u)g(u)du\\
 & \overset{(1)}{=}\left(e^{du_{t,a}}-1\right)\int_{u'}^{1-u_{t,a}}\rho_{a}^{(t-1)}(u)g(u)du+\int_{u'}^{u'+u_{t,a}}\rho_{a}^{(t)}(u)g(u)du\\
 & \quad-\int_{1-u_{t,a}}^{1}\rho_{a}^{(t-1)}(u)g(u)du\\
 & =\left(e^{du_{t,a}}-1\right)\beta_{a}^{(t-1)}-\left(e^{du_{t,a}}-1\right)\int_{0}^{u'}\rho_{a}^{(t-1)}(u)g(u)du+\int_{u'}^{u'+u_{t,a}}\rho_{a}^{(t)}(u)g(u)du\\
 & \quad-e^{du_{t,a}}\int_{1-u_{t,a}}^{1}\rho_{a}^{(t-1)}(u)g(u)du\\
 & \overset{(2)}{\le}\left(e^{du_{t,a}}-1\right)\beta_{a}^{(t-1)}-\left(e^{du_{t,a}}-1\right)\int_{0}^{u'}\rho_{a}^{(t-1)}(u')g(u)du+\int_{u'}^{u'+u_{t,a}}\rho_{a}^{(t)}(u')g(u)du\\
 & \quad-e^{du_{t,a}}\int_{1-u_{t,a}}^{1}\rho_{a}^{(t-1)}(u)g(u)du\\
 & =\left(e^{du_{t,a}}-1\right)\beta_{a}^{(t-1)}+\rho_{a}^{(t)}(u')\underbrace{\left(\int_{u'}^{u'+u_{t,a}}g(u)du-\left(e^{du_{t,a}}-1\right)\int_{0}^{u'}g(u)du\right)}_{(\star)}\\
 & \quad-e^{du_{t,a}}\int_{1-u_{t,a}}^{1}\rho_{a}^{(t-1)}(u)g(u)du
\end{align*}
where in (1) we use that $g(u+u_{t,a})=e^{du_{t,a}}g(u)$ by definition
of $g$ and in (2) we use $\rho_{a}^{(t)}(u)=\rho_{a}^{(t)}(u')$
for $u\in[u',u'+u_{t,a})$ and that $\rho_{a}^{(t-1)}(u)$ is decreasing.
We can evaluate the term
\begin{align*}
(\star) & =\frac{c}{d}\left(e^{d(u'+u_{t,a})}-e^{du'}\right)-\left(e^{du_{t,a}}-1\right)\frac{c}{d}\left(e^{du'}-1\right)\\
 & =\frac{c}{d}\left(e^{du_{t,a}}-1\right).
\end{align*}
Finally, to obtain the bound in the lemma statement, we use that $\rho_{a}^{(t)}(u')=\rho_{t,a}$
and 
\begin{multline*}
e^{du_{t,a}}\int_{1-u_{t,a}}^{1}\rho_{a}^{(t-1)}(u)g(u)du\\
\ge e^{du_{t,a}}g(1-u_{t,a})\int_{1-u_{t,a}}^{1}\rho_{a}^{(t-1)}(u)du=g(1)\int_{1-u_{t,a}}^{1}\rho_{a}^{(t-1)}(u)du.
\end{multline*}
\end{proof}

We can now show Lemma \ref{lem:knapsack-q}. 

\begin{proof}[Proof (Lemma \ref{lem:knapsack-q})]
 Using Lemma \ref{lem:ksubmod-beta-change-1}, we obtain
\begin{align*}
 & u_{t,a}\left(2\rho_{t,a}-\beta_{a}^{(t-1)}\right)+\beta_{a}^{(t)}-\beta_{a}^{(t-1)}\\
 & \le u_{t,a}\left(2\rho_{t,a}-\beta_{a}^{(t-1)}\right)+\left(e^{du_{t,a}}-1\right)\beta_{a}^{(t-1)}+\rho_{t,a}\frac{c}{d}\left(e^{du_{t,a}}-1\right)-g(1)\int_{1-u_{t,a}}^{1}\rho_{a}^{(t-1)}(u)du\\
 & =\underbrace{\left(e^{du_{t,a}}-1-u_{t,a}\right)}_{\ge0}\beta_{a}^{(t-1)}+\rho_{t,a}\left(\frac{c}{d}e^{du_{t,a}}-\frac{c}{d}+2u_{t,a}\right)-g(1)\int_{1-u_{t,a}}^{1}\rho_{a}^{(t-1)}(u)du\\
 & \le u_{t,a}\rho_{t,a}\left(\frac{e^{du_{t,a}}-1}{u_{t,a}}\left(\frac{c}{d}+1\right)+1\right)-g(1)\int_{1-u_{t,a}}^{1}\rho_{a}^{(t-1)}(u)du
\end{align*}
where we use that $\beta_{a}^{(t-1)}\le\rho_{t,a}$. By the definition
of $c$,
\begin{align*}
c & =\frac{e^{d\epsilon}-1+\epsilon}{\epsilon e^{d}-\frac{1}{d}\left(e^{d\epsilon}-1\right)}\\
\iff\frac{e^{d\epsilon}-1}{\epsilon}\left(\frac{c}{d}+1\right)+1 & =ce^{d}\\
\implies\frac{e^{du_{t,a}}-1}{u_{t,a}}\left(\frac{c}{d}+1\right)+1 & \le ce^{d}=g(1)
\end{align*}
since $\frac{e^{d\epsilon}-1}{\epsilon}\le\frac{e^{du_{t,a}}-1}{u_{t,a}}$.
We thus obtain
\[
Q=ce^{d}=\frac{e^{d\epsilon}-1+\epsilon}{\epsilon e^{d}-\frac{1}{d}\left(e^{d\epsilon}-1\right)}e^{d}.
\]
\end{proof}

Our approximation ratio as a function of $d$ is therefore 
\[
\frac{f(\S)}{f(\S^{*})}\ge\frac{1-\epsilon}{\frac{e^{d\epsilon}-1+\epsilon}{\epsilon e^{d}-\frac{1}{d}\left(e^{d\epsilon}-1\right)}e^{d}}.
\]
As $\epsilon\to0$, this approaches the approximation ratio $\frac{1-e^{-d}}{d+1}$
in the monotone $k$-submodular case. This term is minimized if $d$
is the solution to the equation $e^{d}-d-2=0$, which shows Theorem
\ref{thm:ksubmod-knapsack}.

\subsection{Common Cardinality Constraint}

\begin{algorithm}
\textbf{Parameters}: $\left\{ g(i)\right\} _{i\in[n]}$

\textbf{Input}:\textbf{ }monotone $k$-submodular function $f$, common
budget $n$

$\S=\left(S_{1},\dots,S_{k}\right)\gets\left(\emptyset,\dots,\emptyset\right)$

$\beta\gets0$

\textbf{for} $t=1,2,\dots,\left|V\right|$:

$\quad$let $w_{t,a}=\Delta_{t,a}f\left({\bf S}\right)$ for all $a\in[k]$

$\quad$let $a=\arg\max_{a\in[k]}\left\{ \Delta_{t,a}f(\S)-\beta\right\} =\arg\max_{a\in[k]}\Delta_{t,a}f(\S)$

$\quad$\textbf{if} $w_{t,a}-\beta\ge0$:

$\quad\quad$$S_{a}\gets S_{a}\cup\left\{ t\right\} $

$\quad\quad$\textbf{if} $\left|\bigcup_{a'}S_{a'}\right|>n$:

$\quad\quad\quad$let $\left(a',t'\right)=\arg\min_{a\in[k],i\in S_{a}}w_{i,a}$

$\quad\quad\quad$$S_{a'}\gets S_{a'}\setminus\left\{ t'\right\} $

$\quad\quad$let $w(i)$ be the $i$-th largest weight in $\left\{ w_{t,a}\colon a\in[k],t\in S_{a}\right\} $
and $w_{a}(i)=0$ for $i>\left|S_{1}\cup\cdots\cup S_{k}\right|$

$\quad\quad$$\beta\gets\sum_{i=1}^{n}w_{a}(i)g(i)$

\textbf{return} ${\bf S}$

\caption{\label{alg:ksubmod-common-cardinality} Monotone $k$-submodular maximization
under a common cardinality constraint $\left|S_{1}\cup\dots\cup S_{k}\right|\protect\leq n$.}
\end{algorithm}
For simplicity, we only present the algorithm for monotone $k$-submodular
maximization under a common cardinality constraint in Algorithm \ref{alg:ksubmod-common-cardinality}.
As before, we can also adapt this algorithm easily to other settings
discussed in this work. The main difference is that we use a single
threshold $\beta$ which we update based on the weights of items allocated
to all parts. The analysis follows analogously to Theorem \ref{thm:ksubmod}.

\section{\label{sec:experiments-appendix} Additional Experiments}

\begin{figure}
\begin{centering}
\includegraphics[width=0.75\textwidth]{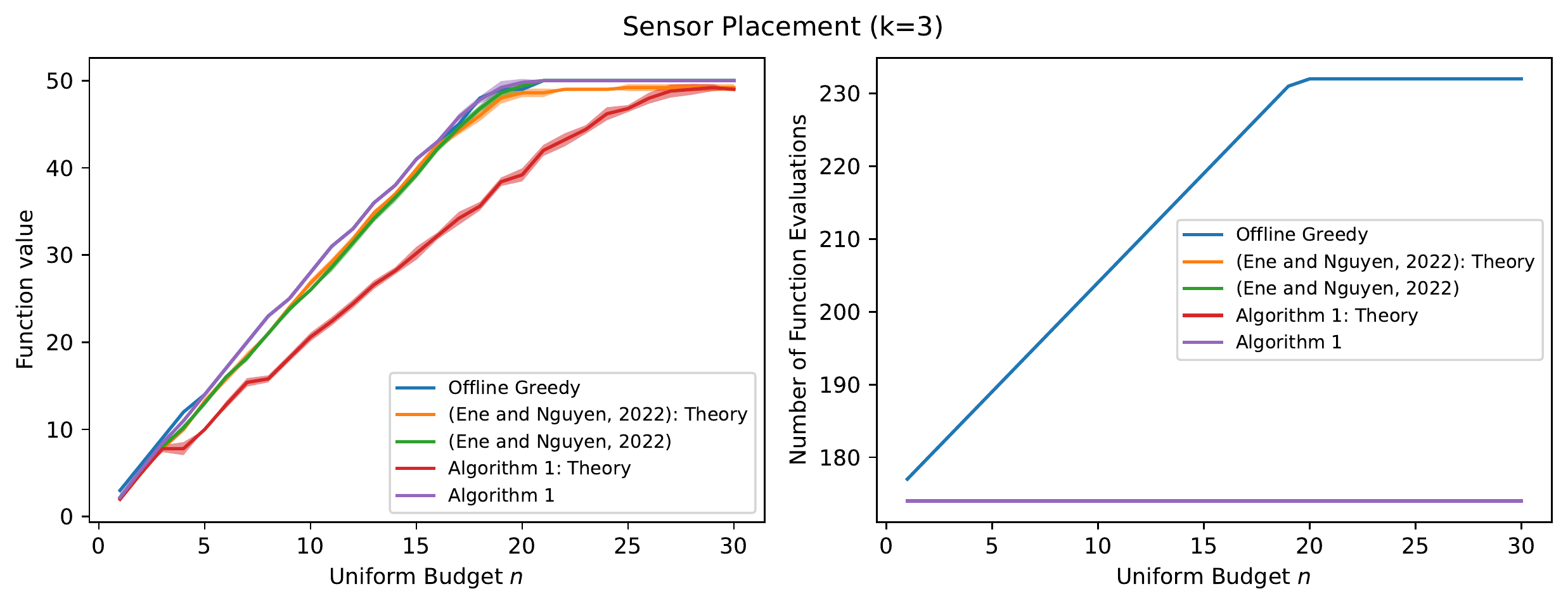}
\par\end{centering}
\caption{\label{fig:icml-experiments-1} Sensor Placement with $k$ Measurements.
We vary a uniform budget $n_{a}=n$ for all $a\in[k]$ and report
mean and standard deviation over 5 runs.}
\end{figure}

In this section, we provide a more detailed description of our experimental
setup and show our results for sensor placement (Figure \ref{fig:icml-experiments-1}).

\paragraph{Ad Allocation}

We consider the problem of allocating ad impressions to $k$ advertisers
\citep{mehta13}. Here, ad impressions $t\in V$ arrive online and
have to be allocated immediately to a single advertiser $a\in[k]$.
We assume that each advertiser $a$ derives value $v_{t,a}\ge0$ from
impression $t$, based on keywords or demographic information. Each
advertiser $a$ is willing to pay for at most $n_{a}$ ad impressions.
We measure advertiser satisfaction through $g_{a}(S_{a})\coloneqq\sqrt{\sum_{t\in S_{a}}v_{t,a}}$.
This function is intended to approximate diminishing returns when
allocating more ads or to enforce a notion of fairness among advertisers,
but not to model any specific real-world scenario. Further, since
$g_{a}$ is the composition of a concave and linear function, it is
also submodular. Our goal is to maximize total advertiser satisfaction
$f(\S)\coloneqq\sum_{a}g_{a}(S_{a})$ while charging each advertiser
for at most $\left|S_{a}\right|\le n_{a}$ ad impressions. 

We use data from a Yahoo dataset \citep{yahoo} and from the iPinYou
ad exchange \citep{zhang14}. We replicate the setup of \citet{lavastida21}
and \citet{spaeh23} to obtain advertiser valuations. Specifically,
the Yahoo dataset yields instances for multiple days where ad valuations
and supply are decided based on the advertiser showing interest into
a keyword. All valuations are in $\{0,1\}$. In order to run the baseline
offline algorithm in reasonable time, we cap the supply of each type
to at most 100 impressions which leaves us with $\approx8500$ instances
per day. Furthermore, we consider only $k=20$ advertisers on $7$
days. The iPinYou dataset contains bids from $k=301$ advertisers
for each impression, which we use as advertiser valuations. We use
the first $3000$ impressions, for each of 7 days. 

\paragraph{Max-$k$-Cut}

In the Max-Cut problem, we are given a graph $G=(V,E)$ and want to
find a subset $S$ maximizing the cut size $\delta_{G}(S)\coloneqq\left|\left\{ \{u,v\}\in E:u\in S,v\notin S\right\} \right|$.
In Max-$k$-cut with cardinality constraints, we are trying to find
$k$ disjoint subsets maximizing the total cut size $f(\S)\coloneqq\sum_{a\in[k]}\delta_{G}(S_{a})$
such that $\left|S_{a}\right|\le n_{a}$ for all $a\in[k]$.  It
can be easily verified that $f$ is non-monotone $k$-submodular. 

We use the Email network from the SNAP database \citep{leskovec14}.
The network contains a total of 1005 nodes and 16706 edges. We use
$k=42$ parts. 
\end{document}